# The Flight Physics Concept Inventory:
## Development of a research-based assessment instrument to enhance learning and teaching

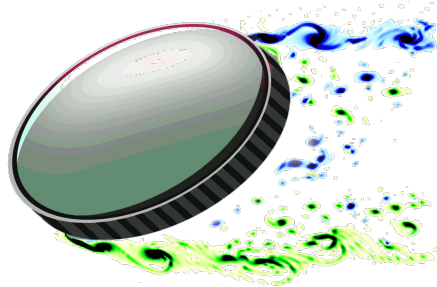

A thesis submitted in fulfillment of the requirements for the degree of
*Doctor rerum naturalium* (Dr. rer. nat. / Ph.D.)
at the
Faculty of Mathematics and Natural Sciences,
University of Cologne
by
Florian Genz
born in Bergisch Gladbach, Germany
(near 50°57'43.7"N, 7°10'31.8"E)

**Contact via:**
florian.genz@alumni.uni-koeln.de

**News & new versions of the Flight Physics Concept Inventory (FliP-CoIn):**
https://www.physport.org/assessments/FliPCoIn
https://flip-coin.uni-koeln.de

<space state="preserved">First reviewer:        Prof. Dr. André Bresges
Second reviewer:       Prof. Dr. Andreas Schadschneider

## Abstract

This work frames the first three publications around the development of the *Flight Physics Concept Inventory* (FliP-CoIn), and elaborates on many aspects in more detail. *FliP-CoIn* is a multiple-choice conceptual assessment instrument for improving fluid dynamics learning and teaching. I give insights into why and how *FliP-CoIn* was developed and how it is best used for improving conceptual learning. Further, this work presents evidence for several dimensions of FliP-CoIn's validity and reliability. Finally, I discuss key insights from the development process, the data analysis, and give recommendations for future research.

## Personal motivation & acknowledgements

The motivation for this work was nurtured from three sources: Science, education research, and love – or more concise: the connectedness to humans and the urge for a sustainable future. In a world of global warming with collapsing ecosystems and drastically changing climatic conditions unprecedented in human history, it is essential to find tools for priming and preparing *paradigm change*.

> How can *efficiency* values evolve to *sufficiency*?
> How can resource *accumulation* shift to *inner growth*?
> How can *sustainability* become the first principle for all conscious players in the ecosystem?

My answer to all questions above is: Adopting deep *conceptual change*. But getting used to, and essentially living, constant *conceptual change*, will be key to the society of the *Anthropocene*. The context and area of where we start to adopt *conceptual change* is not even of first priority here, only the scalability and speed of dissemination of *conceptual change* will decide the fate of humankind.

Global warming is currently one of the biggest, if not, *the* biggest threat for a good life of humanity. Big and efficient levers are necessary to solve this ecological, hyper-complex problem. One lever might become solving the millennium problems in mathematics. The $1.000,000 millennium prize[1] for the *Navier-Stokes existence and smoothness problem* reflects the importance that fluid dynamics have for our global society. A big portion of the global greenhouse gases are emitted in the transportation sector[2], essentially trying to overcome or to use air and water friction. Due to its worldwide importance, even small gains in our understanding of fluid dynamics can lead to enormous greenhouse gas savings, lower resource needs and higher efficiency for generating electric power. Since 201 years (1822) scientists are working on developing and solving the *Navier-Stokes equations*. If one genius, or probably a team of geniuses, is ever to solve them, it will probably happen in an educative ecosystem where *conceptual change* is valued high and adopted on a regular basis. With this work, I will try to do my part towards a society that values *conceptual change*.

I like to thank many persons, that got me and this work where I am now. **André Bresges** who provided patient and valuable feedback of all kind in all phases of this long project. **Andreas Schadschneider** for inspiring discussions and valuable feedback in important moments. **Kathleen Falconer**, who guided me through ample difficult phases and untangled my brain knots so many times. **Rebecca Vieyra**, who got me started at light speed and orchestrated a motivating meeting with **Charles Bolden** at NASA headquarters. **Eleanor Sayre** and **Scott Franklin** for building the **PEER\* community** (*Professional Development for Emerging Education Researchers*[3]). **Ben Archibeque, 李甜甜** and **Pierre-Philippe Ouimet** for keeping me going and also for conducting insightful *equity research* with me. **Dan MacIsaac** for thinking ahead, polishing many projects, and for teaching me important lessons – not only in science. Beyond, I feel deep gratitude towards **Cyrill Slezak, Fritz Kießling, Claudia Ziller**, **Meliha Avcı, Jakob Cordes, Cristal Schult, Christine Quambusch, Stefan Hoffmann, Jacob Beautemps, Bianca Bohn, Maik Schössow, Simon Höfting, Carina Schatz, Jannik Henze, Sascha Therolf, Christian Salinga, Michael Klaas, Simon Loosen, Alexander Rubbert, Sharleen Flowers, Zachary Hazlett, Jennifer Knight, Melissa Dancy, Jon Gaffney, Maria Damas, Sam McKagan, Adrian Madsen, Brian William Lane, Scott McComb, Alfred Ziegler, Graham Wild, Cornelia Genz, Zoë Bohlmann, Christopher Hass, Kristina Vitek, Martina Kramar, Charlotte Kramer, Simon Wessel-Therhorn, John-Luke Ingleson, Irina Lemm, Κάτια Μάρη, Caroline Böning, Kate Godfrey, Liana Kindermann and Rainer Zimmermann** for fostering my data collection, valuable feedback, inspiring discussions, emotional support and important hints at the right moment. This work became significantly better because of you.

---

[1] https://www.claymath.org/millennium-problems/navier%E2%80%93stokes-equation [2023-05-10]

[2] https://www.epa.gov/ghgemissions/global-greenhouse-gas-emissions-data [2023-04-18]

[3] https://peerinstitute.org

I also want to thank all the past and current researchers who trailblazed the path for me – not only the ones that I cited. Here I am especially thinking of the scientifically rigorous but, on a personal level, warm and welcoming research community of the *American Association of Physics Teachers* (AAPT) and the *Deutsche Physikalische Gesellschaft* (DPG).

Finally, one of the most important supports came from my **family** and **friends**. Not only when directly dealing with my doubts and tempers, but also saving me time, caring in many ways, and, last but definitely not least, letting me stay a curious child. This work is also yours!

# Table of Contents

# 1 Introduction

The field of *flight physics* is a complex and ever-evolving area of study, making use of ample challenging and interwoven concepts that students have to master, in order to succeed in their aviation or engineering careers. Due to the huge impact of *flight physics* on many important fields, one could be tempted to assume that evaluation instruments for teaching methods in *flight physics* are much more advanced. However, this is not the case. At the time this research project started, not a single Concept Inventory for *flight physics* existed. Concept Inventories or *research-based assessment instruments* (RBAIs) (Madsen et al., 2017b) were motivated by the observation that learners perform better or worse depending on the teachers' methods, and that exam scores do not seem to be a good predictor for the better methods. The body of research is full of studies demonstrating that typical exams show little to no test fairness[4] (Brügelmann, 2006; Educational Testing Service, 2008), varying validity (Lindell et al., 2006), and low scoring consistency, as well as grade inflation (APA Handbook, 2013b). Hence, traditional knowledge-based exams are poor instruments for reliably indicating the better or worse teaching method. This is where Concept Inventories came into play.

Since the first publication of a Concept Inventory, called the *Force Concept Inventory* in 1992 (Hestenes et al., 1992), the physics education research community has developed many similar assessments for different contexts, and created platforms for their dissemination[5]. This fact reflects the importance of *research-based assessment instruments* for the learning and teaching culture. In 2011 Richard Hake conducted a review of the impact Concept Inventories had on physics education and related disciplines (R. Hake, 2011). However, a Concept Inventory for the physics of aerodynamic flight was yet missing.

The reasons for this are various. Most Concept Inventories only cover introductory topics and less advanced topics. Also, flight physics touches on many different classical canonical topics like mechanics and fluid dynamics. Hence, *flight physics* tends to fall in between these topics and be cut short. Therefore, this work will frame the development of the new *Flight Physics Concept Inventory* (FliP-CoIn). As every Concept Inventory, the FliP-CoIn instrument is a tool for teachers who intend to assess the quality of their teaching, and to inform students about their conceptual learning gains, learning gaps, and general mastery of the subject. This teaching tool provides a comprehensive assessment of student understanding in key flight physics concepts, using a combination of *multiple-choice forced-select questions*, and *ranking tasks*.

Unique to the FliP-CoIn development process is that it was developed simultaneously in English and German. This entailed a lot more pitfalls and challenges to be overcome, but it essentially inspired better items and a precise wording that works in both learning cultures. For example, the term "*banking angle*" is well understood by students in the United States, due to NASCAR racing. But the German word "*Rollwinkel*" is very rare, and considered *special vocabulary*. A direct translation without considering the cultural and language differences would lead to different item difficulties, and, hence, a less valid instrument.

---

[4] test fairness: the test does not discriminate or disadvantage a specific group. For example by gender, socio-economic status, academic background, migration background, language background or and other criteria violating the UN Sustainable Development Goals 4, 5 and 10: https://sustainabledevelopment.un.org

[5] E.g. : https://physport.org & https://learningassistantalliance.org/public/lasso.php

This PhD thesis will present an evaluation of the newly developed Concept Inventory for flight physics. I will examine various dimensions for reliability, validity, and the overall effectiveness of the tool in assessing student understanding. In this work, I will elaborate on how the FliP-CoIn instrument was developed and validated in more detail than in the accompanying three publications (→Section 7 - Contributions and role of accompanying publications). In addition to the accompanying publications, this work intends to describe the development process of the FliP-CoIn instrument in more detail for enabling others to replicate or build on my work.

In addition, I will explore the potential of the Concept Inventory to improve the teaching and learning of *flight physics* in both, the classroom and the underlying theory. This thesis is organized into four longer main sections. Section 2 gives an overview about the state of the art of Concept Inventory development research and relevant literature. Section 3 describes the methods used and provides reasons concerning the most appropriate methodology for the FliP-CoIn instrument. Section 4 describes the results of the statistical analyses – mainly focusing on representability and demographics, the reliability analysis, descriptive results and indicators for *content validity*. Finally, Section 5 discusses key findings and outstanding results.

## 1.1 Motivation for FliP-CoIn: Navier-Stokes equations & Nature of Science

The flow of any viscid fluid – such as air or water – is mathematically well described by the Navier-Stokes equations (McLean, 2012; Tong, 2023). However, it is not yet mathematically proven that smooth solutions always exist in 3-dimensional space[6]. This implies: Flow conditions could – in theory – change drastically with just one condition changing minimally (butterfly effect), and revert to the original flow pattern with another minimal change (e.g., humidity rising from 50.1% to 50.2% and then 50.3%)[7]. One should keep in mind that *any* concept can only be an empirical approximation of the real-world phenomenon as long as the *existence and smoothness problem of the Navier-Stokes equation* is not yet mathematically proven, and analytical solutions are covering all relevant cases.

This is not only a mathematical problem, but also a problem on the conceptual side, since there is no absolute reference (=equation solution) that could neutrally judge the applicability and accuracy of different mental models. As long as our understanding of the Navier-Stokes equation (Tong, 2023) is not sufficient enough for a wide range of analytical solutions, exact predictions cannot be made here. This implies: The relevance, predictive power, and applicability of *any* concept in *flight physics* can be challenged and is up for negotiation.

---

[6] https://www.claymath.org/millennium-problems/navier%E2%80%93stokes-equation [2023-05-10]

[7] However, it is not likely that huge jumps will be found for macroscopic objects of everyday life (like airplanes, cars, wind generators, propellers, etc.). The averaged aerodynamic forces (lift and drag) are thoroughly measured experimentally and can be approximated by computer simulations fairly accurate – in most cases.

However, the concepts probed in the *Flight Physics Concept Inventory* are so basic and empirically verified that one can clearly score the questions as expert-like or naïve. Beyond that, the *Flight Physics Concept Inventory* does not try to challenge the applicability of the Navier-Stokes equations. Instead, it tries to improve our understanding of fluids from the other side: The learning and teaching of concepts. This Concept Inventory was designed to help elicit the logical inaccuracies of student concepts, and inform teachers about which mental models are more helpful than others. The Concept Inventory tries to answer questions like: What implicit knowledge and assumptions do I hold? Which (naïve) concepts and explanations should I contrast with my students? Which perspectives and contexts do they prefer and adopt more easily? Gaining more consciousness about these questions can inspire new and better teaching interventions, and might be especially powerful in teaching the *nature of science* (Galili, 2019).

The *Flight Physics Concept Inventory* can help distill the better learning techniques, improve conceptual understanding, and improve student understanding about the *nature of science*. I hope this will contribute to an *ecosystem of knowledge* in which not only the Navier-Stokes equations become solvable one day.

## 1.2 Purpose and limitations of FliP-CoIn

Concept Inventories are *formative* assessments, and can be a valuable tool for teachers as well as students to gain understanding in their individual teaching/learning process. A further purpose of test results is that students and teachers develop a better intuition for *learning gaps,* and how to give feedback about the conceptual understanding of their students. Beyond that, they can learn about the development of their own understanding. In the best case, this leads to more effectiveness of a learning intervention and all future interventions.

Another motivation that sparked the surge of Concept Inventories derives from a lack of standards and comparability: Much of the differences between teacher grades and assessment performance can be explained by the fact that teachers use their *own* scale. Teachers seem to be quite accurate in rating the competencies of their students, but give different grades for the same level of competency when compared to other teachers (Brügelmann, 2006). Therefore, it is no surprise that an earlier, more rigorous, or competitive selection of students by grade does not necessarily lead to better performance – as international education studies have shown (Brügelmann, 2006).

Hake, as well as Sands and colleagues, emphasize that Concept Inventories should *not* be used for high-stakes exams that are granting or denying privileges. Such usage would violate the validity of the test (R. Hake, 2011; Sands et al., 2018). When state exams are deciding about cashflow for schools, the educators might become seduced to deviate from the prescribed guidelines for a thorough utilization of the Concept Inventory at hand (Madsen et al., 2017a) in order to "*produce*" bigger learning gains in the statistics. At best, by leaving out all other curriculum topics not covered in the Concept Inventory. At worst, by direct *teaching-to-the-test strategies*, giving the Concept Inventory short before, or even giving away the solutions (slate faking of results). Further, when incentives and/or consequences become too high, not only educators, but also students are tempted to try "alternative" success strategies.

Concerning limitations, I like to add: This work will only address aerodynamic theories here as far as they are influencing naïve student concepts in the context of the FliP-CoIn instrument. It does not claim to cover a complete set of all flight physics contents. To understand these choices and limitations made in the development of the FliP-CoIn instrument, the following Section (→Section 2) will sketch the evolution of Concept Inventories and concepts related to *flight physics*.

Even though flapping flight of insects and birds (Feldman et al., 2006; Heavers & Soleymanloo, 2011; Mihail et al., 2008; P. L. Richardson, 2017; Tennekes, 2001) is another motivational context for students, it is not covered in the *Flight Physics Concept Inventory* in order to reduce its complexity and length. Another reason for the exclusion of flapping flight was the fact that there has been recent progress in understanding the modeling methods for flapping flight (Xuan et al., 2020), which might impact the concepts involved and expert discourse. Also, supersonic and helicopter flight are purposely left out by this work.

## 2  State of the art: literature review

This Section will give an overview about the most important literature for the development of the *Flight Physics Concept Inventory*. It starts by elaborating on the *research gap* and the lack of Concept Inventories for *flight physics*, then summarizes premises for conceptual change and the general motivations for *concept inventory development*. In chapter 2.1.1 I describe important definitions and building blocks for my understanding of concepts. Chapter 2.1.2 describes what kind of assessment a Concept Inventory is, and mentions different areas of usage. Chapter 2.1.3 elaborates on Concept Inventories in other languages and the uniqueness of the development process concerning the *Flight Physics Concept Inventory*. Chapter 2.1.4 sketches some important historical backgrounds and the implications for the German education system. The longer Section 2.2 named "*Flight physics related concepts*" was divided into three subchapters elaborating on different explanations of lift (→2.2.1), and their shortcomings (→2.2.1.1). Finally, problems with the free body diagram of airplanes are primed (→2.2.2) which also became an important part of the discussion (→5.5.2):

The *lack of research* for student understanding concerning *flight physics* is huge, since no tools to elicit them exist yet: Neither in recent standard physics education literature (Chaichian et al., 2014; Schecker et al., 2018) nor the biggest Concept Inventory platforms *PhysPort* and *LASSO* (physport.org. & learningassistantalliance.org/public/lasso.php) mention validated test instruments for *flight physics* or *fluid dynamics* (see also Beichner, 2017). Recent overview papers report a lack of research for *research-based assessment instruments* in the field of *statistical mechanics* (Madsen et al., 2017b). *Fluid dynamics* and *flight physics* cover big parts of this field. Previous research from Madsen, McKagan and Sayre shows that many physics faculty are aware that Concept Inventories exist for *introductory physics*, but like to know more about other Concept Inventories for a wider range of topics, including upper-level physics (Madsen et al., 2017b).

Studies looking at teaching methods and Concept Inventories consistently show that research-based teaching methods lead to huge improvements in students' conceptual understanding of physics (R. R. Hake, 1998; Madsen et al., 2017b; Von Korff et al., 2016). Smith and diSessa emphasize that the goal of instruction should not be to simply replace misconceptions for expert concepts, but to provide an experiential basis for the gradual processes of conceptual change (J. P. Smith & DiSessa, 1993). The authors further elaborate that "cognitive conflict" is not a state that leads to the choice of an expert concept over an existing novice conception, but to a more complex pattern of system-level changes that collectively engage many related knowledge elements.

> *"We now need research that focuses on the evolution of expert understandings"*
> (J. P. Smith & DiSessa, 1993)

Psillos and Kariotoglou emphasize that the linking of scientific theories to practical activities is particularly hard for students, and propose the theoretical framework Cosmos–Evidence–Ideas (CEI) to analyse the evolution of didactical activities in teaching-learning sequences in the case of fluids (Kariotoglou & Psillos, 2019; Kariotogloy et al., 1993; Psillos et al., 2004; Psillos & Kariotoglou, 1999). They conclude that scientific inquiry in fluid dynamic lessons is mainly restricted to the representation level and diagnose a lack of intervention practices. In that context, Kariotoglou proposed a laboratory-based teaching learning sequence on fluids for developing primary conceptual and procedural knowledge (Kariotoglou, 2003).

In the late 1990s it was established knowledge that *conceptual change* is key to promote scientific inquiry and methods were developed (Kalman et al., 2004; Kern & Crippen, 2008; Magnusson et al., 1997; Mortimer, 1995). For testing the effectiveness of these new conceptual change methods, it became evident that *research-based assessment instruments* (=Concept Inventories) needed to be developed.

## 2.1 Concept Inventories

> *"The most difficult subjects can be explained to the most slow-witted man [sic!] if he has not formed any idea of them already; but the simplest thing cannot be made clear to the most intelligent man if he is firmly persuaded that he knows already, without a shadow of doubt, what is laid before him."*
> (Leo Tolstoy)

This early observation by Leo Tolstoy primes a central finding of recent *Physics Education Research* (PER) and motivation for Concept Inventories: The fact that one needs to have a cognitive conflict before the mind is ready to integrate new concepts. Derek Muller emphasizes that a good explanation is even worse than no explanation at all, when new information is not contrasted against the existing (and maybe naïve) pre concepts of the audience (Muller, 2012; Muller et al., 2008). This sounds counter-intuitive at first but makes sense when considering that listeners will cherry-pick what they find familiar in an expert explanation, and become more confident in their existing concepts. Conceptual change is prevented or hampered. Therefore, the consideration of an audience's pre concepts is a critical component of establishing conceptual learning (Heckler & Bogdan, 2018). Especially in fluid dynamics, naïve concepts are ample and have a significant impact on customer choice, and, hence, the design of everyday items such as cars and rooftop boxes, and, ultimately, fuel efficiency[8].

### 2.1.1 Concepts: Definitions and building blocks

The most central aspect of a Concept Inventory are the concepts itself. Therefore, I will elaborate on the most important definitions and relations around concepts as they are understood in physics education sciences: In the 1980s the predominant theory was the *conceptual change model* (Posner et al., 1982). In the 1990s Andrea diSessa and colleagues challenged that model, refined it, and coined the term *phenomenological primitive* (*p-prim*), which they see as the base of concepts (DiSessa, 1988, 1993). *P-prims* are ideas and cannot be learned explicitly. Also, p-prims are neither correct nor incorrect within themselves. P-prims can only be *applied* (in)appropriately in a given context (Harlow & Bianchini, 2020). Hammer built on diSessa's *knowledge-in-pieces idea* (Hammer, 2004) coining the terms *resources* and *manifold ontology*. In Hammer's work, *resources* are used almost synonymously to diSessas *p-prims* and *manifold ontology* can be summarized as the understanding of the many-part entities existing in one's mind. Based on that, Hammer and Sikorski criticized existing learning progression frameworks as oversimplistic (Hammer & Sikorski, 2015). They argue that existing learning progression frameworks dismiss variations in students' ideas as conceptually insignificant noise and recommend teaching methods that more closely attend to students' prior ideas.

In this work, I will use the term *concept* to refer to the higher level-thinking structure emerging from the use and application of *p-prims/resources* in a given context. Or, defined reversely, one could see *p-prims/resources* as the building blocks for *concepts*. The sub terms *naïve concept, misconcept, alternative concept, pre concept, prior* or *early concept* will be used more or less interchangeably and treated as synonyms. However, there is an important nuance between the two terms *naïve concept / misconcept* and all the other terms. The prefixes "*naïve*" or "*mis*" contain a judgment in the sense that they are less applicable or correct. So when communicating to students, avoiding these judgmental terms can be advantageous at first. However, for the later discussion, written for the research community, I will use these judgmental terms frequently to enhance the clarity of expert positions. The definition of the central term *naïve concepts* in this work is: Ideas about the world that are inconsistent with expert ideas (McCloskey, 1983).

### 2.1.2 Assessment type and usage domain of Concept Inventories

A Concept Inventory is a multiple-choice research-level instrument designed to test students' conceptual understanding (Lindell et al., 2006; Sands et al., 2018), teachers' pedagogical content knowledge (Maries & Singh, 2019) or course evaluation (R. R. Hake, 1998). Based on a number of key concepts from the subject (Jorion et al., 2015), each question, or item, has one or more correct answer(s) and a number of incorrect answers, known as distractors. Optimally, distractors are based on common student misconceptions (Sadler et al., 2009).

---

[8] https://teslamag.de/news/komisch-effizient-tesla-model-3-dachbox-falsch-herum-montiert-27673 [2023-05-02]

Concept Inventories are a special kind of assessment that aims to elicit and confront students with their implicit assumptions and concepts about the world. Therefore, Concept Inventories need to be formative assessments for students and teachers – given with little or no incentives or consequences. Under these conditions, Concept Inventories can be a valuable tool to evaluate the following questions:

1. Where do I have knowledge gaps or alternative conceptions that differ from expert concepts?
2. How do I learn new concepts? Where has my learning been effective? (for students)
3. Which teaching method or educational ecosystem works best for my students? (teachers)

As formative assessments, Concept Inventories unfold their learning potential after they are conducted - but educational context and implementation matters. As the inventors of the FCI emphasize: Their Concept Inventory is not a test of ability. The applicability of Concept Inventories can be thought of in three main categories (Hestenes et al., 1992):

a) As a diagnostic tool to identify and classify naïve, alternative, students or mis-concepts.
b) For evaluating own instruction.
c) As a placement exam: limited value depending on for which target group the Concept Inventory was developed. Also, the context and stakes connected with the testing outcomes need to be considered. A Concept Inventory is not an ability assessment.

Thinking beyond these uses, the use of Concept Inventories for high-stakes exams (e.g. entrance examinations) or governmental assessment of schools and whole districts contains great risk for educational short-term survival strategies like rote memorization techniques ("bulimic learning") for students (Hilgard et al., 1953), or teaching-to-the-test temptations for teaching personnel. This is especially true when funding depends on these outcomes.

### 2.1.3 Bilingual development & physics Concept Inventories in other languages

There are several translations for Concept Inventories in physics available at PhysPort.org. However, these translations vary in linguistic quality and cultural understanding, as well as the physics education background of the translator. Only a few Concept Inventories on PhysPort.org were re-evaluated in the translated version – according to own investigations. Some re-evaluations found differences between two language versions.
No Concept Inventory yet has been developed bilingually from scratch. This makes the development of the *Flight Physics Concept Inventory* unique (see also Appendix →9.7).

Further, the finding, that no other bilingually developed Concept Inventory exists yet, was backed up by a search in the scientific search engines ERIC and Google Scholar (https://eric.ed.gov , https://scholar.google.de, 2022-01-15), returning no result for Concept Inventories or conceptual assessments which were truly developed in more than one language. Inspired by the PRISMA statement for systematic review the following search terms were used (Liberati et al., 2009):
- bilingual AND concept AND inventory

- "English and" AND concept AND inventory
- "and English" AND concept AND inventory
- "English language and" AND concept AND inventory
- "and English language" AND concept AND inventory
- bilingual AND conceptual AND inventory
- "English and" AND conceptual AND assessment
- "and English" AND conceptual AND assessment
- "English language and" AND conceptual AND assessment
- "and English language" AND conceptual AND assessment

### 2.1.4 Historical development and critiques

The first Concept Inventory (Force Concept Inventory) was motivated by the observation, that students could state Newton's laws but not apply them in problem-solving (Hestenes et al., 1992). This instrument started its development in 1985 where Halloun and Hestenes published an Instrument measuring conceptual understanding in mechanics (I. A. Halloun & Hestenes, 1985; Härtig, 2014). The success of Concept Inventories in physics has spawned Concept Inventories in chemistry, biology, astronomy, materials science and mathematics (Sands et al., 2018). The official funding of Concept Inventory platforms like PhysPort or LASSO (https://physport.org , https://learningassistantalliance.org/public/lasso.php) are the latest peak of this development.

However, success never comes without critiques. Constructive critiques improved and fostered the development of improved Concept Inventory versions and new CIs (Dietz et al., 2012; Eaton et al., 2019; I. Halloun & Hestenes, 1996; Heller & Huffman, 1995; C. Henderson, 2002; R. Henderson et al., 2019; Hestenes et al., 1992; Hudson & Munley, 1996; Huffman & Heller, 1995; Lasry et al., 2011, 2012; Laverty & Caballero, 2018; T. F. Scott & Schumayer, 2019; Wallace, 2012). There is also a more general testing critique emphasizing the following point: Traditionally, science-based assessment instruments focus on the three criteria

1. Objectivity
2. Reliability
3. Validity

In contrast to that, teachers and other practitioners in education focus more on fairness within the group, economy, usability, and on credibility as well as coherence (Brügelmann, 2006).

One of the latest developments for more test fairness seems detectable by the fact that the research community became widely interested in gender differences in Concept Inventories and physics (Coletta et al., 2012; Dancy, 2000; Dickens et al., 2020; Dietz et al., 2012; Docktor & Heller, 2008; Kalender et al., 2020; Kost et al., 2009; Madsen et al., 2013; C. T. Richardson & O'Shea, 2013; Traxler et al., 2018).

#### 2.1.4.1 *Historical implications in Germany*

In Germany, the teaching of *flight physics*, *fluid dynamics* and neighboring disciplines was banned by the *Allied Forces* and later tabooed for the years to follow the *Second World War*. Despite its importance and relevance, *flight physics* never recovered fully into German curricula. To recover from that historic cesura, the *Deutsche Physikalische Gesellschaft* (DPG) reintroduced *fluid mechanics* as a *base concept* and *flight physics* as a possible context in their curricular recommendations in 2016 (DPG, 2016a, 2016b) The lack of curriculum coverage of *flight physics* probably contributed to the rise of *alternative concepts* (also called naïve, non-expert, alternative, pre or student concepts) in that domain.

## 2.2 Flight physics related concepts

Before diving into the methodology of the instrument's development, I would like to give the reader some insights which historical, scientific, and pedagogical problems motivated the scope of the *Flight Physics Concept Inventory*. I will focus on the concept of aerodynamic lift and problems with the free body diagram, since many naïve concepts are found in these domains or even derive from them.

### 2.2.1 Explanations of aerodynamic lift

Water and air are both viscid fluids, which are very similar to describe in physics. In both fluids *laminar* or *turbulent* flow can be observed. Both of these flow states have very different properties but before Ludwig Prandtl came around with the new *boundary layer* or *transition layer concept* in 1904 (Prandtl, 1905) these relatively distinct flow regimes were rather unknown and the theory concerning aerodynamic lift was not well grounded in physical principles. In short, Prandtl's boundary layer concept assumes

> "*the no-slip condition at the surface —and that frictional effects were experienced only in a boundary layer, a thin region near the surface. Outside the boundary layer, the flow was essentially the inviscid flow that had been studied for the previous two centuries*" (J. D. Anderson, 2005)(→ Fig. 1).

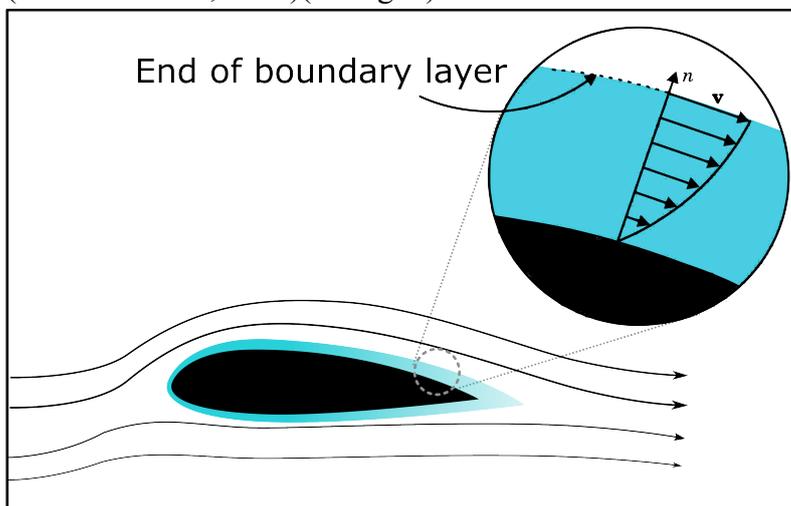

*Fig. 1: Ludwig Prandtl's boundary layer concept: A fluid flow consists of two flow regimes: be viewed as comprising two parts. Within the thin boundary layer adjacent to the surface (light blue), the effects of friction are dominant. Outside the boundary layer, the flow is inviscid, and friction does play a minor role* (adapted from J. D. Anderson, 2005)

In hindsight, it seems surprising that when Ludwig Prandtl held his revolutionary ten-minute presentation about the new *boundary layer* or *transition layer concept* in 1904 (Prandtl, 1905) and published it in the accompanying proceedings, it went fairly unrecognized at first (J. D. Anderson, 2005). It took almost two decades and further publications of Prandtl and his former students Heinrich Blasius and Theodore von Kármán until the boundary layer theory grasped wide attention and inspired engineering practice. This, despite the fact that Prandtl's work solved a long dispute about the role of viscosity in flowing liquids.

This example should demonstrate that it is often difficult for new, revolutionary concepts to emerge, but also how important it is to create a culture where mental concepts and ideas can float freely and inspire to solve complex phenomena. The generation of aerodynamic lift is such a complex phenomenon – even for simple cases. Therefore, it should not surprise that misconceptions concerning aerodynamic lift are ample even though the theory of aerodynamics is well described by the Navier-stokes equations and the numeric applications are well established in economy, there is little consensus about the interpretation of the concept of aerodynamic lift (McLean, 2012). It is simply hard to understand, interpret and communicate complex phenomena. Beyond, inaccurate language use is fostering misconcepts especially well in this context, and it is very common in aerodynamics teaching. For example: NASA's Glenn Research Center states on one of their websites: „*The magnitude of the lift generated by an object depends on the shape of the object and how it moves through the air*"[9]. This statement is not completely wrong, especially not, since angle of attack – usually the bigger contributor to lift -– was mentioned before. But students with the naïve concept *"only the shape of an object is responsible for lift"* will become more confident in their naïve concept after reading this sentence. Therefore, knowing and contrasting the learners' misconceptions is essential to teaching complex phenomena (Muller, 2008; Muller et al., 2008)

Extensive work was done by the research community concerning naïve concepts in the domain of aerodynamic lift and learning interventions have been designed accordingly (Babinsky, 2003; Schecker et al., 2018; Send, 2001; N. F. Smith, 1972; Vieyra et al., 2015; Waltham, 1998; Weltner, 1987, 1990b, 1990a; Wodzinski, 1999; Wodzinski & Ziegler, 2000). Dough McLean elaborated with great rigor and scientific detail on almost every naïve concept concerning aerodynamic lift in the book "Understanding Aerodynamics" (McLean, 2012) and further journal articles (McLean, 2018a, 2018b). In their book, "Understanding Flight", David Anderson and Scott Eberhardt worked out many important basic principles of flight, including the crucial difference between an accurate concept of how the wing transfers gravitational momentum away from the plane and the false notion that the earth would essentially "support the airplane" (D. F. Anderson & Eberhardt, 2010).

Insightful basic research was also done by Besson concerning students' conceptions and reasonings about fluids in static situations by surveying N=924 pupils, students and teachers-in-training of three European countries (Besson, 2004). Some of the naïve concepts in dynamic situations might go back to naïve concepts in fluid statics, as shown in the later discussion.

### 2.2.1.1 Shortcomings of existing explanations and supporting visualizations

Still recent educational standard literature fails to explain the concept of aerodynamic lift concisely, while accounting for and contrasting historically grown misconceptions: For example the Pilot's Handbook of Aeronautical Knowledge (FAA, 2016) in chapter *"Theories about*

---

[9] https://www.grc.nasa.gov/WWW/K-12/airplane/incline.html [2020-07-14]

*the production of Lift*" does *not* mention a definition of aerodynamic lift, but worse, states the following sentence:

> "*Lift from the wing that is greater than the force of gravity, directed opposite to the direction of gravity, enables an aircraft to fly.*" (FAA, 2016)

Even though, strictly speaking, this statement is not wrong, it still fosters the misconcept that "aerodynamic lift *always* points upwards" and "lift must be bigger than weight during flight". The first statement is only correct for horizontal steady flight. The second statement is simply wrong for steady climb flight (→Section 5.5.2), banked flight, and/or descending flight states. The FAA book continues to mention Newton's Laws and the Bernoulli principle in general terms, without explaining *how* these two should actually contribute to lift. The only sentence dealing with this – indirectly – is:

> "*As the wing moves through the air, the flow of air across the curved top surface increases in velocity creating a low-pressure area.*" (FAA, 2016)

This suggests to students: Curvature would be most important to generate lift. (Fosters misconcept: "the airfoil shape *alone* generates lift"). Further, it gives rise to the false notion that curvature of the top surface would increase the velocity difference. The importance of an angle of attack (AoA), or the displacement of air, is neglected. Another implicit assumption of this statement might be: Air particles from the upper and lower surface would rejoin at the tail edge of the airfoil. These misconcepts are reconfirmed when the reader stumbles on the following:

> "*Notice that there is a difference in the curvatures (called cambers) of the upper and lower surfaces of the airfoil. The camber of the upper surface is more pronounced than that of the lower surface, which is usually somewhat flat.*" (FAA, 2016)

From a scientific perspective, it is more than questionable, and from a pedagogical standpoint, it is simply wrong to bend analogies in such simplification to the Bernoulli principle for explaining aerodynamic lift.

> "*However, my concern about using this explanation is that it introduces misconceptions about why aerofoil shapes generate lift, it uses a nonsensical physical argument and it often includes an erroneous 'demonstration' of Bernoulli's equation.*" (Babinsky, 2003)

In fact, the low-pressure areas on the top side of a wing also occur with angled flat plates in a flow. For a slightly angled flat plate, the Bernoulli principle alone would even predict *slowing down* air particles above the wing (→Fig. 2, blue area).

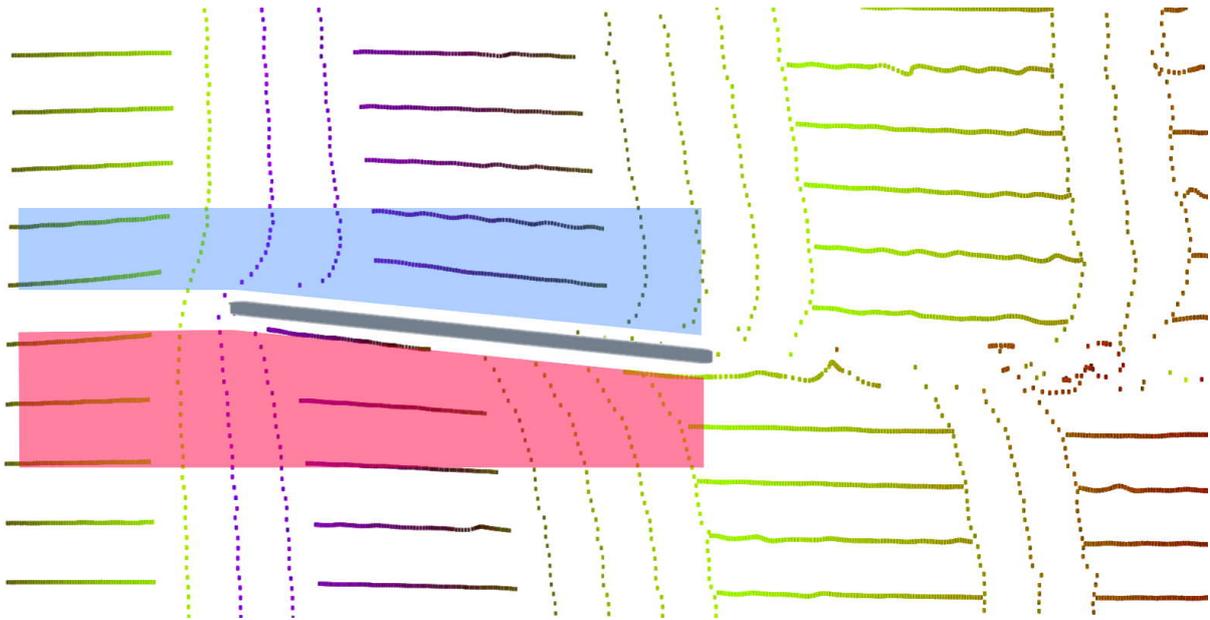

*Fig. 2: The Bernoulli principle analogy fails to explain the pressure and velocity distribution for the given airfoil. In a learner's mind, the bottleneck is under the wing, and air should accelerate there. Above the airfoil, the imaginary nozzle widens, pressure should increase and horizontal particle velocity should decrease. This contradicts qualitative observations and measurements. The colored dots represent selected air particles that were marked at the same time when appearing on the left side of the screen. The flow direction is left to right.*

Drawing parallels to the Bernoulli principle suggests to learners that there must be a bottleneck and/or nozzle somewhere for the air (Send, 2001). This has been shown to be incorrect (McLean, 2012)[10]. However, (wrong) analogies are sticky (Muller, 2008) and learners may still try to rescue the Bernoulli analogy by imagining that the influence area of the imaginary nozzle extends in front of the airfoil (→ Fig. 3).

At least, this mental model results in a high velocity above the leading edge and a lower velocity below the leading edge and, hence, is much closer to the observable phenomena. However, it still fails to explain why air particles above the wing keep accelerating *along* the airfoil and also why the air below keeps lagging behind relative to the air above.

---

[10] See also: https://www.grc.nasa.gov/WWW/K-12/airplane/wrong3.html [2020-07-14]

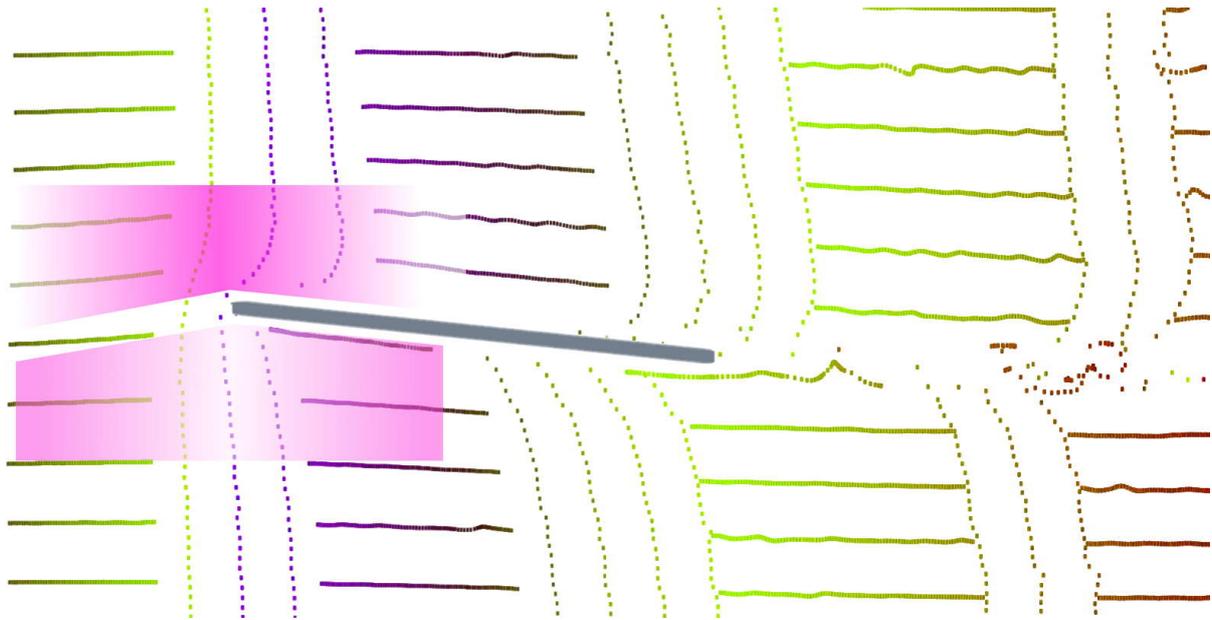

*Fig. 3: Bernoulli principle analogy (part2): Despite contradicting observations, learners may try to rescue the incorrect Bernoulli principle analogy by expanding an imagined influence area of the airfoil way before the air arrives at the leading edge. An increased velocity above the leading edge (pinker) and a decreased velocity below the leading edge (less colored) is the result. This is closer to the observations but still incorrect – especially for the observations along the wing.*

The increase of air speed above the wing can be more intuitively explained to students by the asymmetry of a particle's mean free path in the "wind shadow" above the wing (→ Fig. 4):

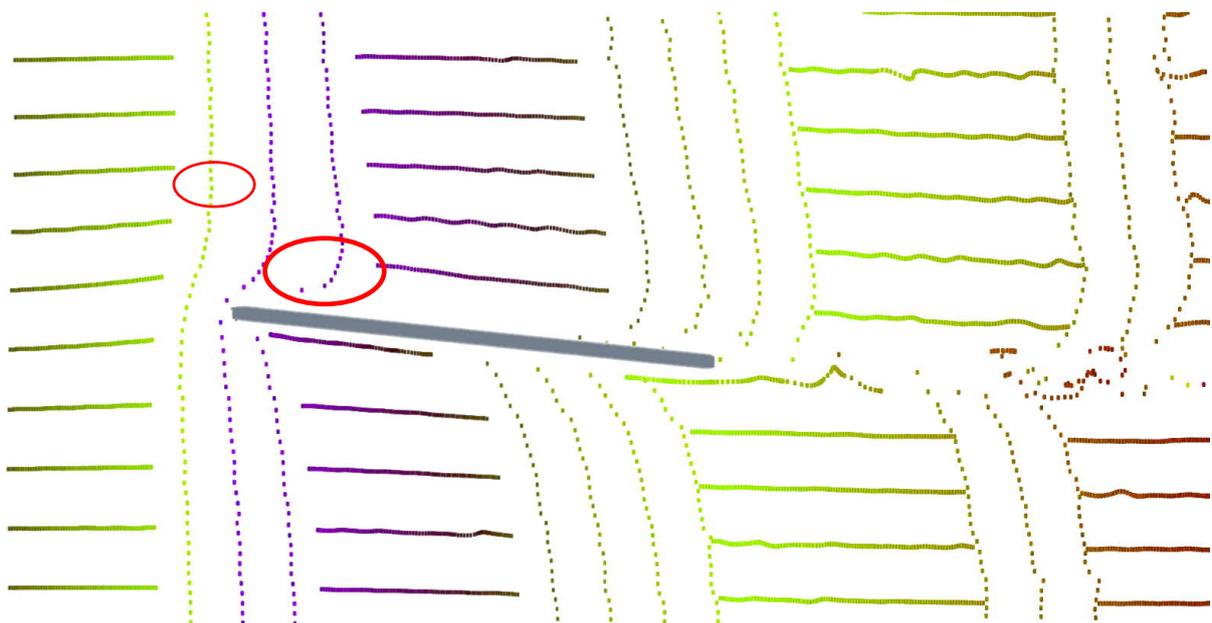

*Fig. 4: Acceleration of particles above a flat airfoil. Often misleadingly explained with the Bernoulli principle. The asymmetry of the mean free path for particles above the airfoil is a more fundamental concept and intuitive for students.*

A particle above the airfoil is more likely to collide with a particle on the left side (or above) than from tailward (or below), and is, therefore, accelerated to the right and down over time (→ Fig. 4). Also, it is intuitively clear to students why particles keep accelerating into low-pressure areas and following the upper wing surface – without even bothering with the concept of viscosity, Coandă effect or boundary friction. Focusing on the mean free path of particles, students can also easily derive why too drastic changes in airfoil curvature / AoA must lead to stall. With high AoAs the collisions with particles from left and top become less probable, and asymmetry of a particle's mean free path shrinks. During my 10-years teaching career, I have made more positive experiences with the mean free path approach when students struggled with explaining the acceleration of the air particles above the wing.

Eastlake highlights that the generation of lift *can* be correctly explained by Newton's Laws as well as by the Bernoulli principle, but also acknowledges that simplifications (for school) might ruin the accuracy of both approaches (Eastlake, 2002). McLean, however, argues that neither the Bernoulli (J. D. Anderson, 1989) nor the Newton approaches alone (D. F. Anderson & Eberhardt, 2010; N. F. Smith, 1972; Waltham, 1998) adequately explain all of the essential cause-and-effect relationships of aerodynamic lift (McLean, 2018b). In fact, the lift force, the pressure field, and the velocity field are in simultaneous interaction with each other (Auerbach, 1988, 2000; Hoerner, 1985; McLean, 2012).

How a wing of an airplane actually generates lift has been theorized, and was passionately discussed in the past, and is still a point of negotiation among scientists, in literature but also in the media with the general public[11]. The confusion about lift also was amplified because in scientific literature different approaches were used that emphasize different aspects of the complex phenomenon. I argue that it is not satisfactory to end the discussion with the conclusion that no model is ever perfect, and that these approaches are simply different perspectives of equal value because some confusion is really based on logic mistakes or semantic inaccuracies. An example: Wolfgang Send once described the generation of aerodynamic lift as follows:

> „*The low pressure on the topside of the airfoil is the result of a [numerical] solution, which occurs when stagnation point line moves below the chord line – in flow direction […]. Therefore, more mass travels over the airfoil and, hence, the velocity must be bigger[12]*" (Send, 2001)

This explanation is problematic for students in many ways. Apart from the complicated grammar and language, it seems to contribute to the "*naïve nozzle theory*" mentioned before. Further, it mixes up and obscures cause and effect vs. mutual influence of the observable phenomena. Beyond that, it leaves students with the question of why the velocity must be bigger on the topside of the airfoil. This provokes circular reasoning in student minds. Later, in the same publication, the author asks the question "*Now, where does lift come from?*" and only mentions the pressure distribution on the border between boundary layer (with significant viscosity) and peripheral flow (without significant viscosity)[13]. This explanation seems to differ significantly from Send's former one and requires a lot of previous knowledge and transfer competencies from students.

---

[11] https://www.tec-science.com/de/mechanik/gase-und-fluessigkeiten/anwendungen-des-bernoulli-effekts/ [2023-04-17]

[12] Translated from German: "*Der Unterdruck auf der Oberseite ist eine Folge der Lösung, bei der sich die teilende Staupunktstromlinie stromauf unterhalb der Mittellinie ergibt […]. Auf der Oberseite wird deshalb mehr Masse durchgesetzt und die Geschwindigkeit ist folglich höher.*"

[13] Paraphrased from German original: „*Wo kommt nun der Auftrieb […] her? Er stammt aus der Druckverteilung an der Grenze zwischen Außenströmung ohne Zähigkeit und Grenzschicht mit Zähigkeit*"

David Robertson also made interesting contributions to flight physics education but, as well, a tiny inaccuracy slipped into one publication when concluding that an aircraft is ultimately supported by its weight force distributed over the ground beneath it (Robertson, 2014). To understand the inaccuracy, I like the reader to imagine the following thought experiment: A plane with angled wings enters a gas cloud in outer space. The wings would still experience lift – due to inertia of the gas. Of course, this would also displace the gas with no return to the original "height" or "counter impulse by the earth surface" but also displace the airplane from its original trajectory. This thought experiment shows that no ground is needed to generate lift.

To sum this up: Most flaws concerning lift explanations fit into the following categories:
1. The correct explanation is not contrasted against learners' naïve/pre concept
2. The explanation is erroneous.
3. The explanation is pedagogically misleading.
4. Phenomenon's contribution to lift is enumerated in a misleading order, ranked wrong or given too much space in a chapter.

### 2.2.2 Problems with the free body diagram

When starting to teach flight related topics many make use of the *four forces diagram* of an airplane indicating aerodynamic lift, drag, thrust and weight. But often, these representations only show horizontal, unaccelerated flight. This can lead to the naïve notions that:
- Lift is always pointing vertically upwards.
- Drag is always horizontal.
- Thrust is always as big as drag.
- Thrust is always pointing exactly in the direction of travel.

Another problem is that the direction of thrust (the resulting force of the engine's propulsion) is often only defined for horizontal flight[14] or the description of the thrust's direction is avoided when thrust is defined[15]. This is understandable for pedagogical reasons, since one would need to introduce many other definitions first (e.g. flight path direction, flight path angle, angle of attack, thrust angle) for a global definition of thrust in aerodynamic flight. However, since airplanes are optimized for horizontal, unaccelerated flight, and many representations only deal with this case, students often think of thrust as parallel to the flight path, perpendicular to lift, and/or antiparallel to drag. All of these three definitions contradict expert definitions. In steady climbing flight, for example, thrust needs to be bigger than drag. Further, thrust is usually inclined at a positive angle with respect to the flight path direction (J. D. Anderson, 1989)(see also → Fig. 5).

---

[14] https://www.aircraftsystemstech.com/p/thrust-and-drag.html [2023-05-27]
[15] https://www1.grc.nasa.gov/beginners-guide-to-aeronautics/what-is-thrust/ [2023-05-27]

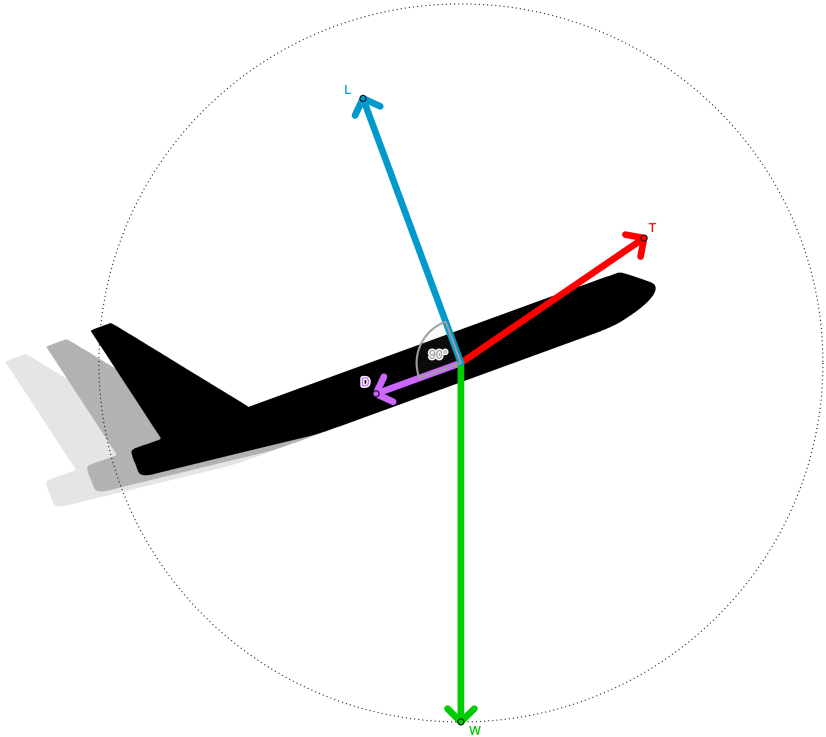

*Fig. 5: Forces of flight in steady climb flight: Thrust (red T) is neither horizontal nor in flight path direction nor only opposing drag (purple D). Lift (blue L) needs to be smaller than weight (green W)*

Further, often no differentiation is made between Center of Mass (CoM), Center of Pressure (CoP) and aerodynamic center (AC). This leads to the false assumption that all forces acting on the airplane can be seen as if they attack at the Center of Mass – including the aerodynamic forces. In practice, CoM and CoP should be close together, but especially for flight states where CoM and CoP divert, it is crucial to differentiate where each of these forces attack.

## 3 Methodology

The iterative development process of the *Flight Physics Concept Inventory* (FliP-CoIn) was structured by design-based research (DBR) (Easterday et al., 2014; Genz & Bresges, 2017; E. E. Scott et al., 2020) and embraced a mixed-methods approach (→ Fig. 6):

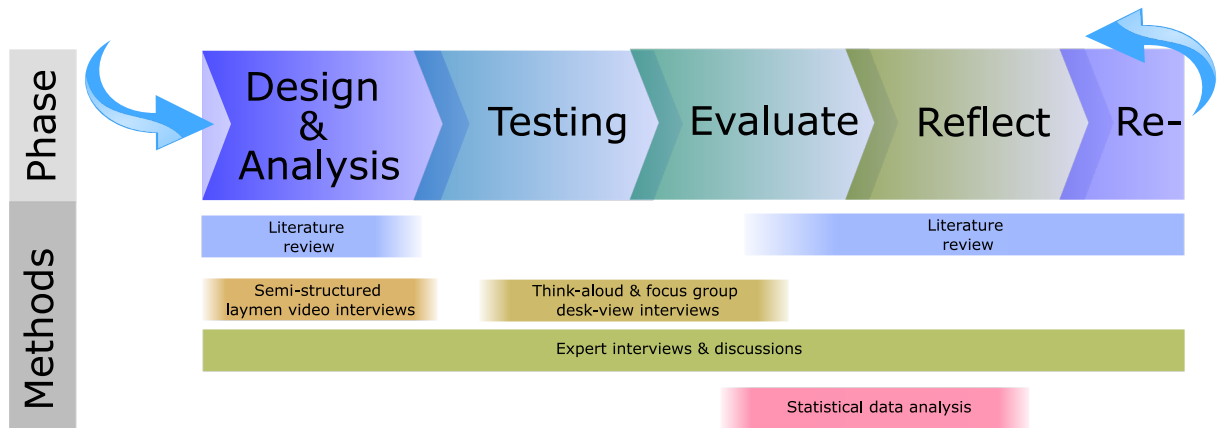

*Fig. 6: Design-Based Research (DBR) phases & methods used: Different interview methods, a literature review, and statistical analysis were used during the cyclic DBR phases – namely Design and analysis, testing, evaluation, and reflection (picture adapted based on E. E. Scott et al., 2020).*

A triangulation of data was ensured by a systematic literature review, different interview formats and different methods for data analysis. *Literature review* (→3.5.2) was mainly used during the *design phase* but was also consulted to enrich and validate findings during the *evaluation* and *reflection* phases. *Expert interviews and discussions* (→3.5.1) were employed in all DBR phases. In the first design cycles, *expert interviews* were primarily focussed on collecting naïve concepts and determining the scope of the instrument. Later, the focus of the *expert interviews* shifted towards discussing expert comments on several item versions and discussions about the interpretation of the data. *Think-aloud and focus group interviews* were mainly used in the testing and evaluation phases to improve item wording and detect items that do not work. *Semi-structured layman video interviews* were exclusively used in the *design phase* for collecting naïve concepts (→3.5.3) and during the test construction process (→3.4). The *statistical data analyses* started out with a qualitative approach – like analyzing comments and visual inspection of descriptive data. Later in that process, correlations between items and subpopulations were analyzed. Also, an item analysis was conducted, including an *internal reliability analysis*. All procedures above resulted in a final set of items whose statistical properties are presented in Section 4 and in more detail in the Appendix.

The currently established main criteria for test quality are objectivity, reliability and validity (Bühner, 2011; Wess et al., 2021). In the following Sections, I will elaborate to what extent these criteria could be satisfied and the methods used to achieve this goal. Further, I will sketch dimensions for future research.

### 3.1 Objectivity

In our case, a test's objectivity means the independence from influences of the administering person (Miller et al., 2021). Bühner differentiates between implementation objectivity, evaluation objectivity and interpretation objectivity (Bühner, 2011).

For a high implementation objectivity, the "*FliP-CoIn implementation and troubleshooting guide*" was developed and is downloadable by verified teachers on PhysPort.org[16].
In the implementation guide necessary steps are described and the option for an online implementation is offered. This was primarily done to reduce the evaluation time for teachers, but also to ensure the highest possible *evaluation objectivity* (avoids encoding errors). The third criterion, the *interpretation objectivity*, is ensured by the *FliP-CoIn scoring template* (also available at PhysPort).

## 3.2 Validity

According to the American Educational Research Association (AERA), American Psychological Association (APA) and National Council on Measurement in Education (NCME)

"[validity] refers to the degree to which evidence and theory support the interpretations of the test scores for proposed uses of tests" (AERA et al., 2014).

Unfortunately, different authors name different subcategories for validity, and there is no huge consensus on them. However, Murphy and Davidshofer emphasize that *content validity* is the most relevant one, embracing all other validity dimensions (Murphy & Davidshofer, 2001), since the items of a test define its content and the quality of items hugely influences its validity (Bühner, 2011). Therefore, this work will focus on *content validity* and describe the methods used for their establishment accordingly.

### 3.2.1 Content validity

The APA handbook of testing and assessment in psychology defines content validity as follows: „*Content validity refers to the degree to which the content of a test is representative of the targeted construct and supports the use of a test for its intended purposes*." (APA Handbook, 2013a). Content validity evidence was gathered with expert reviews and judgments regarding the correctness and relevance of each item, and the degree to which the items represent the construct (see also →3.4.1)

## 3.3 Reliability

Reliability alone became a blurry term with the diversification of statistical procedures and tools for test evaluation. Originally, it only meant the *reliability coefficients* in CTT, defined as the test score similarity on two identical or equivalent tests. Nowadays, the pure term *reliability* can also refer to any form of test consistency measure (i.e., reliability coefficients, generalizability coefficients, standard errors, error tolerance ratios, IRT information functions, classification consistency indices…). Therefore, AERA, APA and NCME recommend using the term *reliability/precision* (AERA et al., 2014) when meaning the broader, more general meaning "*score consistency across repeated tests*". Determining the *reliability/precision* is another task for future research, since the data collected at this point does not allow its calculation. In this work, the *internal consistency / reliability coefficient* (→3.3.1) was calculated to estimate one easy and tangible dimension of reliability. Further, *reliability/precision* is not to be confused with *intra-rater reliability, inter-rater reliability* or *scoring relevance.*

---

[16] https://www.physport.org/assessments/assessment.cfm?A=FliPCoIn [2023-05-08]

### 3.3.1 Internal consistency / reliability coefficient

The internal consistency method calculating the Cronbach's alpha reliability coefficient (Cronbach & Meehl, 1955) is a powerful tool when data can only be collected once in a given population. It is estimated based on the variance (s²) of the item sum and the variances of the items:

$$\alpha_{Cronbach's} = \frac{k}{k-1} * \left(1 - \frac{\sum_{i=1}^{k} s_i^2}{s_Y^2}\right)$$

k: number of items
$s_i^2$: variance for item i
$s_Y^2$: variance of the measurement

For dichotomously scored items (as in our case), its calculation is identical to the Kuder-Richardson Formula 20 (KR-20). By definition, the Cronbach alpha coefficient can range from 0 to 1. A high Cronbach alpha value indicates homogeneity between the items. Current conventions consider a value of 60%-70% sufficient for early-stage research to be satisfying. For applied research, values in the range of 80% are recommended (Nunnally & Bernstein, 1994), since higher values usually harm the usability and/or the validity of the instrument (attenuation paradox)(Lord, 1954). According to current conventions, it is not necessary to manually compute Cronbach's alpha anymore, as it can be computed automatically and more reliably with statistical software (Danner, 2016).

## 3.4 The test construction process

In the following chapters, the path of the FliP-CoIn development is described in order to establish inter-subjectivity and content validity for a first range of items and concept domains:
Before Cronbach and Furby (Cronbach & Furby, 1970), as well as ROID and HYLADYNA (1980) discussed the identification of behaviors that represent a construct, the breakdown of mental models into test items remained rather informal, subjective and undocumented (Algina & Crocker, 1986). Thus led into subjective and idiosyncratic definitions of the mental models. In order to gain more objective and generic measures, the test developer engaged in the following activities suggested by (Algina & Crocker, 1986), p.66:
1. Identifying ideas that represent the construct
2. Domain sampling
3. Preparing test specifications

In the following subchapters, I will elaborate on all three of them.

### 3.4.1 Identifying ideas that represent the construct

In this chapter, I will give an overview of how the scope of the FliP-CoIn instrument was determined and how items were developed. The need for a Concept Inventory emerged from the idea to improve the lab experience of physics education students at the University of Cologne.

In that context it was obvious that the concepts of aerodynamic lift, drag and angle of attack needed to be addressed, since the experiments demanded measuring these entities and creating different airfoil polar plots for a later discussion about profile modifications. In that context, also naïve concepts of aerodynamic drag were diagnosed and frequently led to misunderstandings during the lab's oral pre-exam and revised lab reports. Later, due to expert feedback and written comments on early instrument versions, it became obvious that the concept *center of mass* also needed to be addressed in addition to *aerodynamic stall* and properties of 2-dimensional *streamline pictures*. Finally, to control for practical flight experience, some items were introduced that are more likely to be correctly answered by pilots, since they are part of their study program. In this work I call these items *pilot discriminating items* or *flight experience items*.

In very early versions of the instrument, open-ended questions were posted to physics education students about the constructs of interest. The students' responses were sorted into topical categories to get good *distractors*. In this context, distractors mean incorrect answer options. In the best case, the form and format of a distractor answer does not differ from the correct answer. Finding and refining good distractors was an iterative and parallel process with 20 students and test versions over four semesters, which were repeated seven times. Also, expert feedback and the focus group interviews contributed to some refinements of single distractors.

Early and interim item versions were also inspired by challenging concepts and ideas from *literature*[17] (D. F. Anderson & Eberhardt, 2010; J. D. Anderson, 1989; Bauman & Schwaneberg, 1994; Crane, 1993; de Obaldia et al., 2016; Eastlake, 2002; FAA, 2016; Faulkner & Ytreberg, 2011; Granlund et al., 2014; Guerra et al., 2005; Højgaard Jensen, 2013; Houghton et al., 2012; Kaplowitz et al., 2012; Koumaras & Primerakis, 2018; McLean, 2012, 2018a, 2018b; Regis, 2020; Robertson, 2014; Send, 2001; Suarez et al., 2017; Tennekes, 2001; Torenbeek & Wittenberg, 2009; Waltham, 1998; Waltham et al., 2003; Wang, 2008; Weltner, 1987, 1990a, 1990b, 2002; Wild, 2023; Wodzinski, 1999; Wodzinski & Ziegler, 2000) as well as own data and research (Genz, 2015; Genz et al., 2023; Genz & Falconer, 2021; Genz & Vieyra, 2015) and expert questioning. During four semesters of guiding students of physics education through aerodynamic experiments, the author identified more naïve concepts by direct observation or engaging students in conversations than were found documented in the literature. Some misconceptions were first discovered with the help of virtual flight simulations (namely "Wind Tunnel" and "X-Plane10"), since they immersed the students into realistic scenarios with a decent amount of pressure to action. For example, the visualization of the *parabolic flight path* during the stall of a virtual airplane led some students to express their (wrong) idea of weightlessness during the stall.

Another source of ideas and item improvements was *expert judgement*. Eleven experts in aerodynamics and physics education were interviewed and also asked to comment on several versions of the FliP-CoIn. Also, they were encouraged to add missing constructs to be tested. This was a highly parallel and iterative process that took more than four years. Later versions of the FliP-CoIn instruments were brought to two of the annual national meetings of the *American Association of Physics Teachers* (AAPT) and the *Physics Education Research Conference* (PERC) where a booth for feedback was set up in the convention center. There, conference participants could read through the printed-out instrument and provide oral feedback or written comments.

---

[17] https://www.grc.nasa.gov/WWW/K-12/airplane, https://www1.grc.nasa.gov/beginners-guide-to-aeronautics [2023-05-12]

For *refining instruction objectives,* a list of instructional objectives was given to several experts in fluid dynamics, general physics and physics education. They were asked to comment on it. In this case, an instructional objective is defined as an intended observable student behavior after a learning intervention. The expert feedback was used for the FliP-CoIn implementation guide[18] and for the lesson plan accompanying the third publication connected with this thesis (Genz et al., 2023).

### 3.4.2 Domain sampling

To make an instrument better interpretable and inform teachers about possible learning gaps in a more fine-grained way, Crooker and Algina suggest dividing an instrument up into domains. In our case, the *domains of interest* are areas of expertise the students struggle with. Crooker and Algina suggest the following two steps for defining the *domain of interest* (Algina & Crocker, 1986, p. 71):

1. Identifying a domain of behaviors of practical importance
2. Constructing representative items

> *"[A]s the domain is restricted, the extent to which it is likely to be considered 'practically important' may be reduced."* (Algina & Crocker, 1986, p. 71):

The domain of practical importance was narrowed down carefully due to extensive expert feedback during each new version of an item (see also →3.4.1). Later in this process, the domain was named "*flight physics*". The test was developed in close relation to the learning interventions in the *Physics education lab for advanced student experiments* and *ZuS-ScienceLab* at the University of Cologne.

This iterative process led to the final scope of the instrument with the following concept domains:
- Aerodynamic lift,
- drag,
- stall,
- center of mass,
- streamlines,
- angle of attack, and
- flight experience

---

[18] Available at https://www.physport.org/assessments/assessment.cfm?A=FliPCoIn [2023-05-11]

### 3.4.3 Preparing test specifications

*"On any particular topic, the range of alternative conceptions seen in a particular population tends to be fairly limited. Often, two or three specific ideas account for most observed student responses"*
(Bao & Redish, 2006)

In flight physics, concepts need to be understood quite differently compared to other physics domains. The main variables of interest (e.g., lift, drag and the center of pressure) influence but also presuppose each other. For example, trying to isolate the concept of aerodynamic lift from drag would have resulted in abstract and less valid questions.

Also, the question formats needed to be quite diverse: They often divert from the classical multiple-choice forced-select format with four answer options because a too strict rules for *item specification* result in the production of similar-looking items, which are susceptible to response patterns. *"[A] test could consist of a much greater proportion of faulty items than might occur if item writers had not been constrained to follow [item] specifications."* (Algina & Crocker, 1986, p. 72).

Since coordinated item development and refinement was conducted by only one author, specifying test specifications beforehand was not necessary. Instead, an iterative process derived from *design-based research* was adopted, allowing for rapid improvements between design cycles (Easterday et al., 2014; Genz & Bresges, 2017; Guisasola et al., 2017). Feedback loops from experts, students and focus groups inspired items and design changes (see also → 3.5) For example, question #4 was changed from a ranking task to 13 "A or B" questions and 3-dimensional representations of the drag bodies were added as well as notes concerning flow velocity in different units. Another finding during the item refinement process was that multiple-select items were often mistaken for forced-select items. Hence, for the final set of items, the test specifications became the following:
- Use sorting tasks (with max. four objects) and multiple-choice forced-select questions only (with two to max. eight distractors).
- Use a question mix that aims at recalling knowledge (5-10%), applying knowledge (15-25%) and predicting outcomes (70-80%)
- Use a mix of text-based and picture-based questions.
- Some picture-based items should have a text-based sibling.
- Distractors need to be based on (naïve) student concepts.
- Distractors should have random positions.
- Every answer option should have a similar length and grammar.

(See also →Fig. 9).

## 3.5 Methods

### 3.5.1 Expert- and experience-based development

In order to gain a valid first-hand understanding of the concepts and the reasoning of the students, the author engaged in conducting and improving the wind tunnel experiment embedded in the advance lab class for pre-service teacher students at a large German university over the course of four semesters. The lab experience started out as a traditional recipe lab for re-tracking airfoil polar diagrams (Genz & Vieyra, 2015) and then was iteratively revised and refined due to student feedback, own observation, expert discussions and analysis of the lab reports.

Experts of fluid dynamics, flight and education from several institutes and organizations were involved during the development process, namely: RWTH Aachen, Technische Hochschule Köln (TH Köln), University of Cologne, Kansas State University (KSU), Utah Valley University (UVU), Rochester Institute of Technology (RIT), Goethe University Frankfurt, American Association of Physics Teachers (AAPT), National Aeronautics and Space Agency (NASA), Deutsches Zentrum für Luft- und Raumfahrt (DLR), Deutsche Physikalische Gesellschaft (DPG) and Buffalo State College, State University of New York (BSC, SUNY). Written comments on preliminary items, informal discussions, and written correspondence were used to determine the scope of items, collect expert opinions on important misconceptions and to detect problems with the wording and representations in the questions.

### 3.5.2 Literature review

The literature review (→Section 2) was conducted before and parallel to the expert- and experience-based development (design-based research process) (→3.5.1), containing:
   A) Expert explanations concerning aerodynamic phenomena,
   B) their didactic shortcomings, and
   C) known alternative concepts of learners

The literature review focused on the science of flight physics and then shifted towards the pedagogical, linguistic and psychological aspects of the encountered problems. At first, very broad and general. Later, more specific in the context of flight physics and fluid dynamics. However, this still was a massively paralleled, nonlinear and iterative process, and the chronology in these topics can only be seen as a rough road map.

### 3.5.3 Interviews: think-aloud, focus group, and semi-structured laymen video interviews

The importance of the third pillar for validating a new Concept Inventory was rarely better highlighted as by Bao and Redish:

> *"In constructing a measurement of student conceptual understanding, there is often a 'communication' problem; students can use the same terminology (or a statement) as used by an expert but with a different understanding. A simple word or a statement often fails to extract the actual underlying reasoning, which usually can only be obtained by analyzing how students apply their knowledge"* (Bao & Redish, 2006)

Therefore, student and laymen feedback was formally collected and analyzed in three ways in order to improve the inventory and establish validity:

A) Three think-aloud video interviews were conducted with university students of physics education filling out the complete instrument for their first time. The interviews took 66 min, 55 min and 107 min and resulted in changes concerning item wording and graphical representations. Items that caused confusion or ambiguity were completely revised or deleted.

B) Two focus group interviews were conducted with two groups of three physics education students in their master semesters. Focus group interview #1 took 75 min and the group of three consisted of one female and two male identifying members. Focus group interview #2 took 79 min with a group consisting of three males. The interviews took place after all members of the group had filled out the latest version of the inventory separately just before and then had their answers at hand to discuss the differences in answer patterns or their understanding of the questions. The reflection about each focus group interview resulted in changes concerning item wording, hints and representation.

C) Four semi-structured video interviews were conducted with laymen (non-physics students) in order to validate if the most dominant alternative concepts and explanations are still prevailing and are still relevant in the general German population in 2019. The interviews took 35 min, 25 min, 15 min and 5 min. All four interviews were accompanied by 3D-printed models of a flat wing and an asymmetrically cambered wing (→ Fig. 7).

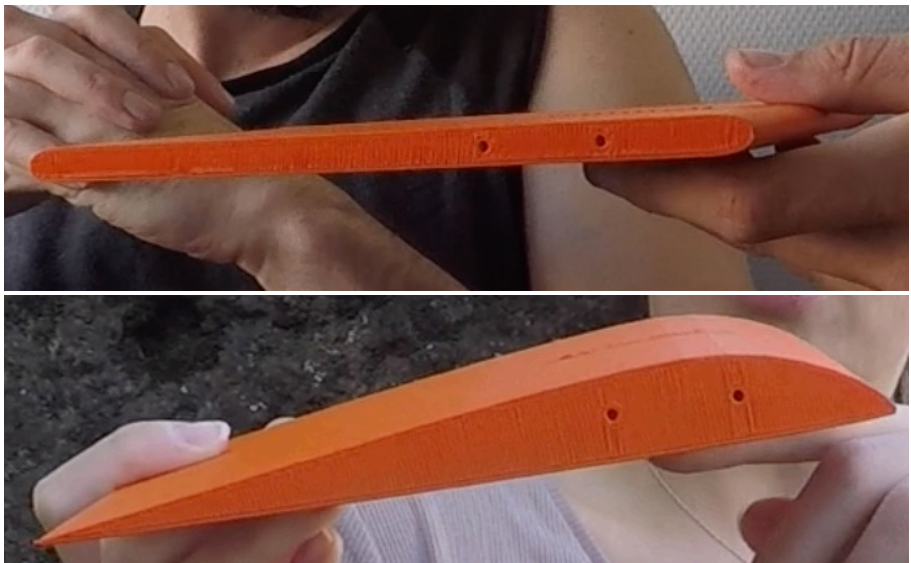

*Fig. 7: Flat airfoil profile (top) and asymmetrically cambered airfoil profile (bottom) used for the video interviews with laymen.*

### 3.5.4 Labelling of questions and factor limitations

Arvey and colleagues highlight that the concept domain labels for each item should be viewed as rationally constructed (Arvey et al., 1990). Originally, the questions were structured according to their respective concept domain in a mind map (→ Fig. 8) and labeled with two tags:

A) The predominant concept domain of the question (lift, drag, turbulent vs laminar flow, center of mass, Bernoulli confusion, and key questions to identify trained pilots – abbreviated with "*pilot discr.*")
B) contributing factors (iconic, definition, predicting)

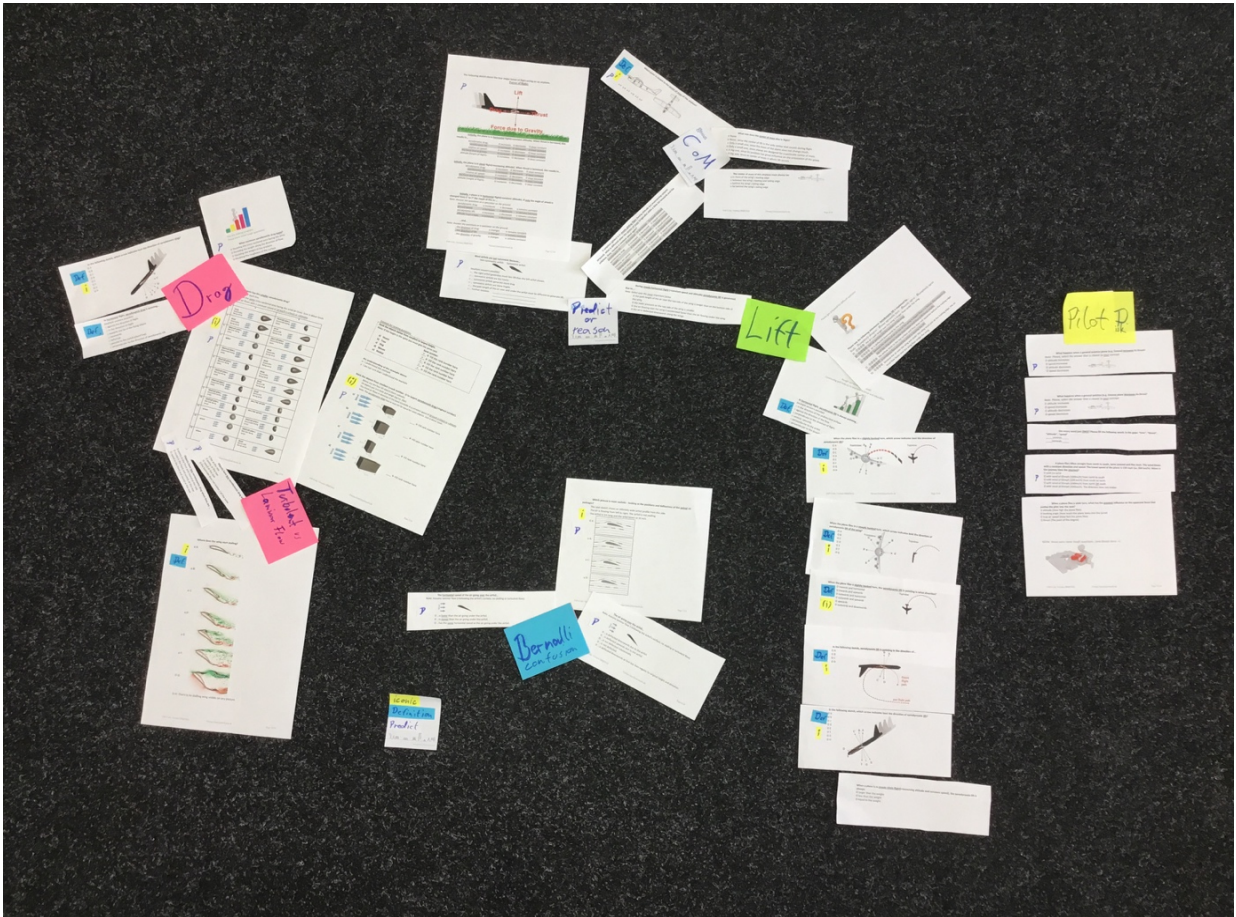

*Fig. 8: Original labels and their items arranged in a mind map.*

The *predicting* label "P" was given to questions which demand a prediction or complex reasoning about what factors to account for. *Iconic* labels "i" were given to all questions for which the iconic representation is central for the question were labeled with "i". The *definition* label "Def." was given to all questions which directly test for the understanding of the Definition of a physical term. However, this was just the first approach.

Due to expert interviews as well as discussions with colleagues and co-authors, an advanced structure was developed in an iterative process containing factor labels, question types and item levels. Also, an overarching concept domain with subdomains was agreed on (→Fig. 9):

| Question No. | QuestionID | Level | | | Type | | Labels | | | | | | | Domain | Sub domain |
|---|---|---|---|---|---|---|---|---|---|---|---|---|---|---|---|
| | | Knowledge | Application | Prediction | text-based | picture-based | lift | drag | stall | CoM | AoA | strL | exp. | | |
| 1 | QID20 | | x | | | x | | 1 | | | 1 | | | Aerodynamic Drag | Direction |
| 2 | QID6b | | | x | | x | 1 | | | | | 1 | | Aerodynamic Lift: | Cause |
| 4 | QID001b | | | x | | x | | 1 | | | | | | Aerodynamic Drag | Factors |
| 5.01 | QID1c_01 | | | x | | x | | 1 | | | | | | Aerodynamic Drag | Factors |
| 5.02 | QID1c_02 | | | x | | x | | 1 | | | | | | Aerodynamic Drag | Factors |
| 5.03 | QID1c_03 | | | x | | x | | 1 | | | | | | Aerodynamic Drag | Factors |
| 5.04 | QID1c_04 | | | x | | x | | 1 | | | | | | Aerodynamic Drag | Factors |
| 5.05 | QID1c_05 | | | x | | x | | 1 | | | | | | Aerodynamic Drag | Factors |
| 5.06 | QID1c_06 | | | x | | x | | 1 | | | | | | Aerodynamic Drag | Factors |
| 5.07 | QID1c_07 | | | x | | x | | 1 | | | | | | Aerodynamic Drag | Factors |
| 5.08 | QID1c_08 | | | x | | x | | 1 | | | | | | Aerodynamic Drag | Factors |
| 5.09 | QID1c_09 | | | x | | x | | 1 | | | | | | Aerodynamic Drag | Factors |
| 5.10 | QID1c_10 | | | x | | x | | 1 | | | | | | Aerodynamic Drag | Factors |
| 5.11 | QID_1c_11 | | | x | | x | | 1 | | | | | | Aerodynamic Drag | Factors |
| 5.12 | QID1c_12 | | | x | | x | | 1 | | | | | | Aerodynamic Drag | Factors |
| 5.13 | QID1c_13 | | | x | | x | | 1 | | | | | | Aerodynamic Drag | Factors |
| 6 | QID29 | | | x | x | | | 1 | | | | | | Aerodynamic Drag | Factors |
| 7 | QID30 | | | x | x | | | 1 | | | | | | Aerodynamic Drag | Factors |
| 8 | QID30b | | | x | x | | | 1 | | | | | | Aerodynamic Drag | Factors |
| 9 | QID27a | x | | | | x | | | | 1 | | | | Center of Mass | relative position |
| 11.1 | QID_27e_A | | | x | x | | 1 | | | 1 | | | 1 | Center of Mass | Lift |
| 11.2 | QID27e_B | | | x | x | | | 1 | | 1 | | | 1 | Center of Mass | Drag |
| 11.3 | QID27e_C | | | x | x | | | | | 1 | | | 1 | Center of Mass | Weight |
| 11.4 | QID27e_D | | | x | x | | | | | 1 | | | 1 | Center of Mass | (relative air) speed |
| 11.5 | QID27e_E | | | x | x | | | | | 1 | | | 1 | Center of Mass | sinking velocity |
| 11.6 | QID27e_F | | | x | x | | | | | 1 | 1 | | 1 | Center of Mass | Angle of Attack |
| 12 | QID006c | | | x | | x | 1 | | | | | 1 | | Airfoil profiles & str | air displacement |
| 13 | QID6 | | | x | | x | 1 | | | | | 1 | | Airfoil profiles & str | air displacement |
| 14 | QID27d | x | | | | x | | | | 1 | | | 1 | Center of Mass | relative position |
| 15 | QID17 | | x | | | x | 1 | | | | 1 | | | Aerodynamic Lift: | Direction: |
| 16 | QID18 | | x | | | x | 1 | | | | 1 | | | Aerodynamic Lift: | Direction: |
| 17 | QID19 | | x | | | x | 1 | | | | | | | Aerodynamic Lift: | Direction: |
| 18 | QID19b | | x | | | x | 1 | | | | | | | Aerodynamic Lift: | Direction: |
| 19 | QID21c | | x | | x | | 1 | | | | | | | Aerodynamic Lift: | Magnitude |
| 20.4 | QID21b_D | | | x | | x | | | | | | | 1 | Flight experience | force due to gravity |
| 20.5 | QID21b_E | | | x | | x | | | | | | | 1 | Flight experience | altitude (height of flight) |
| 21.4 | QID21_D | | | x | x | | | | | | | | 1 | Flight experience | altitude (height of flight) |
| 21.5 | QID21_E | | | x | x | | | | | | | | 1 | Flight experience | force due to gravity |
| 22.2 | QID24_B | | | x | x | | | | | | 1 | | 1 | Flight experience | Ground speed |
| 22.4 | QID24_D | | | x | x | | | | | | 1 | | 1 | Flight experience | altitude (height of flight) |
| 23.1 | QID24b_A | | | x | x | | | 1 | | | 1 | | 1 | Flight experience | Drag |
| 23.2 | QID24b_B | | | x | x | | 1 | | | | 1 | | 1 | Flight experience | Lift |
| 23.3 | QID24b_C | | | x | x | | | | | | 1 | | 1 | Flight experience | direction of F_g |
| 24 | QID35 | | x | | | x | | | | | | | 1 | Flight experience | Wind influence |
| 25 | QID36 | x | | | | x | | | | | | | | Flight experience | Forces on pilot |
| 26 | QID33 | | | x | | x | | | 1 | | | | | Aerodynamic stall | Start of stall |

*Fig. 9: Underlying structures of the FliP-CoIn: Question levels, types and labels as well as concept domain and subdomain. Each column can give insights in which area of expertise the students might lack, or which question types students struggle with.*

However, these concept domains and item labels should not be seen as mathematically distinct factors. In contrast to many other Concept Inventories, most questions in flight physics need to touch more than one concept (domain), when trying to be of practical relevance. This is due to the highly entangled nature of the physical factors influencing each other in non-linear ways. Therefore, an exploratory factor analysis based on a dichotomous item scoring does not promise to converge into more information, nor it is likely to find the multilayered and highly entangled structure of the inventory (see also → 3.6 Factor Analysis limitations).

Further, the different question formats and varying distractor amounts are another influence obscuring underlying factors. However, different question formats became necessary to account for the different amount of naïve concepts around each item and sometimes also to ensure a symmetry of answer options. For example, Q17 (ID#19) contains eight distractors, and despite C to H having very low prevalence, in sum distractors C to H usually account for more than 1/5$^{th}$ of the answers (ca. 22% in our three piloting studies). This reduces the noise of test-wise and randomly picking students at answer options A and B. Moreover, the symmetry avoids hinting the *approximate* direction of aerodynamic lift to students (see Fig. 10):

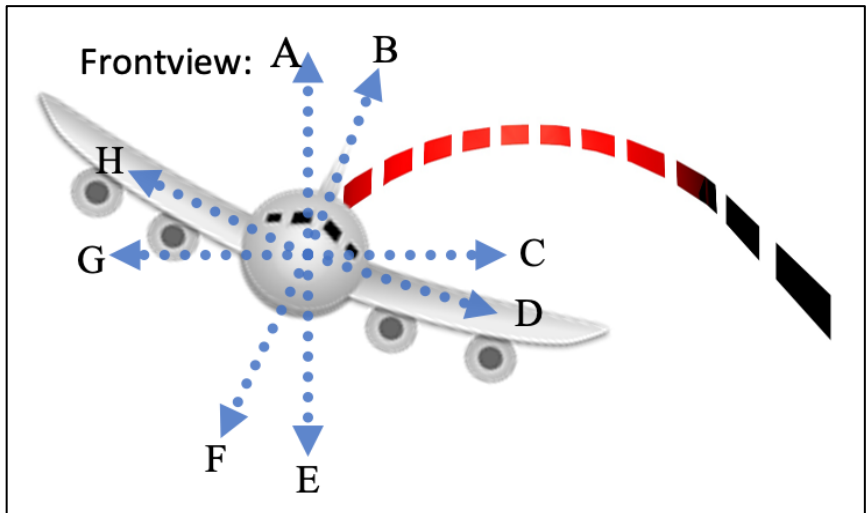

*Fig. 10: Symmetric answer options at Q17. The symmetry avoids giving away hints about the approximate direction of lift. Hence, it reduces the noise of test-wise and randomly picking students at the correct answer.*

## 3.6   Factor Analysis limitations

Some research suggested the use of an *exploratory factor analysis* (EFA) for validation purposes of *Concept Inventories* (Huffman & Heller, 1995). This approach has been criticized in statistics literature, especially in the case of dichotomous question scoring (Mislevy, 1986). Learners with a little coherence in their physics concepts still occasionally mark correct answers, whereas experts almost never fall for the distractors. The symmetry of the r correlation statistic – which is used in exploratory factor analysis – is blind to that fact and does not account for it (Halloun & Hestenes, 1996).

*Factor analysis* is a statistical method widely used in psychology research (Brown, 2006).
It became used in some Concept Inventories, but it is crucial to understand where factor analysis is applicable and where not:
*"Basically, factor analysis extracts information from a correlation matrix usually built from students' scores on different test items. The factors (eigenvectors) extracted from a correlation matrix provide a measure of how different test items may be related in terms of consistencies among student responses […]. In a test instrument, researchers usually design several equivalent questions on a single concept with varying contextual features. This represents the experts' view on how test items are clustered. However, due to the context dependence of learning, the different contextual features of the equivalent students' scores to the cluster of questions that the experts would consider equivalent questions may cause the students to respond differently.*

*In such cases, there will be low-consistency among students' scores to the cluster of questions that the experts would consider equivalent."* (Bao & Redish, 2006)

In contrast to other Concept Inventories, most questions of the FliP-CoIn instrument cover more than one conceptual domain. Further, the concepts themselves are highly dependent upon each other (E.g. aerodynamic lift and drag). In fact, they are different perspectives on a higher and more complex phenomenon of almost infinite particles interacting with each other and the flying object. Even though the scientific definitions for lift and drag as well as laminar and turbulent flow seem sharply defined, they only make sense by looking at the whole picture and the context. One particle can neither make a turbulent nor laminar flow. Lift and drag only make sense with respect to a direction of travel or average flow direction, and all of them influence each other. Therefore, an additional factor analysis seems idle in this case, since one cannot understand the concept of *aerodynamic lift* without understanding the concept of *aerodynamic drag*. Understanding *turbulent flow* without understanding *laminar flow* seems likewise impossible. Therefore, a statistical factor analysis cannot converge into more information or confirm the chosen structure of the inventory (Hestenes & Halloun, 1995).

To sum this up: Factor analysis *can* provide some insights for Concept Inventories in science, but the theoretical factors may also be obscured by strong context-dependence of the students' models and/or many students in mixed-model states. Hence, it was decided that a factor analysis is not a promising method for the highly entangled items of the *Flight Physics Concept Inventory*.

## 3.7   Classical Test Theory & Item Response Theory limitations

This Concept Inventory was developed with Classical Test Theory (CTT), because CTT accounts for the error of the measurement as a whole. Essentially, it throws all sources of error into one error term. This is most convenient for early Concept Inventory development, since, first and foremost, we should have a conservative estimate of the error range.

Item response Theory (IRT) is a powerful tool for a number of important psychometric problems, but IRT requires rather strong assumptions. Even if these assumptions are fulfilled, IRT cannot easily accommodate generalizations to varying measurement conditions (Brennan, 2001). Further, IRT usually requires larger sample sizes than the statistical procedures used for classical test theory (Hancock et al., 2018). IRT might become a feasible choice in the future when big test populations become interested in setting the test difficulty of particular items into context with other Concept Inventories. However, IRT also requires specific, rather stronger assumptions about the data analyzed. Beyond, IRT methods

*"assume certain latent constructs responsible for students' responses on different test items and rely on the measure of consistency among student responses to extract such latent constructs. This does not respond well to the context dependence of student knowledge and works only when students have pure model states. When students have mixed model states, the low consistency among students' responses often causes failure in detecting any latent constructs. In such cases, the results do not yield much insight into students' cognitive states."* (Bao & Redish, 2006)

## 3.8 Survey data collection and analysis

*"[D]ata encoding is often neglected or taken for granted, even though this step forms an important bridge between the data collection and analysis."*
(Springuel et al., 2019, p.2)

This chapter will describe important issues in the process of survey data collection and analysis. Data from the English dataset #1 (DS1) and the German dataset #2 (DS2) was collected online, while data collection for the German dataset #3 (DS3) was a paper-pencil survey conducted in a big lecture hall during the first session of the semester. Of course, cautious scientists can never be sure all possible systematic errors were excluded. Therefore, three quantitative studies were designed in a way that many variables could be controlled for. The studies were administered
1. in two different languages & learning cultures,
2. as online survey and on-site paper-pencil assessment
3. for three slightly different but similar target groups and
4. aggregated demographic data were collected.

In the following subchapters, I will elaborate more on the data, the collection process (→3.8.1), and how I estimated the robustness against scoring errors (→ 3.8.2).

### 3.8.1 Data coding and revalidation

Before the actual data was encoded and compared, the codebooks of all three Datasets (DS1, DS2, DS3) were compared. The process is described in the Appendix (→ 9.6)

The codebook comparison revealed seven minor differences (e.g., QID021: Some minor variable naming variations with no effect on the results or statistics) and five real coding differences. Three of these five real differences were intentional changes. Only two were not intended variations, namely:

- QID024: third distractor "stays constant" was coded with "4" in DS 1&2 whereas it was coded "3" in DS3. Since this question only has three answer options, no conflicts could occur.
- QID021c: differing distractor order and distractor codes in DS1, DS2 and DS3. Possible scoring errors were prevented by catching this error.

All differences, including the minor differences, were considered in the re-evaluation on the data that followed.

#### 3.8.1.1 Datasets 1 and 2 (online surveys)

Data coding for datasets 1 and 2 (DS1 & DS2) was easy to automate, since this data was collected via an online survey service (unipark.de) and coded automatically. The survey encoding by the tool unipark.de was tested by exporting test survey data where the author selected only the first, second, last and "NA" option for each question throughout the whole survey (strategy of dummyuser 1,2 and 3. See →Fig. 11):

| **Codebook check for:** | | |
|---|---|---|
| Dataset 1: | | |
|  | user code: | Strategy: |
| Dummyuser 1 | aaaa9 | always A, 9999, TRUE for all checkboxes, 1st dropdown |
| Dummyuser 2 | bbbb2 | always B, 2222, Only 2nd checkbox of question true, 2nd dropdown |
| Dummyuser 3 | zzzz0 | always last choice, 0000, FALSE for all Checkboxes, LAST dropdown |
| Dummyuser fast | ___ | skips / leaves blank everything |
| Dummyuser perfect | perf37 | answers every question correct |
| Dummyuser allwrong | allw09 | answers no question correct |
| Dataset 2: | | |
|  | user code: | Strategy: |
| Dummyuser 1 | aaaared | always A, 9999, TRUE for all checkboxes, 1st dropdown |
| Dummyuser 2 | bbbbblu | always B, 2222, Only 2nd checkbox of question true, 2nd dropdown |
| Dummyuser 3 | zzzzzer | always last choice, 0000, FALSE for all Checkboxes, LAST dropdown |
| Dummyuser fast | ___ | skips / leaves blank everything |
| Dummyuser perfect | perfect | answers every question correct |
| Dummyuser allwrong | allwrong | answers no question correct |

*Fig. 11: Codebook check strategies for datasets 1 and 2: Six dummyusers were simulated that clicked through the test with different strategies.*

Validating the survey software's (Unipark.de) export file codings:

For making codebook-errors most obvious, the empty survey was clicked through six times with different strategies (→Fig. 11). First, the survey results of the dummyusers were exported and analyzed by a visual inspection of two experts independently. No irregularities were found that contradicted the automatically generated codebook for the survey.

Secondly, all codings were revalidated during the *score validation* (→ compare Section 3.8.2.2)

### 3.8.1.2 Dataset 3 (paper pencil survey)

Data coding validation for dataset 3 (DS3) was more complex, since the survey was a paper pencil survey. 100 % of the paper pencil test booklets were double coded by four independent raters according to the coding manual for dataset 3 (→ see Section 10). One rater coded all 167 booklets, while the remaining three raters shared the rest for the second round of encoding. Each coder had a different encoding volume to be able to control for systematic errors in the process of encoding the data (e.g. test tiredness) (→Fig. 12).

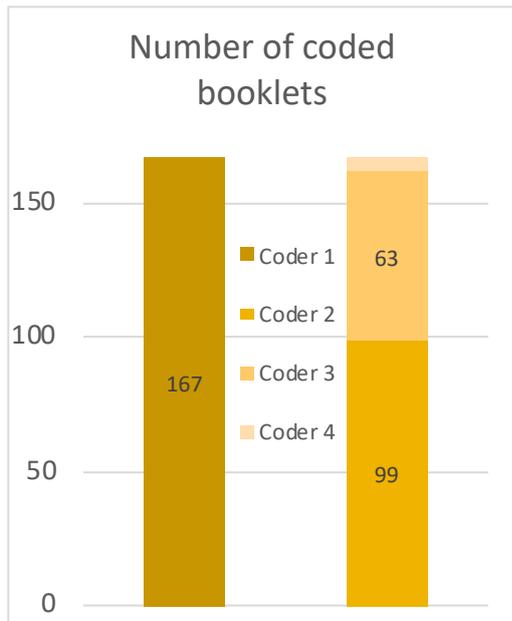

*Fig. 12: Number of coded booklets by coder. Coder 1 coded all 167 booklets. All data was double coded by 3 independent coders.*

In total, 292 (1,96%) of possible 14863 (=167*89 coded variables) real coding differences could be detected between coder 1 and the others. This was calculated with an automated comparison of two spreadsheets using the Formula

"=IF('sheet_of_coder1'A1='sheet_of_coder2'A1)"

in Excel 16.66.1.

Each coding difference was looked up in the original test booklets by coder 4 again, and then corrected. With this method, it also became clear which coder was responsible for which coding difference (→ Fig. 13):

| Detected error rate of coders: | | | | |
|---|---|---|---|---|
| | booklets coded | variables coded | coding differences | in % |
| Coder 1 | 167 | 14863 | 69 | 0,464% |
| Coder 2 | 99 | 8811 | 70 | 0,794% |
| Coder 3 | 63 | 5607 | 151 | 2,693% |
| Coder 4 | 5 | 445 | 2 | 0,449% |
| | | Sum | 292 | 0,982% |

*Fig. 13: Detected error rate of coders: Calculated by coding differences and corrected by an additional lookup of coder 4.*

The probability that, with this method, a coding error remains undetected is equivalent to the probability that coder 1 coincidentally miscoded exactly the same coding variable (item) in the same booklet as another coder. The probabilities and *maximum expectation values* were calculated in →Fig. 14 and sum up to 1,04 expected *undetected* coding differences for the whole dataset. This is sufficient to consider the corrected data as robust, but now it was quantified *how* robust it is.

| (Upper limit) expectation value for undetected coding differences | | |
|---|---|---|
| (coded variables * probability that coder1 makes a mistake at the same variable as the other coder): | | |
| between coder 1 and 2 | 99 * 0,464% * 0,794% = | 0,32 |
| between coder 1 and 3 | 63 * 0,464% * 2,693% = | 0,70 |
| between coder 1 and 4 | 5 * 0,464% * 0,449% = | 0,01 |
| | **Sum** | **1,04** |

*Fig. 14: Maximum expectation value for undetected coding differences*

My talk „*Fails for Concept Inventory (digital) Data Encoding (AAPT WM21 talk)*"[19] describes how and why the *expectation value of undetected coding differences* was calculated in more detail, but it uses slightly simplified numbers to make it easier to focus on the procedure.

### 3.8.2 Data sensitivity / robustness against scoring errors

Comparing the *item-total correlations* of the three *reliability analyses* (RAs) one by one resulted in the discovery of an error in a formula calculating the scores for some items, which had a slight negative influence on the results at that point. This motivated another layer of redundancy, described in the following subchapter (→4.3.2). The summary of all discovered errors, corrections and remaining uncertainties are summed up in the Appendix → 9.8.

#### 3.8.2.1 Dataset 3 (paper pencil survey)

To estimate how robust the final reliability analysis is against *undetected* scoring errors, we conservatively (→3.8.1) assumed that
   a) two undetected coding errors resulted in two undetected scoring errors
   b) both occurred at the most sensitive place for the reliability analysis.

The two most sensitive places - in this case - were the two questions with the highest item total correlations (QID019 and QID024_C_lift, → compare Appendix 9.5) at the highest scoring person[20].

The manipulated reliability analysis with two simulated coding errors resulted in a Cronbach Alpha of .724 instead of .725. This means that the upper limit of uncertainty due to coding errors is less than ±.001. This is more than sufficient to consider the data as robust against scoring errors.

#### 3.8.2.2 Dataset 2 and 1 (online surveys)

The scoring of the encoded data was validated by the author using the strategies of dummyusers "*perfect score*" and "*all wrong*" described in →Fig. 11. With the solutions at hand, the author therefore clicked through copy instances of the online survey two times with each dummy user strategy. All dummyuser answer codes were inserted into a copy of the original spreadsheet that calculated the scores of every single dataset. Through that strategy, one error was found for the formulas calculating the item scores and the total score. This discovery led to a *re-validation* of all formulars. The process is described in detail in the Appendix (→9.8).

---

[19] https://youtu.be/v79XU7_1MU8?t=245 [4:05-6:48] [2022-08-30]

[20] In that case participant with "lfdn SR-A2-67" or cell AW446 in file "2020-08 FC Eval. v31errors2 (…).csv". There are other cases where there might be two highest scoring persons. Then the reliability analysis is most sensitive for scoring errors (0 instead of 1) at the two highest scoring persons at only one question – the question with the highest item-total correlation.

### 3.8.3 Backup calculations for redundancy

To back up the calculations and for deepening the insights about the coding process and the coders' strategies, the analyses of the two figures above (→Fig. 13 and Fig. 14) were differentiated by answer options and re-calculated (→ Fig. 15 and Fig. 16):

| coded test booklets | | Unfoled for questions with N-Answer options: Detected error rate of coders | | | | | | | |
|---|---|---|---|---|---|---|---|---|---|
| | N | 2 | 3 | 4 | 5 | 6 | 7 | 8 | weighted Avs |
| 167 | *Coder1* | 0,444% | 0,503% | 0,547% | 0,399% | 0,200% | 0,599% | 0,000% | **0,464%** |
| 99 | *Coder2* | 0,391% | 1,091% | 0,834% | 2,020% | 0,337% | 4,040% | 0,337% | **0,794%** |
| 63 | *Coder3* | 1,331% | 3,556% | 3,037% | 5,291% | 3,704% | 1,587% | 3,704% | **2,693%** |
| 5 | *Coder4* | 0,000% | 0,000% | 0,870% | 0,000% | 6,667% | 0,000% | 0,000% | **0,449%** |
| | weighted Avs | 0,589% | 1,246% | 1,106% | 1,796% | 0,998% | 1,796% | 0,798% | |
| | Total | | | | 0,982% | | | | |

*Fig. 15: Detected error rate of coders: unfolded for questions with N answer options*

Coder 1 who coded 100% of the test booklets turned out to be very reliable and made very few coding errors (almost six times less than coder 3). Coding answers with 7
answer options seemed to be hardest for most coders, whereas the weaker coders 2 and 3 seemed to struggle especially with five answer option questions. This could also be because there were only three of these kind of questions in the whole test. The fact that coder 4 miscoded one of the three questions wrong with N=6 answer options, must be interpreted carefully, since only five test booklets were coded.

| | Unfolded for questions with N-Answer options: (Upper limit) expectation value for undetected coding differences (coded variables multiplied with the probability that coder1 makes a mistake at the same variable as the other coder): | | | | | | | |
|---|---|---|---|---|---|---|---|---|
| coded test booklets | N | 2 | 3 | 4 | 5 | 6 | 7 | 8 |
| 99 | between coder 1 and 2 | 0,0533 | 0,1358 | 0,1039 | 0,0240 | 0,0020 | 0,0240 | 0,0000 |
| 63 | between coder 1 and 3 | 0,1155 | 0,2817 | 0,2406 | 0,0399 | 0,0140 | 0,0060 | 0,0000 |
| 5 | between coder 1 and 4 | 0,0000 | 0,0000 | 0,0055 | 0,0000 | 0,0020 | 0,0000 | 0,0000 |
| | Sum | 0,1688 | 0,4175 | 0,3499 | 0,0639 | 0,0180 | 0,0299 | 0,0000 |
| | Total | 1,048 | | | | | | |

*Fig. 16: (Upper limit) expectation value for undetected coding differences: Unfolded for questions with N-Answer options*

Fig. 16's different total number (1.048) in contrast to Fig. 14 (Total=1.04) is *not* a rounding error but due to the more sophisticated calculation method. However, it backs up the Fig. 14 calculations in the way that the difference is marginal. Differentiation into each coding variable we considered obsolete therefore.

Another interesting fact: Fig. 16 also holds the information that more than a *third* of all coding errors were caused by coder 3 at questions with N=3 and N=4 answer options. This is surprising, since binary (N=2) questions were more frequent (→Fig. 17).

| Questions with N answer options: | | | | | | | |
|---|---|---|---|---|---|---|---|
| N | 2 | 3 | 4 | 5 | 6 | 7 | 8 |
| Number of questions | 31 | 25 | 23 | 3 | 3 | 1 | 3 |

*Fig. 17: Coded questions with N answer options*

### 3.8.4 Data anonymization

After the digital surveys were closed and the data was exported, all demographic and text fields were manually inspected for potentially identifiable information. Where found, the data was overwritten by the label "##anonymized".

This step became necessary on the following occasions:

- 9 times in DS2 for v_ethnicity and 10 times for v_gender_OtherTXT. These users seemed to have an autofill assistant activated that filled the fields with street names or first names.
- 1 time in DS2 for v_cqr23. An email address was stated here.

(Springuel et al., 2019) distinguish six response types typical for PER assessments:
1. multiple-choice single response (MCSR),
2. multiple-choice multiple response (MCMR),
3. short answer,
4. ranking tasks,
5. Likert scale, and
6. free response.

In early versions, the FliP-CoIn made use of all of them. It was found that MCMR questions were often mistaken for MCSR questions, when embedded between many MCMR questions. Even hints and notes that multiple marks can be made did not help that observation. Also, the option to give MCMR questions as a matrix with three answer options for each line (true, false & NA) does not completely eliminate the risk of students mistaking the *set* of questions as *one* question, marking an answer in *one* line only instead of *every* line.

Further non-exclusive ranking tasks run the risk of being mistaken for exclusive ranking tasks (where double-assigned ranks are *not* possible: e.g.: Rank the bodies from highest to lowest drag - AW: 1,2,2,2). This uncertainty in understanding of QID001b and QID001c was reduced by an example using double-assigned ranks.

### 3.8.5 Incentives for participants

For finishing the survey, the institution for dataset 1 offered some little extra credits equivalent to an extra homework. This was done without endangering the anonymity of the answers (invite token method) and was made transparent to survey participants. The incentive for finishers from dataset 2 was the option to win one of 25 vouchers for shared city bikes (also invite token method). Students from dataset 3 did not get an incentive since the survey was conducted during their lecture time. See also Section →4.1 for more detail and the demographics of the datasets.

### 3.8.6 Complete case analysis versus data imputation

Data imputation can be a valuable tool when big amounts of data (test booklets or questions) would have to be excluded due to a small amount of missing data. However, based on the imputation method and the data structure itself, data imputation also comes along with considerable disadvantages for several analyses types (Little & Rubin, 2002; Rubin, 1976). Some methods of data analysis are only designed for complete-data analyses. Little and Rubin differentiate six missing-data patterns, from which four could be observed before data transformation and filtering for complete cases was conducted (Little & Rubin, 2002, p. 5):

a) univariate nonresponse
b) multivariate two patterns
c) monotone
d) general

It was decided for a complete-case analysis and against imputing the data, since most survey quitters abandoned the survey before page 4 (see also Fig. 18). Modeling their possible answers based on the very few completed items would have been very sketchy. Further, since only 25% of survey starters finished the survey, the risk of biasing the data by imputing 75% of non-finishers was rated as too high.

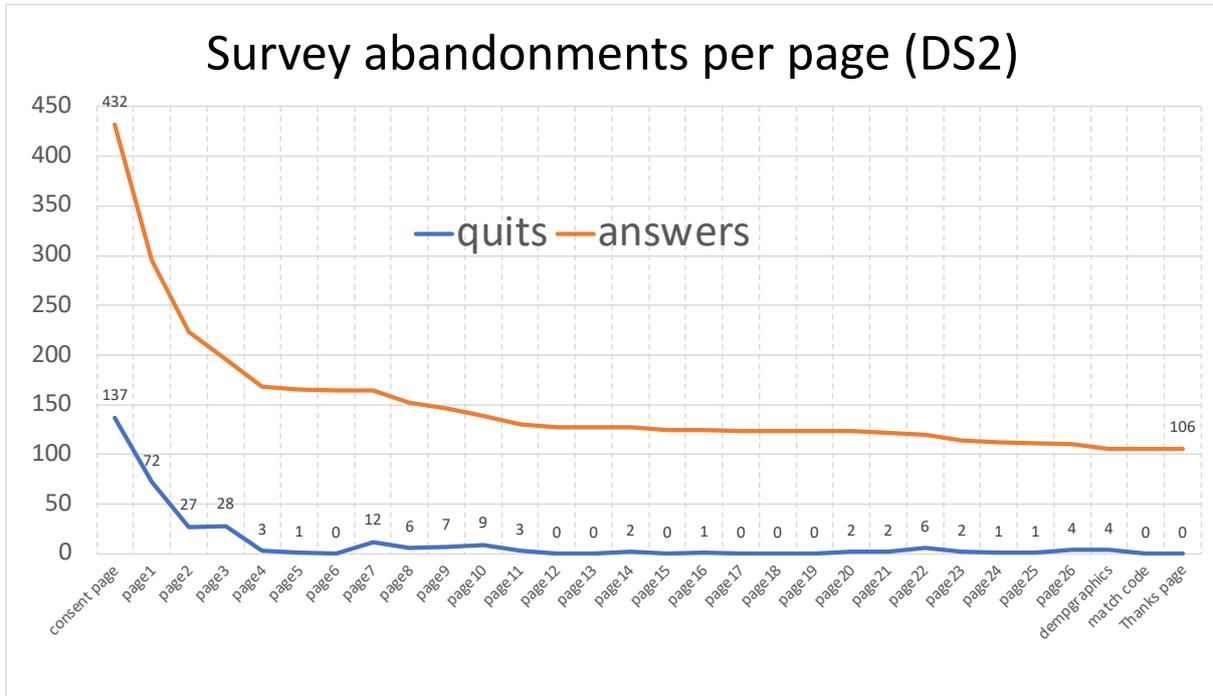

*Fig. 18: Survey abandonments per page (DS2). Most drop-outs happened within the first four survey pages, making data imputation unattractive.*

# 4 Statistical analyses

## 4.1 Representability and demographics

The three datasets collected for determining the final selection of items contained N=543 handed-in survey booklets. The total average age of all survey participants is 25 years. The ratio of female students is 12%, which is not representative of society, but realistic for most present engineering study programs. For data differentiated by dataset, see →Table 1.

Table 1: Properties, demographics and representability data for all three data sets

|  |  | Data set | | |
|---|---|---|---|---|
|  |  | DS1 | DS2 | DS3 |
| Age in years | AV | 28,2 | 22,2 | 22,7 |
|  | SD | 7,9 | 3,5 | 4,3 |
| gender of finishers | female | 7% | 14% | 18% |
|  | male | 90% | 75% | 78% |
|  | other/divers,… | 2%* | 11%* | 4%* |
| response rate |  | .24 | .72 / >.10 | .90 |
| completion rate |  | .50 | NA | .81 |
| Country |  | USA | Germany | Germany |
| Institution | type | University of applied science | | |
|  | & focus | Aviation | Engineering | Engineering |
| Survey format |  | online | on-site / online | on site |
| n_finishers = |  | 270 | 107 | 166 |

*probably overestimated (e.g. often filled with "Reptiloid", "Dolphin", "Apache Attack Helicopter", etc. )

At first, it seems outstanding that the percentage of female students is lower than one fifth in every dataset. However, the proportions seem to be quite accurate for the surveyed populations. An analysis of first names[21] from the whole institution (N=1135) of dataset 1 resulted in a male proportion of 89%. This indicates the absence of a gender mediated non-response bias. Respondents and non-respondents are similar in gender proportions. This was also confirmed by a Pearson-Chi² test (see Appendix →9.2).

The second outstanding number seems to be dataset2 (DS2) containing 11% of "divers/other,…" genders. This number is much higher than in DS1 and DS3. However, these twelve incidences turned out to be "Reptiloid", "Delfin", and ten German first names, which probably were autofilled by the user's browser[22]. In DS1 and DS3 almost all "divers/other, …" genders were specified with fun fillings like "jedi", "Apache attack helicopter" or "EWOK", suggesting that this field was not filled seriously, or challenged by a trending internet meme concerning helicopters.

---

[21] via https://genderapi.io [2023-05-18]

[22] The first names were deleted to ensure anonymity of the data.

The *response rates* and *survey completion rates* varied widely, from 10-90%. This had a lot to do with the different settings. DS3 was given to students during lecture time of a big lecture, whereas DS1 was a university-wide online survey. DS2 combined both approaches. Also, the incentives varied: DS1 survey offered students a homework credit. The online part of DS2 was incentivized by a raffle for 25 vouchers for a yearlong subscription of shared city bikes in Germany. DS3 did not have an incentive since lecture time was used and the possibility was given to ask questions afterward (→Table 2).

*Table 2: Incentives for survey participation*

| Incentives for survey participation | |
|---|---|
| Dataset 1 | Homework credit (for online participation) |
| Dataset 2 | Raffle (for online participation), lecture time (for onsite participation), |
| Dataset 3 | Lecture time (for onsite participation) |

In order to detect possible gender biases, the data of total scores were differentiated by "female", "male" and "divers/other…" as well as by dataset. Fig. 19 does not show any statistically significant performance differences in DS3 but in DS1 and DS2 highly significant differences with *strong effect sizes* (Cohen, 1988):

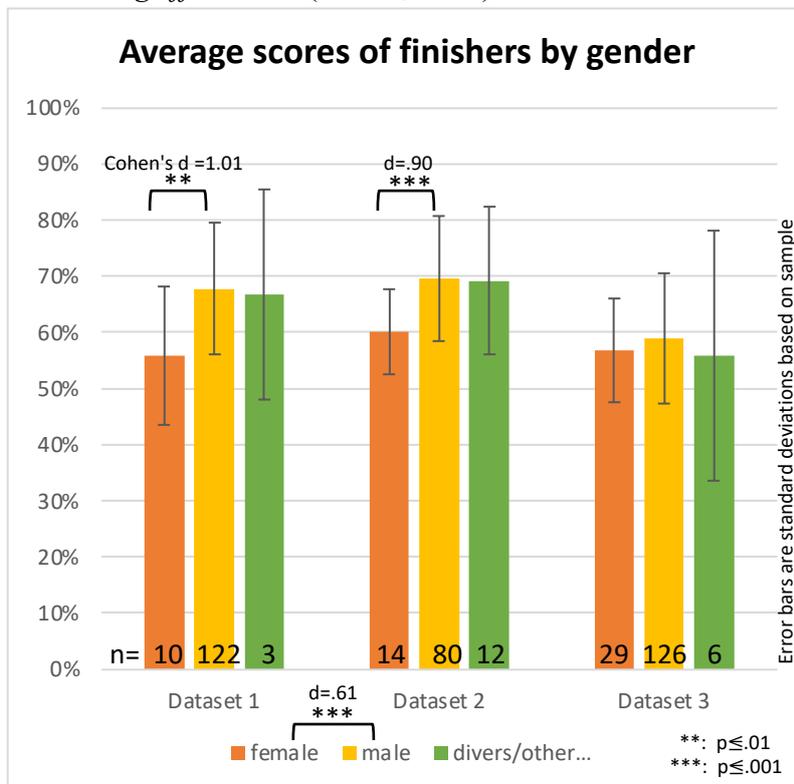

*Fig. 19: Average scores of finishers by gender: Dataset 3 does not show a statistically significant gender gap (according to a Mann-Whitney-U test). This suggests a very equitable learning culture at the institution of dataset 3. Dataset 1 and 2 show statisticalyl significant gender gaps between females and males with strong and medium effect sizes. Cohen's d for differently sized groups is d=1.01 in dataset 1, d=.90 in dataset 2 and d=.61 for all datasets combined. Due to the small sample sizes in the non-male subgroups, all conclusions should be interpreted with caution.*

DS3 does not show a statistically significant gender gap according to paired Mann-Whitney-U tests (→9.1). This suggests a relatively equitable learning culture at the institution of DS3. DS1 and DS2 show statistically significant gender gaps between females and males with strong effect sizes (Hartung et al., 2008). *Cohen's d for differently sized groups* is d=1.01 in DS1, d=.90 in DS2 and d=.61 for all datasets combined (DS123). Of course, all conclusions should be interpreted with caution due to the small sample sizes in the non-male subgroups.

## 4.2 Descriptive results

This chapter will highlight the most surprising and insightful results from analyzing the descriptive data results. The complete test statistics of the descriptive data analysis can be found in the Appendix (→ 9.5).

The distribution of total scores follows approximately a normal distribution in dataset 3 (see Appendix →9.3), and showing no ceiling or floor effect in any dataset (→Fig. 20). The average (AV) total score and median score (MD) is a bit lower in dataset 3 (DS3) when compared to DS1 and DS2. The standard deviation (SD) of total scores in DS2 is slightly smaller than in DS1 and DS3, suggesting a more homogeneous learning group. This all results in the fact that a person with a total score of 29 points would be a mid-scorer in DS1, a low scorer in DS2 and at the threshold of being a high scorer in DS3. This example demonstrates that the classification as a low, mid or high scorer is a relative one depending on the peer group. According to Kelley (Kelley, 1939) a low or high scorer is defined as the bottom or top 27% of students.

The lowest-scoring finisher in all three datasets got seven total points, but also 20 items were skipped here. The lowest-scoring finisher in DS1 scored 12 total points, with 14 items skipped. In general, the data seems to be relatively clean and low in noise. Only two finishing students in DS1 skipped more than 2 items.

The expectation value of a student picking random answer options was calculated to be 16 points. This number takes the varying amount of distractors into account. Therefore, the *data noise* caused by randomly picking students is estimated to be very low (e.g., only 3 students in DS1 reached less than 18 points).

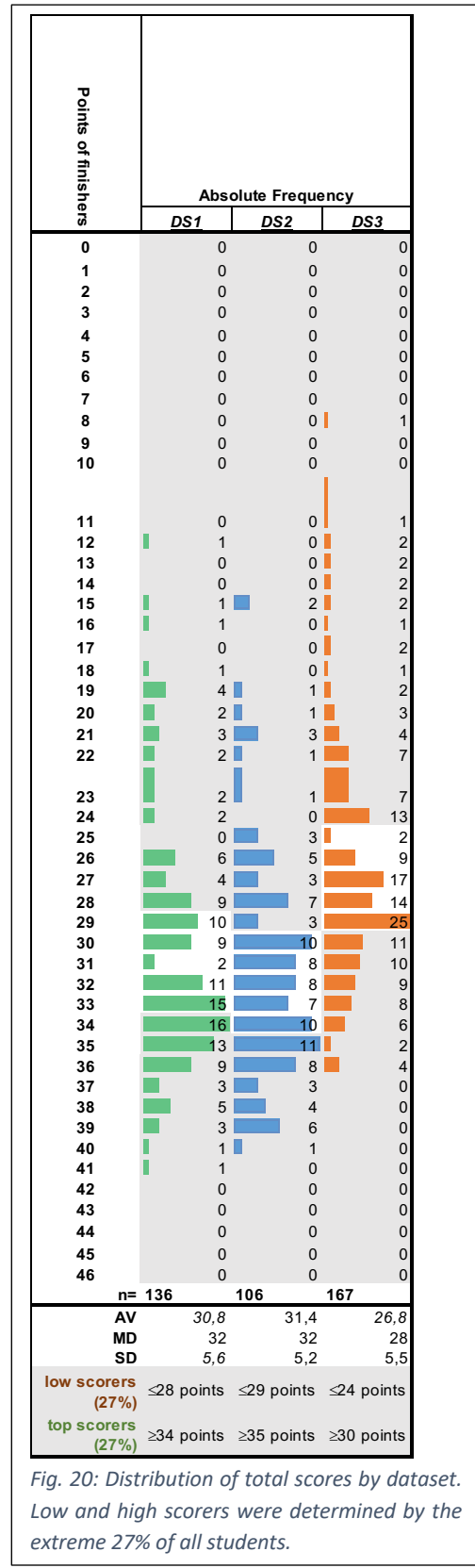

*Fig. 20: Distribution of total scores by dataset. Low and high scorers were determined by the extreme 27% of all students.*

## 4.3 Reliability analysis

After a plausibility check and analysis of the descriptive statistics (→4.2) several internal *reliability analyses* (RAs) were run with different item sets. Only *complete* survey booklets were considered for the reliability analyses. A survey booklet was considered *complete* when 95% of the scored items contained the value 0 or 1 (incorrect or correct).

In the *first iteration round,* an internal reliability analysis with SPSS (Versions 26.0.0.0 and 28.0.1.0) was run with all variables of dataset 1. This was only done to get a first orientation, since some scores appeared two times (e.g., in pts1c_ALLcorrect and pts1c_13). Therefore, in the iteration rounds of level 2, only variables of first order (e.g., pts1c_13) were used and summarizing variables (e.g., pts1c_ALLcorrect) were left out.

In the *second level iteration rounds,* items that did not perform well were re-evaluated for closer inspection. This procedure led to a rescoring of variable pts001b for only the quantitatively correct answer patterns (1342, 1232 and 1243). At first, also answer patterns 1222 and 1234 were considered qualitatively still close enough to consider them "*correct*", but high scorers almost never used these combinations.

*Third level iteration rounds*:
   a) Low-performing items (item-total correlation <.15) were discussed with experts in PER and flight physics. Some low-performing items were removed, modified, and/or re-introduced when they were considered important, understandable and valid.
   b) After discussion, in item #033 only answer patterns B and C were scored as correct. Before this was scored more liberal. Answers B, C, D, and E were considered *qualitatively correct* (correct on a conceptual level).
   c) Item #001b was re-scored only considering the empirically verified answer patterns "1232", "1243", "1342" as correct.

In the final and *fourth level iteration round*, the analysis for internal reliability was run with all remaining items, now including all datasets (DS1-3).

### 4.3.1 Result of reliability analysis

The final analysis of internal reliability, including all datasets (DS1, DS2, and DS3), resulted in a *Cronbach Alpha coefficient* for *internal reliability* of $\alpha=.73$ (see → Fig. 21). This is within the scope of the recommended level for applied research (Nunnally & Bernstein, 1994). The value was determined with *IBM SPSS Statistics version 28.0.1.0 (142)* and based on the scoring data in the Appendix (→9.4). A total of n=274 complete surveys were taken into account (>95% of all scored questions answered). This was 49.9% of all surveys collected.

Fig. 21 shows the results in more detail. The question IDs appear in order of the test given. Column "mean" indicates the ratio of correct answers. A mean of .09 can therefore be seen as a difficulty index of 91% (=1-0.09). The column "*corrected item-total correlation*" indicates how consequently the item was answered, comparing low and high scorers[23]. When most high scorers and few low scorers answer an item correctly, this will result in a high "*corrected item-total correlation*".

Looking at Fig. 21 it is easy to see that some items remained in the final item selection despite low or even negative item-total correlation (e.g., pts27e_A, pts27e_B, pts21c and pts1c_11). This was done to keep an item, including all sub-items, as a whole, since other research showed that changing or deleting one sub-item or distractor can significantly change the difficulty of an item (Kreiten, 2012). Item #21c and #1c_11, which showed a relatively strong negative *item-total correlation,* are discussed in Section →5.5 separately.

For weakly correlating items that made it into the final item selection several criteria played a role:
- A) A high correlation in at least 1 dataset was found,
- B) Item is a sub-question of a matrix question,
- C) a revised item version was created meanwhile or
- D) the item was re-discussed with *physics education research* experts and found valid, understandable and important.

Procedure D was done in particular for the ten weakest correlating items (Appendix →9.5.2) By this procedure, it was also found that the weak correlating items tend to be the difficult ones. The average mean of the ten weakest correlating items is only .39 whereas it is .85 for the ten strongest correlating items (→Fig. 39). This can be seen as "*everybody is guessing – and mostly wrong*", meaning no strong pre-concepts exist, or naïve concepts are widely established at about half of students – independent of their total score. The remaining low correlating but difficult items, mentioned above, were also kept to avoid a ceiling effect for advanced courses. Since these items tend to be the more difficult ones (→Fig. 39, column "AV mean"), it is likely that they work better for student groups who are on a higher level. For more detailed results of the reliability analyses, differentiated by dataset, see Chapter → 9.5.4 in the Appendix.

---

[23] "Corrected" does not indicate that there ever was an incorrect *item-total correlation*. Rather, it is the name of the measure. *"The correlation is corrected in the sense that the value of item i is subtracted from the total for the correlation between the total and item 1. The rationale is that including item i in the total will result in an inflated correlation between item i and that total, as a component of the correlation will be the correlation of item i with itself. This inflation will be more severe if the number of items is low."* (https://www.ibm.com/support/pages/item-total-correlations-spss [2023-01-26])

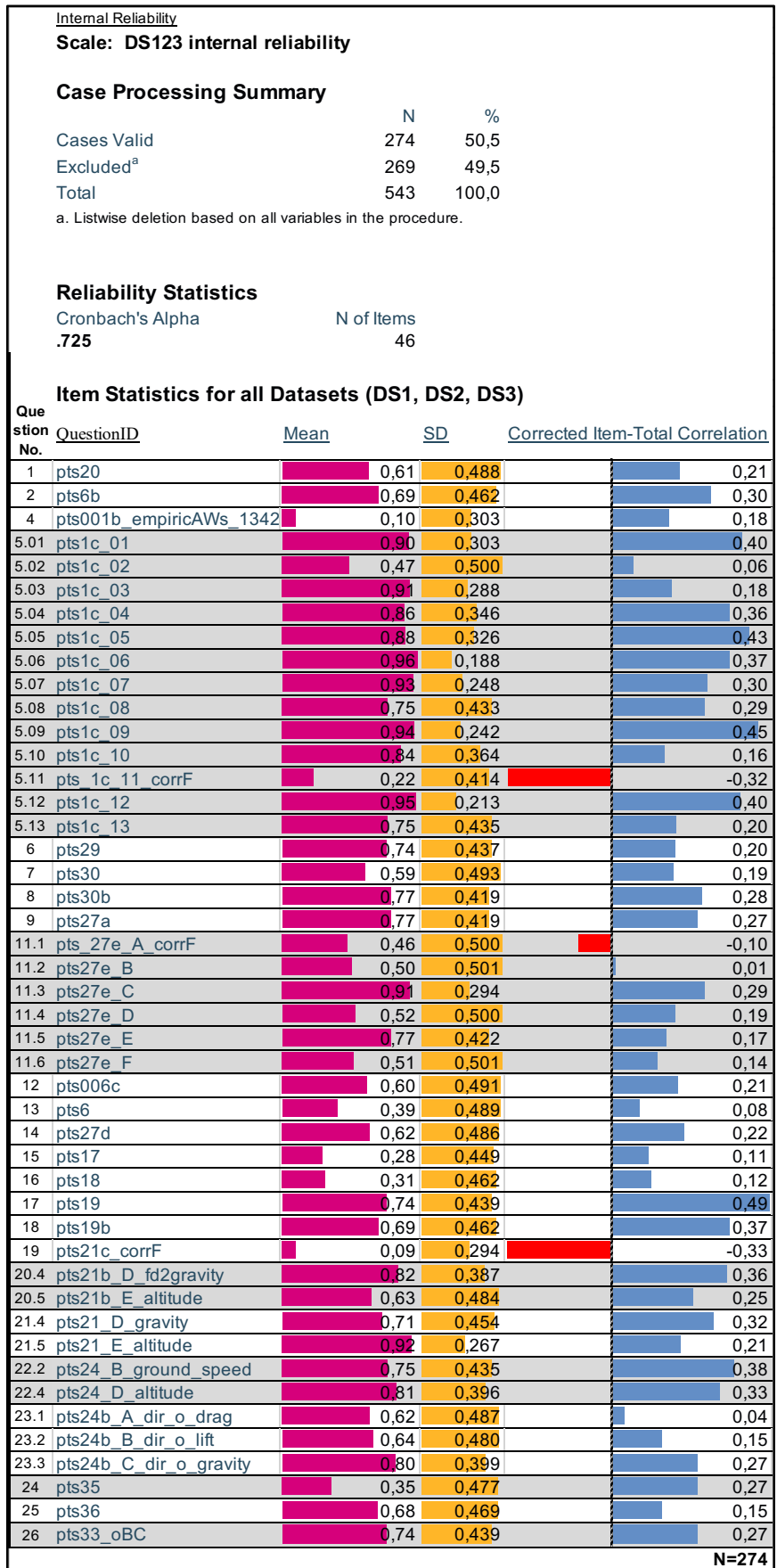

*Fig. 21: Final analysis of internal reliability including all datasets (DS1, DS2, and DS3): Cronbach's α =73%*

### 4.3.2 Scoring error corrections and 4-eye recheck of scoring

During preparation for the third iteration round, an error in the automatic score calculation was discovered at QID0027e. Therefore, a *4-eye recheck of all scorings* was conducted before proceeding. The exact procedure and methods involved in all error corrections are documented in the Appendix (→Chapter 9.8).

In summary, three more errors were found, which could have slightly changed the scoring if not caught by this process.
These errors were:
1. Two alternative ways to calculate the correct scores for variable *QID027e_ALLcorrect*. This did not affect the original codings.
2. The count of *NAs* for variable *pts_034b*. If this *inaccuracy* was not detected, this could have had a slight influence on the data analysis and the scoring on item QID034b. However, even after the correction, QID034b was still not used for the final itemset because it did not satisfy the criteria of the item analysis.
3. An *encoding error* of both coders concerning variable *QID_24b* in one survey-booklet of dataset 3.

In a nutshell, the newly introduced redundancy layer, called *4-eye recheck of scoring*, could only detect one further error with a theoretical effect on the final reliability analysis. This was not a calculation error, but an *encoding error* concerning *QID_24b* in dataset 3. This identical *encoding error* of two coders could be the predicted *undetected coding difference* from chapter →3.8.1. In that chapter, the expec*tation value for undetected coding differences* was calculated to be 1.04 (→Fig. 14). This fulfilled prediction seems like another indicator for robust data, and for reaching saturation concerning further methods for data redundancy.

### 4.4 Evaluating content validity

One of the strongest empirical indicators that a very discrete amount of concepts are underlying the participant answers can be seen in answer pattern analyses of questions #4 and #5 (→Fig. 22):

| Question #5: | | | Dataset 1 | Dataset 2 | Dataset 3 |
|---|---|---|---|---|---|
| AV freq. | AW pattern | | rel. frequency | rel. frequency | rel. frequency |
| 27,8% | A BBBB BBBA AAAB | | 25% | 21% | 37% |
| 13,5% | A ABBB BBBA AAAB | | 12% | 22% | 6% |
| 5,7% | A BBBB BBBA ABAB | | 5% | 5% | 7% |
| 5,5% | A ABBB BBAA AAAA | | 4% | 7% | 5% |
| 3,8% | A BBBB BBBA BAAB | | 3% | 6% | 3% |
| 2,9% | A BBAB BBBA AAAB | | 1% | 5% | 3% |
| | displayed patterns: n= | | 94 | 66 | 93 |
| | all patterns: N= | | 185 | 99 | 155 |

*Fig. 22: Question #5: Answer pattern frequencies*

At question #5, 58% of all answers are covered by only six answer patterns. In total, however, there are $2^{13}$ possible answer combinations. Under a random answer condition, these six patterns should therefore only cover 0.073% of answers at average.

The empirical evidence for underlying concepts becomes even more convincing with the analysis of question #4, which aims at the factors *cross-sectional area, length* and *surface area of drag bodies* (see → Fig. 23):

> QID001b: drag
> 4. Rank the objects from smallest (=rank number: 1)
>    to largest aerodynamic drag (=highest number).
> Below you see four 3D objects (here: cuboids). All objects are 2cm wide and weigh 1kg (ca. 2lbs).
> The initial air flows in direction of the black frontside at a constant speed of 50 km/h (=15m/s or =30mph).

Fig. 23: Excerpt from question #4: Rank the objects from smallest to highest drag

The vast majority, 89% (=421/471), of all given answers are covered by only six particular answer patterns and their inverse patterns (see → Fig. 24) :

| Top six answer patterns with identified concepts | | |
|---|---|---|
| Identified concepts | AW pattern | Absolute frequency |
| Only volume / suface area counts. Direction of wind does not matter. | 1213 | 16 |
| Only the cross-sectional area facing the wind counts | 1222 | 147 |
| 1st factor is a small cross-sectional area. 2nd factor: Avoid length (e.g. a ratio of length:cross-section close to 1). | 1223 | 15 |
| 1st factor is a small cross-sectional area. 2nd factor: Close to a 1:1 ratio of length:cross-section. 3rd: Avoid length. A 2:4 ratio is better than a 4:2 ratio . | 1234 | 15 |
| 1st a small cross-secional area and then shortness counts. | 1324 | 161 |
| **1st factor is a small cross-sectional area. 2nd factor: Avoiding shortness means less drag*** | **1342** | 36 |
| like 1222 pattern | 1444 | 4 |
| inverse of 1222 | 2111 | 7 |
| inverse of 1342 | 4213 | 6 |
| inverse of 1324 | 4231 | 12 |
| inverse of 1234 | 4321 | 2 |
| | Sum | 421 |
| *experimentally confirmed answer pattern | N= | 471 |

Fig. 24: Question#4: Answer pattern analysis: Top six answer patterns with identified concepts (and inverse patterns).

Since question #4 is a sorting task that allows for identical ranks, there are 256 (=$4^4$) possible answer patterns, from which 142 could be identified as *logically feasible*. For example, pattern 3-3-2-2 does not make sense in any logic because neither the first rank (1) nor the last rank (4) is used (compare also →Fig. 41). Under a random answer condition, one would only expect 8% (=11/142) of all answers within the eleven answer patterns aiming at the top six identified concepts (compare Fig. 24). This exceeds the expectation value more than eleven times. I consider this strong empirical evidence that mental concepts do exist here and that they had a strong influence on the answers given.

# 5 Discussions

The goal of this chapter is to work out specific examples for key insights during the development of the *Flight Physics Concept Inventory* (FliP-CoIn) (→ 5.2). Further, I like to discuss an indicator for student consistency (→5.4) on the test and some controversial statistical results (→ 5.5). As a basis for all following discussions, I like to discuss the gender statistics first (→5.1).

## 5.1 Interpreting gender-sensitive statistics

The low fraction of survey participants identifying themselves as *female* motivated a discussion of the gender-sensitive statistics. Is the test biased towards a thinking culture that benefits males, or do discriminating structures, that existed prior to the survey, lead to performance gaps of female-identifying students? Fig. 19 suggests the latter (→ Section 4.1). The performance gap between females and males does not occur in all datasets. Dataset 3, which does not show a *gender gap*, is also the dataset with the highest percentage of females, suggesting that as soon as a learning culture at an educational institution is equitable towards all genders, the test performances equal out. This is well aligned with findings of the research community (Joel & Vikhanski, 2019). The human brain, as the most flexible organ, simply does not know gender from the beginning. Instead, discriminating societal structures and procedures *create* measurable gender gaps in ample areas (Rippon, 2020). Of course, all conclusions should be interpreted with caution, since the sample sizes in the non-male subgroups are small. My preliminary conclusion, based on the data available, is: The FliP-CoIn instrument *can* detect gender performance gaps and might therefore also be a valuable tool for inspiring a more equitable learning culture. However, the test itself does *not* seem to be biased towards any gender, as the distribution of data from dataset 3 in Fig. 19 suggests.

## 5.2 Key developmental changes

### 5.2.1 Key insight #1: Visualizations can confuse, but also lower cognitive load

A key insight we developed along the evolution of many questions, but especially with QID006, was: Visualizations can be of great help in supporting the quality of student understanding but can also confuse students when too much information is given.

Brief text hints below the question that were labeled as "*Hint*:" seemed to ease student understanding in general. Extensive example boxes or example questions seemed to confuse some students, and were often ignored (e.g. Fig. 26).

Further, the way a question is presented, and how streamlines are visualized, have a huge influence on how well a question is understood and its general cognitive load. The evolution of item QID006 is an archetype for this phenomenon. Primarily, QID006 was created to give insights into whether students hold the "different pathlength" misconception of aerodynamic lift, and also to diagnose, whether the student knows about the fact that neighboring air particles in front of an airfoil do *not* rejoin at the trailing edge – when one air particle goes over the topside and one below the airfoil (see also → 2.2.1). In the first versions of the question, it was underestimated that streamlines are a concept in itself, which needs to be known and understood first (→Fig. 25).

Drawings of crossing particle paths were observed, as well as air particles bouncing off the airfoil and each other, when students were asked to draw two particle paths around an asymmetric 2D airfoil:

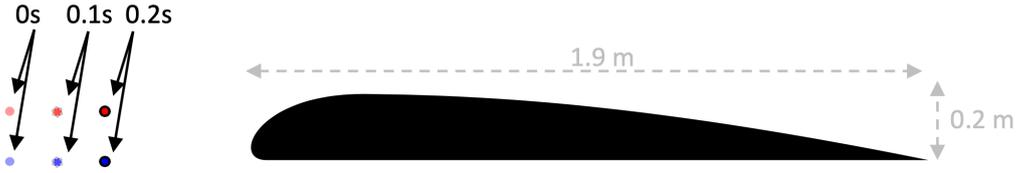

*Fig. 25: QID006 v4: Early versions of the question underestimated that streamlines are a concept in itself, that needs to be known and understood before information can be derived from their direction, distances and flow patterns.*

Later versions (→Fig. 26) tried to ensure the streamline concept is known and understood by introducing an example of streamlines, and the use of pairwise marks for air particles traveling along these streamlines for a symmetric airfoil profile. Test takers were then asked to transfer these example to an asymmetric airfoil profile, and later also to an angled board in a flow field (not displayed in Fig. 26).

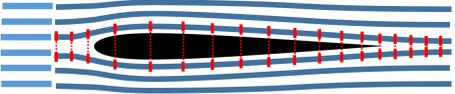

*Fig. 26: QID006 v7: An example should help to ensure the concept of streamlines is understood and ease cognitive load by splitting the task into two parts (Question 6 and 7)*

However, some think-aloud interviews suggested that the question was still not working as intended. Some students were irritated, or overwhelmed, by the unusual question format, the request to draw, and, *especially*, the example box, which was introduced to *ease* the student understanding. Therefore, it was decided to change the representation of the question drastically by using a multiple-select, forced-choice format with real and manipulated photos of pulsed streamlines (→Fig. 27)

**QID006: Which streamline photo is the real one (looking at the positions and deflections of the air packages)?**

The next sketch shows an infinitely wide airfoil profile from the side.
The pulsed smoke is flowing from left to right and is alternating between red and white.
The airfoil is 1m long and the wind blows with 20 m/s.

A:

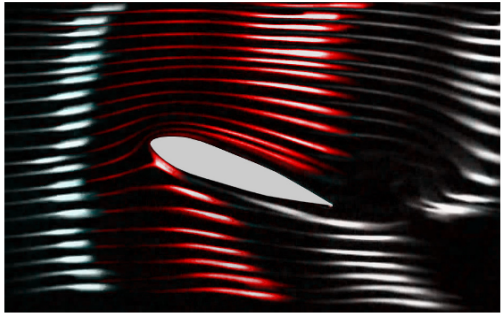

B:

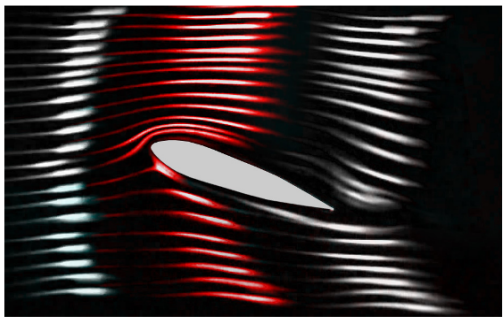

C:

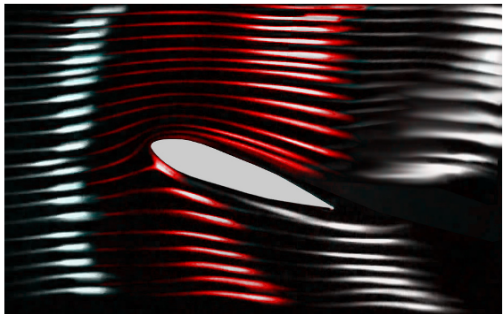

D:

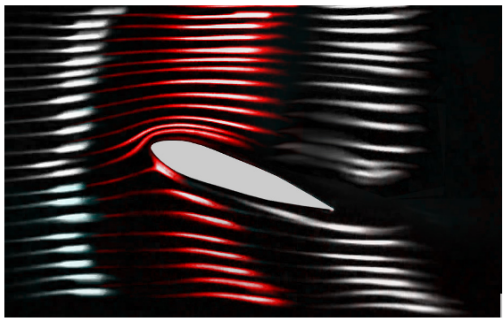

*Fig. 27: QID006 v10: The question format was radically changed in order to ease cognitive load, and also for a barrier-free online implementation.*

Version 10 (QID006 v10) showed pictures of pulsed smoketrails around a symmetric airfoil at an angle of attack. The middle pulse was dyed red in order to ease the cognitive load for the test takers. The original picture (A) shows a significant downward deflection behind the airfoil, and an acceleration of the air flowing over the airfoil. The manipulated pictures (B-D) only show one or none of these two properties. However, think-aloud interviews and expert feedback suggested that an iconic version could reduce the cognitive load drastically, and hence, the following version was created (→Fig. 28):

> **QID006: Which streamline photo is the real one - looking at the positions and deflections of the air packages? Focus on the very right ones who went <u>over</u> the airfoil.**
> The next sketch shows an infinitely wide airfoil profile from the side.
> The pulsed smoke is flowing from left to right.
> The airfoil is 1m long and the wind blows with 20 m/s.
> The airfoil is not stalling.
>
> A:
>
> B:
>
> C:
>
> D:

*Fig. 28: QID006 v17: Iconic version of QID006 v10. The iconic version was created to reduce cognitive load in comparison to real pictures of pulsed smoketrails.*

The iconic version also allowed for a trace-free manipulation of the distractors. No image manipulation needed to be camouflaged. The simpler design led to a much better focus on the two properties (acceleration and deflection of air) according to expert feedback. To further improve the quality of the question
- the wording was slightly eased,
- a higher color contrast for black and white copies was realized, and
- the central streamline leading to the stagnation point was deleted.

The final version of question 13 (QID006) can be seen in Fig. 29:

**QID006: profiles&streamlines, (lift)**
**13. Which picture is most realistic - looking at the positions and deflections of the <u>paired</u> air packages?**
The next sketch shows an infinitely wide airfoil profile from the side.
The air is flowing from left to right. The airfoil is not stalling.
The airfoil is 1m long and the wind blows at 20 m/s.

○ A:

○ B:

○ C:

○ D:

*Fig. 29: QID006 v17: The final version of the question contains a higher color contrast, slightly eased wording and no central streamline leading to the stagnation point.*

To summarize: It was found that iconic representations of 2D streamlines seem to work better than photos. In a way this was surprising because 2D photos were introduced midway with the intention to ease cognitive load. But fewer expressed confusion in think-aloud interviews, and fewer written comments suggest that iconic representations reduce the cognitive load, and improve item understanding significantly – without affecting the *easiness*[24] of the item. This was elaborated with the example of question QID006. To conclude, iconic representations in the *multiple-select forced-choice format* seem to be the best compromise for 2D conceptual streamline questions, which need some form of visualization help.

---

[24] Easiness index of an item: Percentage of test takers answering the item correctly.

Of course, one could directly ask students whether neighboring air particles rejoin at the trailing edge of an angled airfoil, but this only tests their *explicit* knowledge, and still leaves open questions – even when the correct answer "No" is given. E.g., "Does the student think the topside air particles fall behind?", "What is the student's definition of '*rejoining*'?", How close do the air particles need to come for a rejoining event in the eyes of a student?".

### 5.2.2 Key insight #2: Question format should stay constant or differ significantly

In 5.2.1 it was already mentioned that the question format had a big influence on student understanding and cognitive load. Further, I also could observe that the format of the previous question can influence student understanding. Therefore, I reason that the question format should stay the same or differ hugely – e.g. ranking task versus forced-select. Mixing multiple-select questions among forced-select questions led to the observation that multiple-select items are mistaken for forced-select items even when hints and notes, directly placed before the distractors, suggest otherwise. As a result, multiple-select items were left out for the final set of items (E.g. see →Fig. 30):

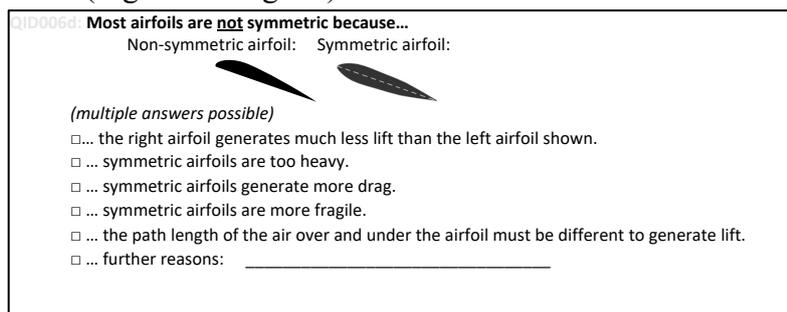

*Fig. 30: Example of a multiple-select item. All multiple select items were deleted or modified to forced-select items in process. The change between forced-select and multiple-select was overseen by too many students.*

### 5.2.3 Key insight #3: Structure and formatting matters

In early versions, question number 11 (QID027e) was much shorter, and no use of formatting was made to highlight or amend information. It was simply worded as follows:

„*What changes when the center of mass of a sailplane shifts forwards?*",

Then the item asked whether aerodynamic lift, drag, weight, speed and angle of attack increases, decreases or stays constant. This led to frequent comments and questions from students and experts asking about details like:

- How far did the weight shift?
- I cannot answer the question. Too little information.
- Does it keep shifting?

To tackle these problems, more information was added (see →Fig. 31) but it led to other questions like:

- What is steady gliding flight?
- How did the weight shift?

Therefore, insertions were made, but put in brackets to ease cognitive load. So the new question version became much longer. Also, the item was consciously broken up into three lines – always at the end of a sense unit (→Fig. 31):

> QID027e: flight ex. , lift, drag
> **11.**
> **Initially**, a sailplane (glider) is in <u>steady gliding flight</u> (losing altitude but maintaining relative airspeed).
> What changes when the <u>center of mass</u> (magically) shifts a little bit forwards
> and the sailplane, then, reaches a <u>constant</u> speed again??

*Fig. 31: Preliminary formatting of Question #11. Working with brackets and underline formatting improved understanding of the item.*

Working with brackets and underline formatting also seemed to improve the student understanding of this longer item. Fewer comments and irritations were observed in think-aloud interviews than in earlier versions. Simultaneously, inserting the words "initially" at an emphasized position in the sentence structure was done to ease cognitive load on the question.

The key learning from the evolution of this, and other items, may be summarized in Albert Einstein's simplicity quote:

> "*It can scarcely be denied that the supreme goal of all theory is to make the irreducible basic elements as simple and as few as possible without having to surrender the adequate representation of a single datum of experience.*"[25]

## 5.3 Rankings & rationales concerning drag (Question 5)

Question 5 of the *FliP-CoIn* instrument asks students which body experiences less aerodynamic drag in 13 pairwise comparisons. All bodies have the same maximum cross-sectional area facing the wind (ø=2cm, see Table 3). According to our data, the most common naïve concept concerning aerodynamic drag is the naïve notion that "*a pointy nose is best*". More than 50% of our students rank B higher than A. This is understandable when listening to their rationales like "*race cars are pointy, too*", "*one has to split the air first*", "*it looks closest to a bullet to me*", and "*the backside is not so important*". Experts know that in subsonic conditions a pointy backside/tail is more usually important than a pointy frontside for avoiding turbulence, and reason more like: "*The high pressure area in front of the body will function as an air cushion, making the body's frontside slippery anyways.*" Of course, an aerodynamically optimized

---

[25] On the Method of Theoretical Physics. Lecture delivered at Oxford, 10 June 1933: https://www.oxfordreference.com/display/10.1093/acref/9780191826719.001.0001/q-oro-ed4-00003988 [2023-02-10]

  Ironically, this quotation is often inaccurately quoted as 'Everything should be made as simple as possible, but not simpler'. There are some rumors that Einstein could have said something more similar to the simple quote above, but no direct sources seem to exist for this claim (https://quoteinvestigator.com/2011/05/13/einstein-simple/ [2023-05-13]) spindle, with a pointy back *and* frontside, is even more slippery, but this is consciously not an option in the assessment.

*Table 3: Drag Bodies sorted from low to high aerodynamic drag: Diameter ø=2cm, Wind speed 25 m/s*

| Drag Bodies (sorted from low to high aerodynamic drag) | | |
|---|---|---|
| a) | drop1 (round side facing wind) | 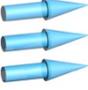 |
| b) | drop2 (sharp side facing wind) | 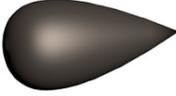 |
| c) | filled half sphere (round side facing wind) | 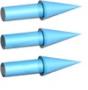 |
| d) | sphere | 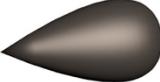 |
| e) | filled half sphere (flat side facing wind) | 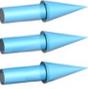 |
| f) | coin / flat cylinder | 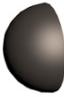 |

By far, the most common misconception is indicated by the fact that usually three quarters of students pick body D over C in direct comparison. This is coherent with the concept of "*avoid sharp edges*", "*a tail is good*", and "*symmetry makes slippiness*". However, it contradicts the measurements. Experts argue that the surface drag is smaller at C, and sometimes also that stable vortexes in the wind shadow behind drag body C form a somewhat pointy tail. Also surprising was the fact that the sphere is often seen as the "*optimal shape*" concerning drag, and is often preferred over all other drag bodies by 14% of students. This might be due to memorized sentences from mathematics or Greek history like "*a sphere is the perfect shape*" bleeding into the physics domain. Since not few students (13%) also hold the naïve concept *"a flat end"* or "*the backside does not matter*", it is particularly hard for these students to decide between F and C or between C and E.

## 5.4 Estimating student consistency: Subitem QID001c_05

In early versions, question 5 (QID001) was a ranking task(→Fig. 32):

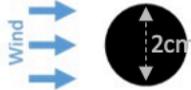

*Fig. 32: Question #5 (QID001 v15): In earlier versions, the item had the format of a ranking task. This caused several problems concerning understanding and scoring (e.g., How to score equal ranks?).*

Later, question 5 (QID001) was broken up into 13 dichotomous sub-items (→Fig. 33):

*Fig. 33: QID001c version final_1.3: Final question 5 (QID001c) was broken up into 13 dichotomous subitems.*

This was done to better identify the underlying student concepts, and spread the cognitive load on more subitems (Brassil & Couch, 2019). Also, it allows measuring the consistency of student thinking and how strongly embedded some naïve concepts are in student minds. Further, it avoids the problem of how to score equal ranks.

Simultaneously, a *repetitive item* was introduced. Sub item QID001c_5 and _1 are identical, but the drag bodies are presented in inverse order. This allowed to estimate another dimension of consistency – the test-retest consistency. Table 4 shows an overall consistency of 92%-95% in datasets 1, 2 and 3. This value seems high at first, but could be higher when considering that this identical question was asked on the *same page* with only three questions in between.

*Table 4: Test retest consistency QID001c_1 vs _5:*

| Test retest consistency QID001c_1 vs _5 | | | | | |
|---|---|---|---|---|---|
| Dataset | 2x correct | 2x incorrect | correct -> incorrect | incorrect -> correct | n= |
| DS1 | 86% | 9% | 3% | 2% | 185 |
| DS2 | 88% | 7% | 5% | 1% | 105 |
| DS3 | 82% | 10% | 6% | 2% | 159 |
| weighted AV | **85%** | **9%** | **4%** | **2%** | **100%** |

Comparing the results of subitem _5 and _1, it stands out that the transition from correct to incorrect answers (4%) is twice as frequent as the reverse transition (2%). This is partially explained by the fact that the *low-scorer ratio* is much higher for the latter category (68% vs 88%). Students in that category are more likely to have guessed (without having any concept in mind). I speculate that being asked the same question again, might have also confused some students, and made them adapt the "survival strategy" of making at least 50% of the points with two inconsistent answers. However, we should not forget that the absolute numbers are very small (only 8 of 449 students changed from incorrect to correct) and other factors as well as random noise might play a role here.

## 5.5 Negatively correlated items

Some items, which showed negative *item-total correlations,* were consciously kept in the final set of items. In this chapter I will discuss why. Ordinarily, one would remove negatively correlating items. A negative *item-total correlation* means that students with a high total score performed poorly at this particular item, and/or students with a low total score performed well on this particular item. Therefore, negative *item-total correlations* usually are interpreted as that these items are worded or scored inversely or that these items are measuring a different construct. However, in the following chapters I will argue why these questions are still valuable and insightful for instructors as well as students, despite tapping into another aspect of lift or drag.

### 5.5.1 Item QID001c_11 / Question Q5.11

The relatively big and negative *item-total correlation* (-.322) of sub item _11 caught our eye before (→Fig. 21 in chapter 4.3.1). Negative *item correlations* usually mean that
   A) the item is scored inversely,
   B) the item wording inverts the meaning or
   C) the items measures a different construct.

Option A and B do not seem to be the case. The item is not word-based, but strongly picture-based[26], and measurements as well as literature (Hoerner, 1965) show that the full sphere experiences more drag than the semi-sphere (round side facing wind)[27]. This seems mysterious, since all other sub items seem to work well, and the question format is essentially the same. However, sub item _11 challenges the naïve concept "*sharp edges cause turbulence*" and "*turbulence is always bad*". A notion that is not challenged so much in the other questions. So sub item _11 touches, in fact, a different and new aspect of aerodynamic drag – which is: Behind the half sphere, standing vortexes can build a cone-like shape (see →Fig. 34). Since these relatively stable vortexes rotate in the direction of the bypassing air stream, both, surface friction and turbulence can be reduced for the surrounding flow.

---

[26] and identical words are used for the drag body descriptions
[27] https://www.engineeringtoolbox.com/drag-coefficient-d_627.html [2023-02-22]

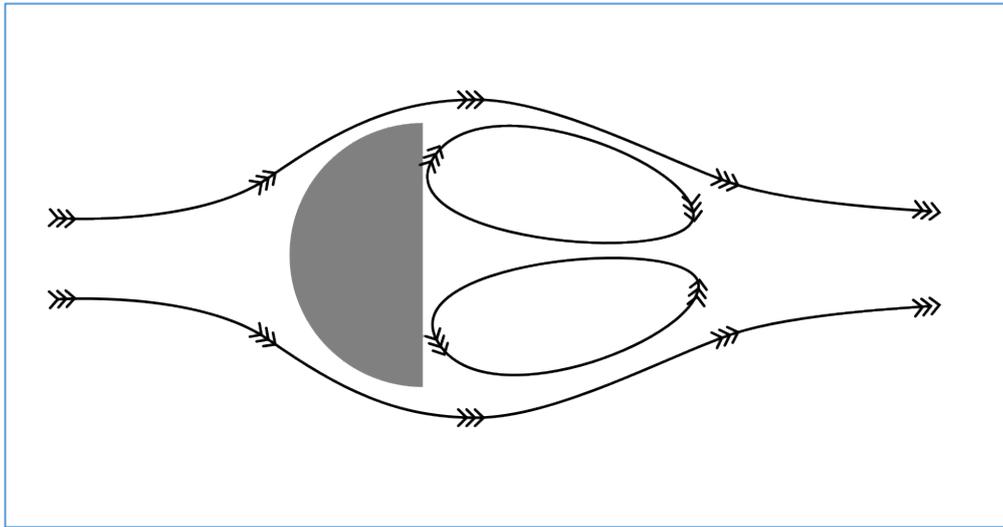

*Fig. 34: Filled semi-sphere: Standing vortexes are dragged along in the wind shadow resulting in a drop-like shape for the surrounding flow. This results in less surface friction, and less turbulence in the surrounding flow.*

Fig. 34 illustrates how stable vortexes can build up behind the semi-sphere, which allow the other air particles to flow around relatively smoothly and laminarly (at the conditions given). The half sphere, plus stable vortexes behind it, forms a hull – similar to a drop shape. This pseudo-drop experiences less drag than the sphere[28] (→Fig. 35). This principle is also used by "*swallow tail*" airfoils (Grasso & Ceyhan, 2015).

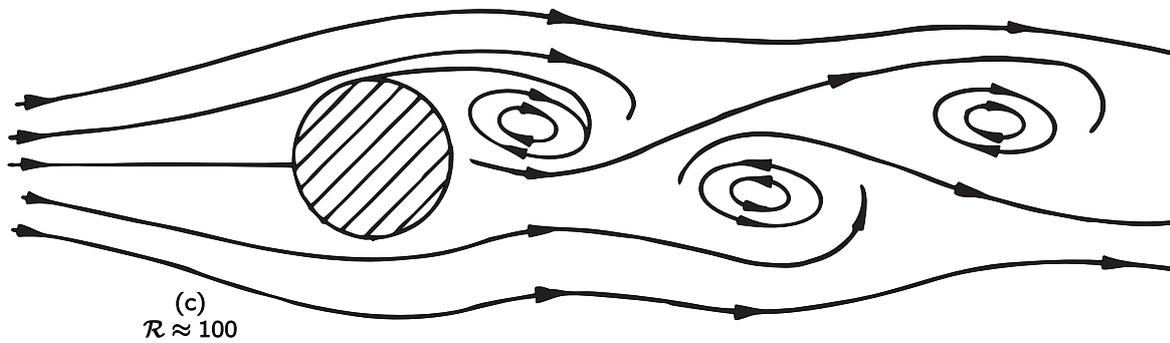

*Fig. 35: Full sphere: Vortexes build up on the backside but leave the periphery of the sphere in alternating motion (Kármán street). This creates more drag than the standing vortexes behind a filled semi-sphere at the airflow conditions of interest. (Picture adapted from Feynman (Feynman, 1964))*

To conclude: It was decided to keep sub item 1c_11, despite lowering the overall internal reliability quite a lot (with subitem QID001c_11 Cronbach's alpha = .72 versus .74 without the subitem). This was done because the subitem 1c_11 touches on an important aspect of aerodynamic drag, which is not tested by any other item. Hence, its negative *item-total correlation* is in accordance with the underlying theory.

---

[28] https://www1.grc.nasa.gov/beginners-guide-to-aeronautics/drag-on-a-sphere/ [2023-03-04]

### 5.5.2 Item QID021c / Question Q19

Due to its outstanding position in the reliability analysis, I would like to discuss Item #21c (Question 19) here. For the original formatting of the question and the answer frequencies, see Fig. 36:

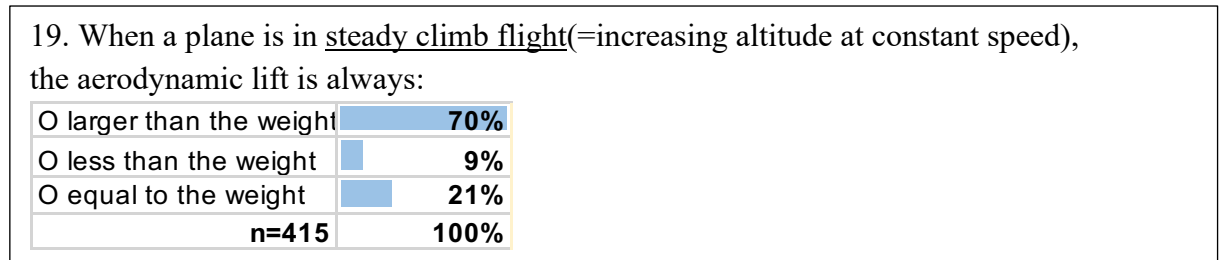

*Fig. 36: Item #21c/Q19 Distractor frequency (combined datasets 1, 2 and 3)*

In the German version of the piloting survey, this question was phrased:
> "*Wenn das Flugzeug im konstanten Steigflug ist (=gewinnt an Höhe bei gleichbleibender Geschwindigkeit), ist der aerodynamische Auftrieb: A) größer als das Gewicht, B) kleiner als das Gewicht, C) genau so groß wie das Gewicht*".

Even though item #21c (Q19) has a relatively strong negative *item-total correlation* (-.25)(see also Fig. 21), it was left in the final item selection after thorough discussion with experts in physics education research and flight physics. Item #21c is the most difficult single question in the inventory, answered by only 9% correctly. In all three datasets (DS1-3) nine high scorers, but 17 low scorers, answered this question correctly. I interpret these data the following: Among n = 409 survey finishers, the difference of 8 (17-9) students could be random noise, and any conclusions should be drawn very carefully. Also, a bigger test tiredness among low-scorers might explain part of the higher prevalence of the correct answer B among them. At this late question in the inventory, low-scorers are likely to be more overwhelmed by the previous questions, have bigger test tiredness, when reaching item #21c (Q19), and might have picked more randomly (and correctly). But when picking at random is a better strategy, it also means there is a highly prevalent naïve conception trapping thoroughly thinking students into the wrong answer:

The reason why distractor A ("*larger than weight*") is so tempting might be due to the fact that free body diagrams are almost exclusively dealing with *horizontal* flight. Therefore, students often forget that there is an upward component of trust in steady climb flight (see Fig. 37). Hence, aerodynamic lift always needs to be smaller than the weight force in steady climb flight. Otherwise, the forces do not balance out, and the flight is unsteady.

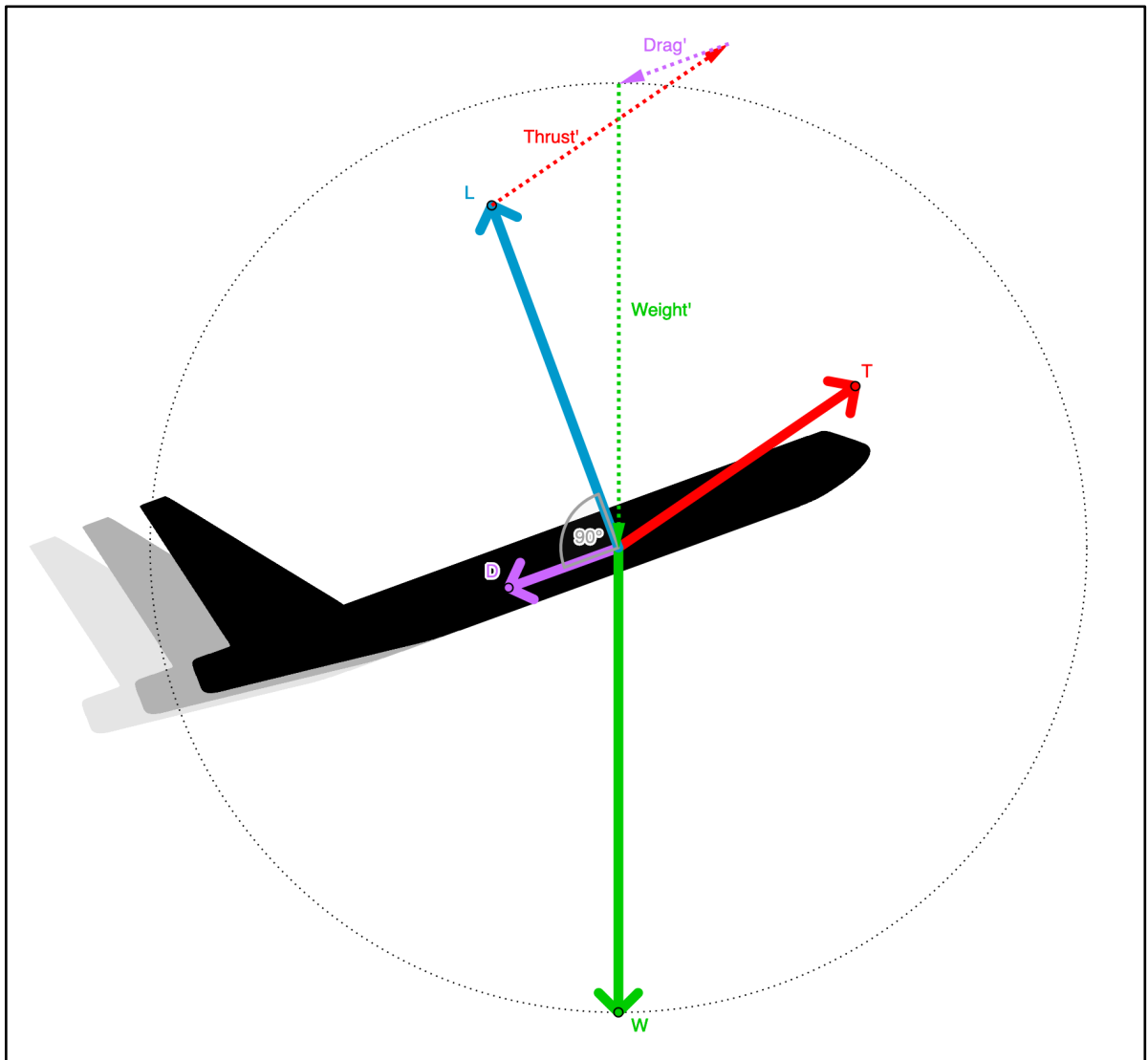

*Fig. 37: Free body diagram: Forces of steady climb flight. All forces need to balance out in total. Thrust (red T), lift (blue L), drag (purple D) and weight (green W) are indicated by solid arrows. The dashed arrows are of identical length and direction as their solid counterparts.*

Moreover, students will pick C ("*equal to the weight*") when only thinking of lift and weight as acting forces, or feel reminded of the elevator Question 17 on the Force Concept Inventory (Eaton et al., 2019; Hestenes et al., 1992). This might indicate that they are in transition to an expert-like understanding of flight physics – holding Newtonian concepts but lack the ability to apply them to all four forces of flight. However, flight is almost never a scenario where only two forces are acting on an airplane. Further, I speculate that in item #21c (Q19) the *lift force* is often confused with the *resulting force,* or *direction of movement*. Beyond, *thrust force* is overlooked as another contributor, since thrust is not explicitly mentioned in the question. Derek Muller found similar effects for the *resulting force* in a scenario where a ball is tossed through the air (Muller, 2008).

To sum this up, retaining the three items (Q5.11, Q11.1 & Q19) with relative strong negative *item-total correlations*, reduces the overall internal reliability of the scale (the overall Cronbach alpha drops from .76 to .73), but it *does* strengthen the scale's *content validity (→3.2.1)*.

## 5.6 Recommendations for further research

Smith and diSessa emphasize that the goal of instruction should not be to simply replace misconceptions for expert concepts, but to provide an experiential basis for the gradual processes of changing the connections between the building blocks (p-prims) for concepts (J. P. Smith & DiSessa, 1993). The foundation for this was laid out for this by developing interactive lecture demonstrations (ILDs) (Sokoloff & Thornton, 1997) based on web quizzes in the third publication accompanying this work (→7.5). The *FliP-CoIn* instrument, will be a great help in improving this newly developed learning intervention further, inspire others, and collect findings about the process of *conceptual change* in the context of *flight physics*.

In the context above, Smith and diSessa also emphasize that "*cognitive conflict*" is not a particular *knowledge state* that leads to the choice of an expert concept over an existing novice conception, but to a more complex pattern of system-level changes that collectively engage many related knowledge elements (J. P. Smith & DiSessa, 1993). This finding calls for a re-validation of similar questions in slightly varying contexts. E.g. marine life, (para-)gliding flight, or insect flight. Also, the validation and feedback process of *the FliP-CoIn instrument* inspired new and modified questions in the given context of *flight physics* – which should be validated in follow-up studies. Beyond, testing *pre/post reliability* and *follow-up reliability* will be a further area of interest, yielding more valuable insights for researchers and practitioners.

*Predictive validity* should be established in *longitudinal studies* and/or matching FliP-CoIn data with other long term predictors like *retention rates*, *growth mindset scores* or *attitudes towards the nature of science* (Galili, 2019). Couse grades are often biased by many other factors and should not be considered a valid proxy. *Concurrent validity*, will become measurable when the FliP-CoIn instrument is given together with other fluid dynamics inventories At best, given in a 2x2 study design, or other study designs minimizing systematic biases. Beyond that, *reproduction studies* from other institutions with bigger populations promise even more insights, especially when analyzing subpopulations and/or *traditionally marginalized groups* in different leaning cultures. In that context, one of the biggest challenges seems to be finding a study program with enough non-white persons, females, and diverse identifying persons in order to investigate biases in migration background and gender on a statistically strong basis.

# 6 Eidesstattliche Erklärung (declaration of honor)

Hiermit versichere ich an Eides statt, dass ich die vorliegende Dissertation selbstständig und ohne die Benutzung anderer als der angegebenen Hilfsmittel und Literatur angefertigt habe. Alle Stellen, die wörtlich oder sinngemäß aus veröffentlichten und nicht veröffentlichten Werken dem Wortlaut oder dem Sinn nach entnommen wurden, sind als solche kenntlich gemacht. Ich versichere an Eides statt, dass diese Dissertation noch keiner anderen Fakultät oder Universität zur Prüfung vorgelegen hat; dass sie - abgesehen von unten angegebenen Teilpublikationen und eingebundenen Artikeln und Manuskripten - noch nicht veröffentlicht worden ist sowie, dass ich eine Veröffentlichung der Dissertation vor Abschluss der Promotion nicht ohne Genehmigung des Promotionsausschusses vornehmen werde. Die Bestimmungen dieser Ordnung sind mir bekannt. Darüber hinaus erkläre ich hiermit, dass ich die Ordnung zur Sicherung guter wissenschaftlicher Praxis und zum Umgang mit wissenschaftlichem Fehlverhalten der Universität zu Köln gelesen und sie bei der Durchführung der Dissertation zugrundeliegenden Arbeiten und der schriftlich verfassten Dissertation beachtet habe und verpflichte mich hiermit, die dort genannten Vorgaben bei allen wissenschaftlichen Tätigkeiten zu beachten und umzusetzen. Ich versichere, dass die eingereichte elektronische Fassung der eingereichten Druckfassung vollständig entspricht.

Versionen:

- Ordnung zur Sicherung guter wissenschaftlicher Praxis der Universität zu Köln vom 25. Januar 2022
- Ordnung zur Untersuchung wissenschaftlichen Fehlverhaltens der Universität zu Köln vom 25. Januar 2022
- Promotionsordnung der Mathematisch-Naturwissenschaftlichen Fakultät der Universität zu Köln vom 12. März 2020

________________

Florian Genz

# 7 Contributions and role of accompanying publications

Parts of this research have been published in condensed form and with differing foci in the three accompanying publications preceding this work. This work elaborates in much more detail about the state of the art in Concept Inventory development, the development process, methodological decisions involved, the data, results, analyses, and, last but not least, discussions about the beforementioned.

The first publication describes how educational practices and goals made the development of the FliP-CoIn instrument necessary (Genz & Vieyra, 2015); the second publication describes how data from the FliP-CoIn instrument inspired new leaning theory (Genz & Falconer, 2021); and the third publication how these research findings translate back to educational practice (Genz et al., 2023). The chapters of sub Section 7 will prime a very brief overview about the publications connected with this dissertation and their role in the development process of the *Flight Physics Concept Inventory*.

The following table gives an overview about each author's contribution to the respective publication (→Table 5):

*Table 5: Author contribution by category and publication[29]*

|  | Publication1: "Evaluating the Use of Flight Simulators for the NASA/AAPT 'Aeronautics for Introductory Physics' Educator Guide" | | Publication2: "Naïve concepts of aerodynamic lift – data lessons from different (learning) cultures" | | Publication3: "The Flight Physics Concept Inventory – Reliably Evaluting Aerodynamic Lift, Drag and Associated (Naïve) Concepts of Flight in Class & In-Game" | | |
|---|---|---|---|---|---|---|---|
|  | Florian Genz | Rebecca E. Vieyra | Florian Genz | Kathleen A. Falconer | Florian Genz | André Bresges | Kathleen A. Falconer |
| Wrote the manuscript | ✓ |  | ✓ |  | ✓ |  |  |
| Participated in conception and/or design of manuscript | ✓ | ✓ | ✓ | ✓ | ✓ |  | ✓ |
| Edited the manuscript | ✓ | ✓ | ✓ | ✓ | ✓ |  | ✓ |
| Consultant |  | ✓ |  | ✓ |  | ✓ | ✓ |
| Financial Support |  |  |  |  |  | ✓ |  |
| Mentor/Sponsor/or Other Support |  |  |  | ✓ |  |  | ✓ |

---

[29] according to CourseSource Physics journal: https://www.coursesource.org/for_authors [2023-02-21]

In addition, my advisor was also involved as helpful consultant, mentor and supporter throughout my whole PhD period.

## 7.1 Publication 1

The first publication, titled

> *"Evaluating the Use of Flight Simulators for the NASA/AAPT 'Aeronautics for Introductory Physics' Educator Guide",*

emerged from the practical lab experience while teaching the advanced lab "FP06 – experiments about flight". The observation that some naïve concepts only become visible when students need to apply their theoretical knowledge led to the development of the interactive inquiry lab described in the publication.

## 7.2 Author contributions for publication 1

The first author in this publication produced the text, all primary data, early piloting versions of the FliP-CoIn instrument, photos, figures, graphs, and the feedback rubric, while using and implementing ideas from the 'Aeronautics for Introductory Physics' educator guide (Vieyra et al., 2015) in the learning intervention described. The second author of this publication improved and edited the manuscript. The second author also provided feedback and ideas as well as worldwide and US comparison data (EPA 2015, IPCC 2014). Further, she was a discussion partner for improving on the structure and framing of the publication together with the first author.

## 7.3 Publication 2

The second publication, named

> *"Naïve concepts of aerodynamic lift – data lessons from different (learning) cultures",*

was driven by observations made while analyzing the survey data from the FliP-CoIn instrument. It describes a new theory of looking at naïve concepts by considering the Model Merging Phenomena (MMP): In short, the MMP describes the observation that when imitations of two coexisting models remain obscure, students try to merge the two models into one. The gap is often filled by naïve model aspects outside the real world phenomenon. The model break remains unconscious for students.

Further, this publication reconfirmed that naïve concepts do coexist with expert concepts and even with other – conflicting – naïve concepts. The key finding of this publication was that the coexistence phenomenon seems to be even more prevalent among high scoring students.

## 7.4 Author contributions for publication 2

The first author in this publication wrote the manuscript, and produced all learning materials and attachments, including photos, figures, and graphs. The design of the manuscript underwent several iterations in discussion with the second author. The second author of this publication contributed by editing the manuscript, providing mentoring support and acted as a consultant.

## 7.5 Publication 3

The third publication, titled

> *"The Flight Physics Concept Inventory – Reliably Evaluating Aerodynamic Lift, Drag and Associated (Naïve) Concepts of Flight in Class & In-Game"*, was written to disseminate the findings and new theory (from the second publication) into practice and to inspire teaching with digital media – especially in tertiary education of big lectures. The publication's motivation is to re-translate the rather abstract research findings into practical application in the classroom again. It sketches three interactive lectures with active student engagement. The lessons described contrast the students' knowledge with naïve concepts in flight physics – but without individual students needing to risk exposing their own naïve concepts before a big group. An important requirement for inclusive teaching (Sheehy, 2017).

## 7.6 Author contributions for publication 3

The first author in this publication wrote the manuscript, and produced all learning materials and attachments, including photos, figures, and graphs. The second author of this publication provided financial support as well as feedback on the manuscript, some edits and ideas. The third author of this publication contributed with editing the manuscript, providing mentoring support and acted as a consultant. Also, the structure of the manuscript experienced several iterations in discussion with the third author.

# 9  Appendix

Also available at

https://drive.google.com/drive/folders/1DOmH38WY3PZc6hZtrOPTKhP3tPCSr3mZ?usp=sharing

## 9.1  Gender bias statistics

The following statistics are exported from the significance tests conducted with SPSS28. They were used to enhance chapter 4.1 and Fig. 19:

DS123 female vs male
Notes

| | | |
|---|---|---|
| Output Created | | 19-MAY-2023 21:13:27 |
| Comments | | |
| Input | Data | 2023-05 gender bias.sav |
| | Active Dataset | 2023-05 gender bias.sav |
| | Filter | <none> |
| | Weight | <none> |
| | Split File | <none> |
| | N of Rows in Working Data File | 543 |
| Missing Value Handling | Definition of Missing | User-defined missing values are treated as missing. |
| | Cases Used | Statistics for each test are based on all cases with valid data for the variable(s) used in that test. |
| Syntax | | NPAR TESTS /M-W= ptsTOTAL BY gender(1 2) /MISSING ANALYSIS /METHOD=EXACT TIMER(1). |
| Resources | Processor Time | 00:00:06.23 |
| | Elapsed Time | 00:00:06.00 |
| | Number of Cases Allowed[a] | 449389 |
| | Time for Exact Statistics | 0:00:06.13 |

a. Based on availability of workspace memory.

Mann-Whitney Test
Ranks

| | gender | N | Mean Rank | Sum of Ranks |
|---|---|---|---|---|
| ptsTOTAL | 1 | 53 | 123.12 | 6525.50 |
| | 2 | 328 | 201.97 | 66245.50 |
| | Total | 381 | | |

Test Statistics[a]

| | ptsTOTAL |
|---|---|
| Mann-Whitney U | 5094.500 |
| Wilcoxon W | 6525.500 |
| Z | -4.846 |
| Asymp. Sig. (2-tailed) | <.001 |
| Exact Sig. (2-tailed) | <.001 |
| Exact Sig. (1-tailed) | <.001 |
| Point Probability | .000 |

a. Grouping Variable: gender

# 9 Appendix

DS123 male vs other

Notes

| | | |
|---|---|---|
| Output Created | | 19-MAY-2023 21:15:30 |
| Comments | | |
| Input | Data | 2023-05 gender bias.sav |
| | Active Dataset | 2023-05 gender bias.sav |
| | Filter | <none> |
| | Weight | <none> |
| | Split File | <none> |
| | N of Rows in Working Data File | 543 |
| Missing Value Handling | Definition of Missing | User-defined missing values are treated as missing. |
| | Cases Used | Statistics for each test are based on all cases with valid data for the variable(s) used in that test. |
| Syntax | | NPAR TESTS /M-W= ptsTOTAL BY gender(2 3) /MISSING ANALYSIS /METHOD=EXACT TIMER(5). |
| Resources | Processor Time | 00:00:00.50 |
| | Elapsed Time | 00:00:00.00 |
| | Number of Cases Allowed[a] | 449389 |
| | Time for Exact Statistics | 0:00:00.48 |

a. Based on availability of workspace memory.

Mann-Whitney Test

Ranks

| | gender | N | Mean Rank | Sum of Ranks |
|---|---|---|---|---|
| ptsTOTAL | 2 | 328 | 173.95 | 57054.50 |
| | 3 | 21 | 191.45 | 4020.50 |
| | Total | 349 | | |

Test Statistics[a]

| | ptsTOTAL |
|---|---|
| Mann-Whitney U | 3098.500 |
| Wilcoxon W | 57054.500 |
| Z | -.772 |
| Asymp. Sig. (2-tailed) | .440 |
| Exact Sig. (2-tailed) | .443 |
| Exact Sig. (1-tailed) | .221 |
| Point Probability | .000 |

a. Grouping Variable: gender

# 9 Appendix

DS1 female vs male

Notes

| | | |
|---|---|---|
| Output Created | | 19-MAY-2023 21:19:29 |
| Comments | | |
| Input | Data | 2023-05 gender bias.sav |
| | Active Dataset | 2023-05 gender bias.sav |
| | Filter | dataset=1 (FILTER) |
| | Weight | <none> |
| | Split File | <none> |
| | N of Rows in Working Data File | 270 |
| Missing Value Handling | Definition of Missing | User-defined missing values are treated as missing. |
| | Cases Used | Statistics for each test are based on all cases with valid data for the variable(s) used in that test. |
| Syntax | | NPAR TESTS /M-W= ptsTOTAL BY gender(1 2) /MISSING ANALYSIS /METHOD=EXACT TIMER(5). |
| Resources | Processor Time | 00:00:00.03 |
| | Elapsed Time | 00:00:00.00 |
| | Number of Cases Allowed[a] | 449389 |
| | Time for Exact Statistics | 0:00:00.03 |

a. Based on availability of workspace memory.

Mann-Whitney Test

Ranks

| | gender | N | Mean Rank | Sum of Ranks |
|---|---|---|---|---|
| ptsTOTAL | 1 | 10 | 31.90 | 319.00 |
| | 2 | 122 | 69.34 | 8459.00 |
| | Total | 132 | | |

Test Statistics[a]

| | ptsTOTAL |
|---|---|
| Mann-Whitney U | 264.000 |
| Wilcoxon W | 319.000 |
| Z | -2.984 |
| Asymp. Sig. (2-tailed) | .003 |
| Exact Sig. (2-tailed) | .002 |
| Exact Sig. (1-tailed) | .001 |
| Point Probability | .000 |

a. Grouping Variable: gender

# 9 Appendix

DS2 female vs male
Notes

| | | |
|---|---|---|
| Output Created | | 19-MAY-2023 21:23:42 |
| Comments | | |
| Input | Data | 2023-05 gender bias.sav |
| | Active Dataset | 2023-05 gender bias.sav |
| | Filter | dataset=2 (FILTER) |
| | Weight | <none> |
| | Split File | <none> |
| | N of Rows in Working Data File | 106 |
| Missing Value Handling | Definition of Missing | User-defined missing values are treated as missing. |
| | Cases Used | Statistics for each test are based on all cases with valid data for the variable(s) used in that test. |
| Syntax | | NPAR TESTS /M-W= ptsTOTAL BY gender(1 2) /MISSING ANALYSIS /METHOD=EXACT TIMER(5). |
| Resources | Processor Time | 00:00:00.03 |
| | Elapsed Time | 00:00:00.00 |
| | Number of Cases Allowed[a] | 449389 |
| | Time for Exact Statistics | 0:00:00.02 |

a. Based on availability of workspace memory.

Mann-Whitney Test
Ranks

| | gender | N | Mean Rank | Sum of Ranks |
|---|---|---|---|---|
| ptsTOTAL | 1 | 14 | 23.68 | 331.50 |
| | 2 | 80 | 51.67 | 4133.50 |
| | Total | 94 | | |

Test Statistics[a]

| | ptsTOTAL |
|---|---|
| Mann-Whitney U | 226.500 |
| Wilcoxon W | 331.500 |
| Z | -3.551 |
| Asymp. Sig. (2-tailed) | <.001 |
| Exact Sig. (2-tailed) | <.001 |
| Exact Sig. (1-tailed) | <.001 |
| Point Probability | .000 |

a. Grouping Variable: gender

DS3 female vs male
Notes

# 9 Appendix

| Output Created | | 19-MAY-2023 21:24:52 |
|---|---|---|
| Comments | | |
| Input | Data | 2023-05 gender bias.sav |
| | Active Dataset | 2023-05 gender bias.sav |
| | Filter | dataset=3 (FILTER) |
| | Weight | <none> |
| | Split File | <none> |
| | N of Rows in Working Data File | 167 |
| Missing Value Handling | Definition of Missing | User-defined missing values are treated as missing. |
| | Cases Used | Statistics for each test are based on all cases with valid data for the variable(s) used in that test. |
| Syntax | | NPAR TESTS /M-W= ptsTOTAL BY gender(1 2) /MISSING ANALYSIS /METHOD=EXACT TIMER(5). |
| Resources | Processor Time | 00:00:00.33 |
| | Elapsed Time | 00:00:01.00 |
| | Number of Cases Allowed[a] | 449389 |
| | Time for Exact Statistics | 0:00:00.31 |

a. Based on availability of workspace memory.

Mann-Whitney Test
Ranks

| | gender | N | Mean Rank | Sum of Ranks |
|---|---|---|---|---|
| ptsTOTAL | 1 | 29 | 66.57 | 1930.50 |
| | 2 | 126 | 80.63 | 10159.50 |
| | Total | 155 | | |

Test Statistics[a]

| | ptsTOTAL |
|---|---|
| Mann-Whitney U | 1495.500 |
| Wilcoxon W | 1930.500 |
| Z | -1.527 |
| Asymp. Sig. (2-tailed) | .127 |
| Exact Sig. (2-tailed) | .128 |
| Exact Sig. (1-tailed) | .064 |
| Point Probability | .000 |

a. Grouping Variable: gender

# 9 Appendix

DS1 male vs other

Notes

| | | |
|---|---|---|
| Output Created | | 19-MAY-2023 21:28:49 |
| Comments | | |
| Input | Data | 2023-05 gender bias.sav |
| | Active Dataset | 2023-05 gender bias.sav |
| | Filter | dataset=3 (FILTER) |
| | Weight | <none> |
| | Split File | <none> |
| | N of Rows in Working Data File | 167 |
| Missing Value Handling | Definition of Missing | User-defined missing values are treated as missing. |
| | Cases Used | Statistics for each test are based on all cases with valid data for the variable(s) used in that test. |
| Syntax | | NPAR TESTS /M-W= ptsTOTAL BY gender(2 3) /MISSING ANALYSIS /METHOD=EXACT TIMER(5). |
| Resources | Processor Time | 00:00:00.03 |
| | Elapsed Time | 00:00:00.00 |
| | Number of Cases Allowed[a] | 449389 |
| | Time for Exact Statistics | 0:00:00.02 |

a. Based on availability of workspace memory.

Mann-Whitney Test

Ranks

| | gender | N | Mean Rank | Sum of Ranks |
|---|---|---|---|---|
| ptsTOTAL | 2 | 126 | 66.36 | 8361.00 |
| | 3 | 6 | 69.50 | 417.00 |
| | Total | 132 | | |

Test Statistics[a]

| | ptsTOTAL |
|---|---|
| Mann-Whitney U | 360.000 |
| Wilcoxon W | 8361.000 |
| Z | -.197 |
| Asymp. Sig. (2-tailed) | .844 |
| Exact Sig. (2-tailed) | .849 |
| Exact Sig. (1-tailed) | .425 |
| Point Probability | .002 |

a. Grouping Variable: gender

# 9 Appendix

DS2 male vs other

Notes

| | | |
|---|---|---|
| Output Created | | 19-MAY-2023 21:30:57 |
| Comments | | |
| Input | Data | 2023-05 gender bias.sav |
| | Active Dataset | 2023-05 gender bias.sav |
| | Filter | dataset=2 (FILTER) |
| | Weight | <none> |
| | Split File | <none> |
| | N of Rows in Working Data File | 106 |
| Missing Value Handling | Definition of Missing | User-defined missing values are treated as missing. |
| | Cases Used | Statistics for each test are based on all cases with valid data for the variable(s) used in that test. |
| Syntax | | NPAR TESTS /M-W= ptsTOTAL BY gender(2 3) /MISSING ANALYSIS /METHOD=EXACT TIMER(5). |
| Resources | Processor Time | 00:00:00.04 |
| | Elapsed Time | 00:00:00.00 |
| | Number of Cases Allowed[a] | 449389 |
| | Time for Exact Statistics | 0:00:00.03 |

a. Based on availability of workspace memory.

## Mann-Whitney Test

Ranks

| | gender | N | Mean Rank | Sum of Ranks |
|---|---|---|---|---|
| ptsTOTAL | 2 | 80 | 46.13 | 3690.50 |
| | 3 | 12 | 48.96 | 587.50 |
| | Total | 92 | | |

Test Statistics[a]

| | ptsTOTAL |
|---|---|
| Mann-Whitney U | 450.500 |
| Wilcoxon W | 3690.500 |
| Z | -.343 |
| Asymp. Sig. (2-tailed) | .732 |
| Exact Sig. (2-tailed) | .737 |
| Exact Sig. (1-tailed) | .369 |
| Point Probability | .002 |

a. Grouping Variable: gender

# 9 Appendix

DS3 male vs other
Notes

| | | |
|---|---|---|
| Output Created | | 19-MAY-2023 21:31:58 |
| Comments | | |
| Input | Data | 2023-05 gender bias.sav |
| | Active Dataset | 2023-05 gender bias.sav |
| | Filter | dataset=3 (FILTER) |
| | Weight | <none> |
| | Split File | <none> |
| | N of Rows in Working Data File | 167 |
| Missing Value Handling | Definition of Missing | User-defined missing values are treated as missing. |
| | Cases Used | Statistics for each test are based on all cases with valid data for the variable(s) used in that test. |
| Syntax | | NPAR TESTS /M-W= ptsTOTAL BY gender(2 3) /MISSING ANALYSIS /METHOD=EXACT TIMER(5). |
| Resources | Processor Time | 00:00:00.02 |
| | Elapsed Time | 00:00:00.00 |
| | Number of Cases Allowed[a] | 449389 |
| | Time for Exact Statistics | 0:00:00.02 |

a. Based on availability of workspace memory.

Mann-Whitney Test
Ranks

| | gender | N | Mean Rank | Sum of Ranks |
|---|---|---|---|---|
| ptsTOTAL | 2 | 126 | 66.36 | 8361.00 |
| | 3 | 6 | 69.50 | 417.00 |
| | Total | 132 | | |

Test Statistics[a]

| | ptsTOTAL |
|---|---|
| Mann-Whitney U | 360.000 |
| Wilcoxon W | 8361.000 |
| Z | -.197 |
| Asymp. Sig. (2-tailed) | .844 |
| Exact Sig. (2-tailed) | .849 |
| Exact Sig. (1-tailed) | .425 |
| Point Probability | .002 |

a. Grouping Variable: gender

# 9 Appendix

## 9.2 Gender distribution of DS1: Pearson-Chi² test

**NPar Tests**

| | Notes | | |
|---|---|---|---|
| Output Created | | | 23-MAY-2023 16:12:59 |
| Comments | | | |
| Input | Active Dataset | | DataSet1 |
| | Filter | | <none> |
| | Weight | | <none> |
| | Split File | | <none> |
| | N of Rows in Working Data File | | 136 |
| Missing Value Handling | Definition of Missing | | User-defined missing values are treated as missing. |
| | Cases Used | | Statistics for each test are based on all cases with valid data for the variable(s) used in that test. |
| Syntax | | | NPAR TESTS /CHISQUARE=gender_2male_1nonmale /EXPECTED=15 120 /MISSING ANALYSIS /METHOD=EXACT TIMER(5). |
| Resources | Processor Time | | 00:00:00.02 |
| | Elapsed Time | | 00:00:00.00 |
| | Number of Cases Allowed[a] | | 786432 |
| | Time for Exact Statistics | | 0:00:00.01 |

a. Based on availability of workspace memory.

## Chi-Square Test
### Frequencies

**gender_2male_1non-male**

| | Observed N | Expected N | Residual |
|---|---|---|---|
| non-male | 13 | 15.1 | -2.1 |
| male | 123 | 120.9 | 2.1 |
| Total | 136 | | |

**Test Statistics**

| | gender_2male_1non-male |
|---|---|
| Chi-Square | .332[a] |
| df | 1 |
| Asymp. Sig. | **.565** |
| Exact Sig. | **.592** |
| Point Probability | .097 |

a. 0 cells (0.0%) have expected frequencies less than 5. The minimum expected cell frequency is 15.1.

# 9 Appendix

## 9.3 Total score distributions

**DS1 Frequencies**

### Notes

| | | |
|---|---|---|
| Output Created | | 23-MAY-2023 19:08:11 |
| Comments | | |
| Input | Active Dataset | 2023-05 total_points_distribution - Chi tests.sav |
| | Filter | <none> |
| | Weight | <none> |
| | Split File | <none> |
| | N of Rows in Working Data File | 130 |
| Missing Value Handling | Definition of Missing | User-defined missing values are treated as missing. |
| | Cases Used | Statistics are based on all cases with valid data. |
| Syntax | | FREQUENCIES VARIABLES=frequency_of_point_category_in_dataset1 /STATISTICS=STDDEV MEAN MEDIAN SKEWNESS SESKEW KURTOSIS SEKURT /HISTOGRAM NORMAL /ORDER=ANALYSIS. |
| Resources | Processor Time | 00:00:00.15 |
| | Elapsed Time | 00:00:00.00 |

### Statistics

frequency_of_point_category_in_dataset1

| N | Valid | 130 |
|---|---|---|
| | Missing | 0 |
| Mean | | 3.95 |
| Median | | 4.00 |
| Std. Deviation | | 1.308 |
| Skewness | | -.659 |
| Std. Error of Skewness | | .212 |
| Kurtosis | | -.048 |
| Std. Error of Kurtosis | | .422 |

### frequency_of_point_category_in_dataset1

| | | Frequency | Percent | Valid Percent | Cumulative Percent |
|---|---|---|---|---|---|
| Valid | 18to22points | 10 | 7.7 | 7.7 | 7.7 |
| | 23to25points | 6 | 4.6 | 4.6 | 12.3 |
| | 26to29points | 26 | 20.0 | 20.0 | 32.3 |
| | 30to33points | 37 | 28.5 | 28.5 | 60.8 |
| | 34to37points | 41 | 31.5 | 31.5 | 92.3 |
| | 38to42points | 10 | 7.7 | 7.7 | 100.0 |
| | Total | 130 | 100.0 | 100.0 | |

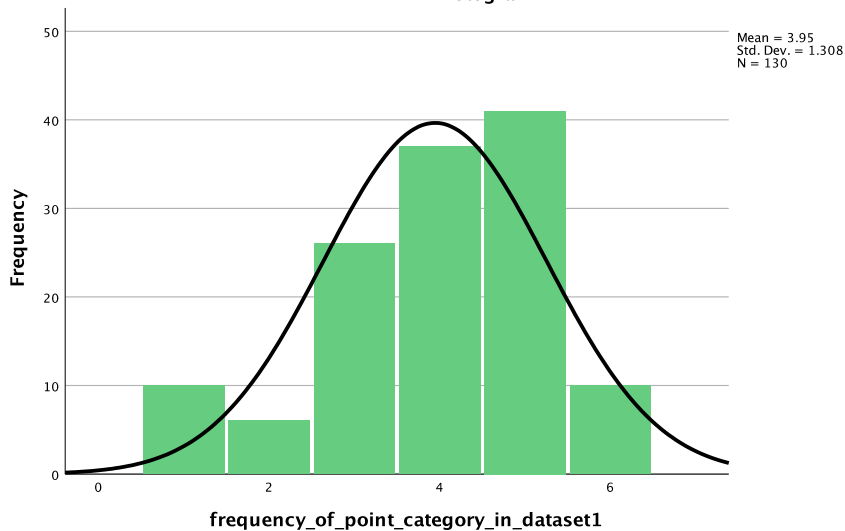

Histogram

Mean = 3.95
Std. Dev. = 1.308
N = 130

# 9 Appendix

### Notes

| | | |
|---|---|---|
| Output Created | | 23-MAY-2023 18:54:43 |
| Comments | | |
| Input | Active Dataset | 2023-05 total_points_distribution - Chi tests.sav |
| | Filter | <none> |
| | Weight | <none> |
| | Split File | <none> |
| | N of Rows in Working Data File | 167 |
| Missing Value Handling | Definition of Missing | User-defined missing values are treated as missing. |
| | Cases Used | Statistics for each test are based on all cases with valid data for the variable(s) used in that test. |
| Syntax | | NPAR TESTS /CHISQUARE=frequency_of_point_category_in_dataset1 /EXPECTED=3.189 13.352 31.161 40.536 29.392 11.879 /MISSING ANALYSIS. |
| Resources | Processor Time | 00:00:00.01 |
| | Elapsed Time | 00:00:00.00 |
| | Number of Cases Allowed[a] | 786432 |

a. Based on availability of workspace memory.

**Chi-Square Test**

**frequency_of_point_category_in_dataset1**

| | Observed N | Expected N | Residual |
|---|---|---|---|
| 18to22points | 10 | 3.2 | 6.8 |
| 23to25points | 6 | 13.4 | -7.4 |
| 26to29points | 26 | 31.3 | -5.3 |
| 30to33points | 37 | 40.7 | -3.7 |
| 34to37points | 41 | 29.5 | 11.5 |
| 38to42points | 10 | 11.9 | -1.9 |
| Total | 130 | | |

**Test Statistics**

| | frequency_of_point_category_in_dataset1 |
|---|---|
| Chi-Square | 24.545[a] |
| df | 5 |
| Asymp. Sig. | **<.001** |

a. 1 cells (16.7%) have expected frequencies less than 5. The minimum expected cell frequency is 3.2.

# 9 Appendix

## DS2 Frequencies

**Notes**

| | | |
|---|---|---|
| Output Created | | 23-MAY-2023 19:10:06 |
| Comments | | |
| Input | Active Dataset | 2023-05 total_points_distribution - Chi tests.sav |
| | Filter | <none> |
| | Weight | <none> |
| | Split File | <none> |
| | N of Rows in Working Data File | 104 |
| Missing Value Handling | Definition of Missing | User-defined missing values are treated as missing. |
| | Cases Used | Statistics are based on all cases with valid data. |
| Syntax | | FREQUENCIES VARIABLES=frequency_of_point_category_in _dataset2 /STATISTICS=STDDEV MEAN MEDIAN SKEWNESS SESKEW KURTOSIS SEKURT /HISTOGRAM NORMAL /ORDER=ANALYSIS. |
| Resources | Processor Time | 00:00:00.12 |
| | Elapsed Time | 00:00:00.00 |

**Statistics**

frequency_of_point_category_in_dataset2

| N | Valid | 104 |
|---|---|---|
| | Missing | 0 |
| Mean | | 4.11 |
| Median | | 4.00 |
| Std. Deviation | | 1.238 |
| Skewness | | -.643 |
| Std. Error of Skewness | | .237 |
| Kurtosis | | .209 |
| Std. Error of Kurtosis | | .469 |

**frequency_of_point_category_in_dataset2**

| | | Frequency | Percent | Valid Percent | Cumulative Percent |
|---|---|---|---|---|---|
| Valid | 18to22points | 5 | 4.8 | 4.8 | 4.8 |
| | 23to25points | 5 | 4.8 | 4.8 | 9.6 |
| | 26to29points | 18 | 17.3 | 17.3 | 26.9 |
| | 30to33points | 33 | 31.7 | 31.7 | 58.7 |
| | 34to37points | 32 | 30.8 | 30.8 | 89.4 |
| | 38to42points | 11 | 10.6 | 10.6 | 100.0 |
| | Total | 104 | 100.0 | 100.0 | |

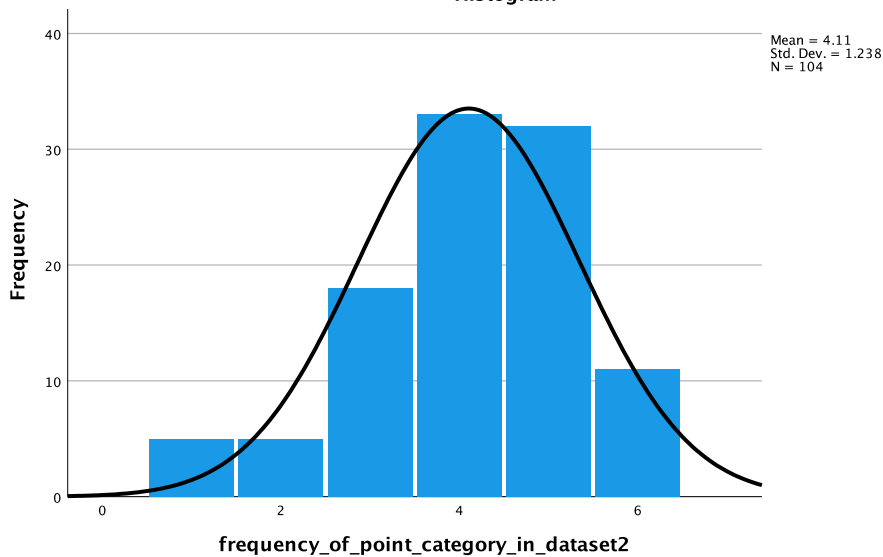

Histogram

Mean = 4.11
Std. Dev. = 1.238
N = 104

# 9 Appendix

|  | Notes |  |
|---|---|---|
| Output Created | | 23-MAY-2023 18:57:08 |
| Comments | | |
| Input | Active Dataset | 2023-05_total_points_distribution - Chi tests.sav |
| | Filter | <none> |
| | Weight | <none> |
| | Split File | <none> |
| | N of Rows in Working Data File | 167 |
| Missing Value Handling | Definition of Missing | User-defined missing values are treated as missing. |
| | Cases Used | Statistics for each test are based on all cases with valid data for the variable(s) used in that test. |
| Syntax | | NPAR TESTS /CHISQUARE=frequency_of_point_category_in_dataset2 /EXPECTED=1.428 7.842 22.421 33.382 25.882 10.450 /MISSING ANALYSIS. |
| Resources | Processor Time | 00:00:00.01 |
| | Elapsed Time | 00:00:00.00 |
| | Number of Cases Allowed[a] | 786432 |

a. Based on availability of workspace memory.

**Chi-Square Test**

**frequency_of_point_category_in_dataset2**

| | Observed N | Expected N | Residual |
|---|---|---|---|
| 18to22points | 5 | 1.5 | 3.5 |
| 23to25points | 5 | 8.0 | -3.0 |
| 26to29points | 18 | 23.0 | -5.0 |
| 30to33points | 33 | 34.2 | -1.2 |
| 34to37points | 32 | 26.5 | 5.5 |
| 38to42points | 11 | 10.7 | .3 |
| Total | 104 | | |

**Test Statistics**

| | frequency_of_point_category_in_dataset2 |
|---|---|
| Chi-Square | 11.944[a] |
| df | 5 |
| Asymp. Sig. | .036 |

a. 1 cells (16.7%) have expected frequencies less than 5. The minimum expected cell frequency is 1.5.

## DS3 Frequencies

### Notes

| | | |
|---|---|---|
| Output Created | | 23-MAY-2023 19:06:17 |
| Comments | | |
| Input | Active Dataset | 2023-05 total_points_distribution - Chi tests.sav |
| | Filter | <none> |
| | Weight | <none> |
| | Split File | <none> |
| | N of Rows in Working Data File | 154 |
| Missing Value Handling | Definition of Missing | User-defined missing values are treated as missing. |
| | Cases Used | Statistics are based on all cases with valid data. |
| Syntax | | FREQUENCIES VARIABLES=frequency_of_point_category_in_dataset3 /STATISTICS=STDDEV MEAN MEDIAN SKEWNESS SESKEW KURTOSIS SEKURT /HISTOGRAM NORMAL /ORDER=ANALYSIS. |
| Resources | Processor Time | 00:00:00.15 |
| | Elapsed Time | 00:00:00.00 |

### Statistics

frequency_of_point_category_in_dataset3

| | | |
|---|---|---|
| N | Valid | 154 |
| | Missing | 0 |
| Mean | | 3.08 |
| Median | | 3.00 |
| Std. Deviation | | 1.003 |
| Skewness | | -.092 |
| Std. Error of Skewness | | .195 |
| Kurtosis | | -.257 |
| Std. Error of Kurtosis | | .389 |

### frequency_of_point_category_in_dataset3

| | | Frequency | Percent | Valid Percent | Cumulative Percent |
|---|---|---|---|---|---|
| Valid | 18to22points | 10 | 6.5 | 6.5 | 6.5 |
| | 23to25points | 29 | 18.8 | 18.8 | 25.3 |
| | 26to29points | 65 | 42.2 | 42.2 | 67.5 |
| | 30to33points | 38 | 24.7 | 24.7 | 92.2 |
| | 34to37points | 12 | 7.8 | 7.8 | 100.0 |
| | Total | 154 | 100.0 | 100.0 | |

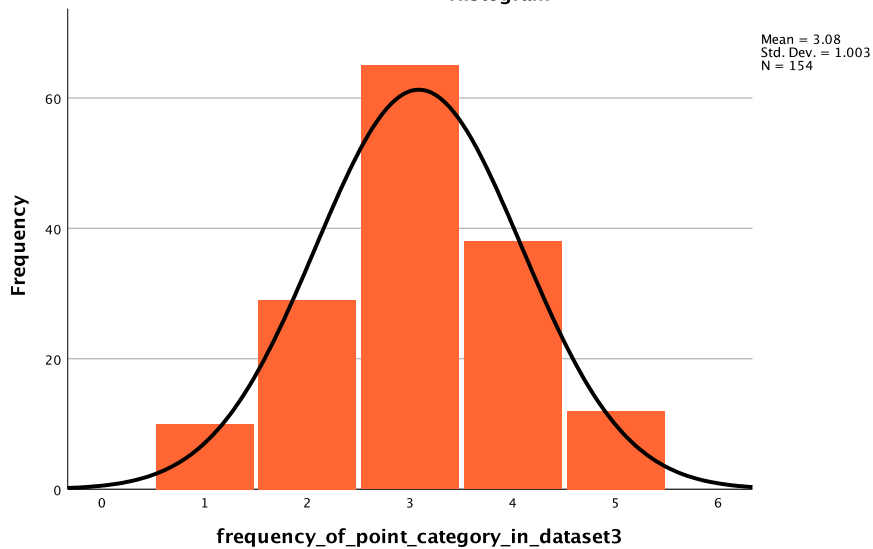

Histogram

Mean = 3.08
Std. Dev. = 1.003
N = 154

# 9 Appendix

**Notes**

| | | |
|---|---|---|
| Output Created | | 23-MAY-2023 18:59:52 |
| Comments | | |
| Input | Active Dataset | 2023-05 total_points_distribution - Chi tests.sav |
| | Filter | <none> |
| | Weight | <none> |
| | Split File | <none> |
| | N of Rows in Working Data File | 167 |
| Missing Value Handling | Definition of Missing | User-defined missing values are treated as missing. |
| | Cases Used | Statistics for each test are based on all cases with valid data for the variable(s) used in that test. |
| Syntax | | NPAR TESTS /CHISQUARE=frequency_of_point_category_in_dataset3 /EXPECTED=7.133 34.505 61.059 40.219 9.804 /MISSING ANALYSIS. |
| Resources | Processor Time | 00:00:00.01 |
| | Elapsed Time | 00:00:00.00 |
| | Number of Cases Allowed[a] | 786432 |

a. Based on availability of workspace memory.

**Chi-Square Test**

**frequency_of_point_category_in_dataset3**

| | Observed N | Expected N | Residual |
|---|---|---|---|
| 18to22points | 10 | 7.2 | 2.8 |
| 23to25points | 29 | 34.8 | -5.8 |
| 26to29points | 65 | 61.6 | 3.4 |
| 30to33points | 38 | 40.6 | -2.6 |
| 34to37points | 12 | 9.9 | 2.1 |
| Total | 154 | | |

**Test Statistics**

| | frequency_of_point_category_in_dataset3 |
|---|---|
| Chi-Square | 2.865[a] |
| df | 4 |
| Asymp. Sig. | .581 |

a. 0 cells (.0%) have expected frequencies less than 5. The minimum expected cell frequency is 7.2.

## 9.4 Scoring data

File: "2023-04-21 SPSS DS123 IA v8(for EXPORT).sav"
See digital Appendix.

# 9 Appendix

## 9.5 Internal reliability statistics

### 9.5.1 Formatted Excel export data: sorted by question order

| | | | | | | | | | | | | | | Internal Reliability | | | | |
|---|---|---|---|---|---|---|---|---|---|---|---|---|---|---|---|---|---|---|
| | | | | | | | | | | | | | | Scale: DS123 internal reliability | | | | |
| | | | | | | | | | | | | | | Case Processing Summary | | | | |
| | | | | | | | | | | | | | | | | N | % | |
| | | | | | | | | | | | | | | Cases Valid | | 274 | 50,5 | |
| | | | | | | | | | | | | | | Excluded[a] | | 269 | 49,5 | |
| | | | | | | | | | | | | | | Total | | 543 | 100,0 | |
| | | | | | | | | | | | | | | a. Listwise deletion based on all variables in the procedure. | | | | |
| | | | | | | | | | | | | | | Reliability Statistics | | | | |
| | | | | | | | | | | | | | | Cronbach's Alpha | | N of Items | | |
| | | | | | | | | | | | | | | .725 | | 46 | | |
| Level | | | Type | | | | | | | | | | | Item Statistics for all Datasets (DS1, DS2, DS3) | | | | |
| Knowl edge | Applic ation | Predic tion | text-based | picture-based | lift | drag | stall | CoM | AoA | strL. | exp. | Domain | Sub domain | Que stion No. | QuestionID | Mean | SD | Corrected Item-Total Correlation | Cronbach's Alpha if Item Deleted |
| | x | | | x | | | | | | | | Aerodynamic Drag | Direction | 1 | pts20 | 0,61 | 0,488 | 0,21 | 0,720 |
| | | x | | x | 1 | | | | | | 1 | Aerodynamic Lift: | Cause: | 2 | pts6b | 0,69 | 0,462 | 0,30 | 0,715 |
| | x | | | x | | 1 | | | | | | Aerodynamic Drag | Factors | 4 | pts001b_empiricAWs_1342 | 0,10 | 0,303 | 0,18 | 0,721 |
| | | | | | | 1 | | | | | | Aerodynamic Drag | Factors | 5.01 | pts1c_01 | 0,90 | 0,303 | 0,40 | 0,713 |
| | | | | | | 1 | | | | | | Aerodynamic Drag | Factors | 5.02 | pts1c_02 | 0,47 | 0,500 | 0,06 | 0,728 |
| | | | | | | 1 | | | | | | Aerodynamic Drag | Factors | 5.03 | pts1c_03 | 0,91 | 0,288 | 0,18 | 0,721 |
| | | | | | | 1 | | | | | | Aerodynamic Drag | Factors | 5.04 | pts1c_04 | 0,86 | 0,346 | 0,36 | 0,714 |
| | | | | | | 1 | | | | | | Aerodynamic Drag | Factors | 5.05 | pts1c_05 | 0,88 | 0,326 | 0,43 | 0,712 |
| | | | | | | 1 | | | | | | Aerodynamic Drag | Factors | 5.06 | pts1c_06 | 0,96 | 0,188 | 0,37 | 0,718 |
| | x | | | x | | 1 | | | | | | Aerodynamic Drag | Factors | 5.07 | pts1c_07 | 0,93 | 0,248 | 0,30 | 0,718 |
| | | | | | | 1 | | | | | | Aerodynamic Drag | Factors | 5.08 | pts1c_08 | 0,75 | 0,433 | 0,29 | 0,716 |
| | | | | | | 1 | | | | | | Aerodynamic Drag | Factors | 5.09 | pts1c_09 | 0,94 | 0,242 | 0,45 | 0,714 |
| | | | | | | 1 | | | | | | Aerodynamic Drag | Factors | 5.10 | pts1c_10 | 0,84 | 0,364 | 0,16 | 0,722 |
| | | | | | | 1 | | | | | | Aerodynamic Drag | Factors | 5.11 | pts_1c_11_corrF | 0,22 | 0,414 | -0,32 | 0,744 |
| | | | | | | 1 | | | | | | Aerodynamic Drag | Factors | 5.12 | pts1c_12 | 0,95 | 0,213 | 0,40 | 0,716 |
| | | | | | | 1 | | | | | | Aerodynamic Drag | Factors | 5.13 | pts1c_13 | 0,75 | 0,435 | 0,20 | 0,720 |
| | x | | x | | | 1 | | | | | | Aerodynamic Drag | Factors | 6 | pts29 | 0,74 | 0,437 | 0,20 | 0,720 |
| | x | | x | | | 1 | | | | | | Aerodynamic Drag | Factors | 7 | pts30 | 0,59 | 0,493 | 0,19 | 0,721 |
| | x | | x | | | 1 | | | | | | Aerodynamic Drag | Factors | 8 | pts30b | 0,77 | 0,419 | 0,28 | 0,716 |
| x | | | | x | | | | 1 | | | | Center of Mass | relative posi | 9 | pts27a | 0,77 | 0,419 | 0,27 | 0,717 |
| | | | | | 1 | | | 1 | | | 1 | Center of Mass | Lift | 11.1 | pts_27e_A_corrF | 0,46 | 0,500 | -0,10 | 0,737 |
| | | | | | | 1 | | 1 | | | 1 | Center of Mass | Drag | 11.2 | pts27e_B | 0,50 | 0,501 | 0,01 | 0,731 |
| | x | | x | | | | | 1 | | | 1 | Center of Mass | Weight | 11.3 | pts27e_C | 0,91 | 0,294 | 0,29 | 0,718 |
| | | | | | | | | 1 | | | 1 | Center of Mass | (relative air) | 11.4 | pts27e_D | 0,52 | 0,500 | 0,19 | 0,721 |
| | | | | | | | | 1 | | | 1 | Center of Mass | sinking veloc | 11.5 | pts27e_E | 0,77 | 0,422 | 0,17 | 0,722 |
| | | | | | | | | 1 | 1 | | 1 | Center of Mass | Angle of Atta | 11.6 | pts27e_F | 0,51 | 0,501 | 0,14 | 0,724 |
| | x | | | x | 1 | | | | | 1 | | Airfoil profiles & str | air displacem | 12 | pts006c | 0,60 | 0,491 | 0,21 | 0,720 |
| | x | | | x | 1 | | | | | 1 | | Airfoil profiles & str | air displacem | 13 | pts6 | 0,39 | 0,489 | 0,08 | 0,727 |
| x | | | | x | | | | 1 | | | 1 | Center of Mass | relative pos. | 14 | pts27d | 0,62 | 0,486 | 0,22 | 0,719 |
| | x | | | x | 1 | | | | 1 | | | Aerodynamic Lift: | Direction: | 15 | pts17 | 0,28 | 0,449 | 0,11 | 0,725 |
| | x | | | x | 1 | | | | 1 | | | Aerodynamic Lift: | Direction: | 16 | pts18 | 0,31 | 0,462 | 0,12 | 0,724 |
| | x | | | x | 1 | | | | | | | Aerodynamic Lift: | Direction: | 17 | pts19 | 0,74 | 0,439 | 0,49 | 0,705 |
| | x | | | x | 1 | | | | | | | Aerodynamic Lift: | Direction: | 18 | pts19b | 0,69 | 0,462 | 0,37 | 0,711 |
| | x | | x | | 1 | | | | | | | Aerodynamic Lift: | Magnitude | 19 | pts21c_corrF | 0,09 | 0,294 | -0,33 | 0,738 |
| | | x | | x | | | | | | | 1 | Flight experience | force due to | 20.4 | pts21b_D_fd2gravity | 0,82 | 0,387 | 0,36 | 0,713 |
| | | | | | | | | | | | 1 | Flight experience | altitude (hei | 20.5 | pts21b_E_altitude | 0,63 | 0,484 | 0,25 | 0,717 |
| | | x | x | | | | | | | | 1 | Flight experience | altitude (hei | 21.4 | pts21_D_gravity | 0,71 | 0,454 | 0,32 | 0,714 |
| | | | | | | | | | | | 1 | Flight experience | force due to | 21.5 | pts21_E_altitude | 0,92 | 0,267 | 0,21 | 0,720 |
| | | x | x | | | | | | | | 1 | Flight experience | Ground spee | 22.2 | pts24_B_ground_speed | 0,75 | 0,435 | 0,38 | 0,711 |
| | | | | | | | | | | | 1 | Flight experience | altitude (hei | 22.4 | pts24_D_altitude | 0,81 | 0,396 | 0,33 | 0,714 |
| | | | | | 1 | | | | | | 1 | Flight experience | Drag | 23.1 | pts24b_A_dir_o_drag | 0,62 | 0,487 | 0,04 | 0,729 |
| | | | | | 1 | | | | | | 1 | Flight experience | Lift | 23.2 | pts24b_B_dir_o_lift | 0,64 | 0,480 | 0,15 | 0,723 |
| | | | | | | | | | | | 1 | Flight experience | direction of | 23.3 | pts24b_C_dir_o_gravity | 0,80 | 0,399 | 0,27 | 0,717 |
| | x | | x | | | | | | | | 1 | Flight experience | Wind influe | 24 | pts35 | 0,35 | 0,477 | 0,27 | 0,717 |
| x | | | x | | | | | | | | 1 | Flight experience | Forces on pil | 25 | pts36 | 0,68 | 0,469 | 0,15 | 0,723 |
| | x | | | x | | | 1 | | | | | Aerodynamic stall | Start of stall | 26 | pts33_oBC | 0,74 | 0,439 | 0,27 | 0,717 |
| | | | | | | | | | | | | | | | | | | N=274 | |

*Fig. 38: Formatted internal reliability statistics: sorted by question order*

# 9 Appendix

## 9.5.2 Formatted Excel export data: sorted by correlation value

| **Reliability Statistics** | | | | | |
|---|---|---|---|---|---|
| Cronbach's Alpha | N of Items | | | | |
| **.725** | 46 | | | | |
| | *Sorted by Correlation* | | | | |

| | **Item Statistics for all Datasets (DS1, DS2, DS3)** | | | | Cronbach's Alpha if Item Deleted |
|---|---|---|---|---|---|
| Question No. | QuestionID | Mean | AV mean | Corrected Item-Total Correlation | |
| 17 | pts19 | 0,74 | | 0,49 | 0,705 |
| 5.09 | pts1c_09 | 0,94 | | 0,45 | 0,714 |
| 5.05 | pts1c_05 | 0,88 | | 0,43 | 0,712 |
| 5.01 | pts1c_01 | 0,90 | | 0,40 | 0,713 |
| 5.12 | pts1c_12 | 0,95 | | 0,40 | 0,716 |
| 22.2 | pts24_B_ground_speed | 0,75 | **0,85** | 0,38 | 0,711 |
| 5.06 | pts1c_06 | 0,96 | | 0,37 | 0,718 |
| 18 | pts19b | 0,69 | | 0,37 | 0,711 |
| 5.04 | pts1c_04 | 0,86 | | 0,36 | 0,714 |
| 20.4 | pts21b_D_fd2gravity | 0,82 | | 0,36 | 0,713 |
| 22.4 | pts24_D_altitude | 0,81 | | 0,33 | 0,714 |
| 21.4 | pts21_D_gravity | 0,71 | | 0,32 | 0,714 |
| 2 | pts6b | 0,69 | | 0,30 | 0,715 |
| 5.07 | pts1c_07 | 0,93 | | 0,30 | 0,718 |
| 5.08 | pts1c_08 | 0,75 | | 0,29 | 0,716 |
| 11.3 | pts27e_C | 0,91 | | 0,29 | 0,718 |
| 8 | pts30b | 0,77 | | 0,28 | 0,716 |
| 23.3 | pts24b_C_dir_o_gravity | 0,80 | | 0,27 | 0,717 |
| 24 | pts35 | 0,35 | | 0,27 | 0,717 |
| 26 | pts33_oBC | 0,74 | | 0,27 | 0,717 |
| 9 | pts27a | 0,77 | | 0,27 | 0,717 |
| 20.5 | pts21b_E_altitude | 0,63 | | 0,25 | 0,717 |
| 14 | pts27d | 0,62 | | 0,22 | 0,719 |
| 21.5 | pts21_E_altitude | 0,92 | | 0,21 | 0,720 |
| 1 | pts20 | 0,61 | | 0,21 | 0,720 |
| 12 | pts006c | 0,60 | | 0,21 | 0,720 |
| 5.13 | pts1c_13 | 0,75 | | 0,20 | 0,720 |
| 6 | pts29 | 0,74 | | 0,20 | 0,720 |
| 11.4 | pts27e_D | 0,52 | | 0,19 | 0,721 |
| 7 | pts30 | 0,59 | | 0,19 | 0,721 |
| 5.03 | pts1c_03 | 0,91 | | 0,18 | 0,721 |
| 4 | pts001b_empiricAWs_1342_123 | 0,10 | | 0,18 | 0,721 |
| 11.5 | pts27e_E | 0,77 | | 0,17 | 0,722 |
| 5.10 | pts1c_10 | 0,84 | | 0,16 | 0,722 |
| 23.2 | pts24b_B_dir_o_lift | 0,64 | | 0,15 | 0,723 |
| 25 | pts36 | 0,68 | | 0,15 | 0,723 |
| 11.6 | pts27e_F | 0,51 | | 0,14 | 0,724 |
| 16 | pts18 | 0,31 | | 0,12 | 0,724 |
| 15 | pts17 | 0,28 | | 0,11 | 0,725 |
| 13 | pts6 | 0,39 | | 0,08 | 0,727 |
| 5.02 | pts1c_02 | 0,47 | **0,39** | 0,06 | 0,728 |
| 23.1 | pts24b_A_dir_o_drag | 0,62 | | 0,04 | 0,729 |
| 11.2 | pts27e_B | 0,50 | | 0,01 | 0,731 |
| 11.1 | pts_27e_A_corrF | 0,46 | | -0,10 | 0,737 |
| 5.11 | pts_1c_11_corrF | 0,22 | | -0,32 | 0,744 |
| 19 | pts21c_corrF | 0,09 | | -0,33 | 0,738 |
| | | | | **N=274** | |

*Fig. 39: Formatted internal reliability statistics: sorted by correlation value. Including AV mean of the top10 and bottom10*

# 9 Appendix

## 9.5.3 Formatted Excel export data: sorted by mean (easiness of item)

| Reliability Statistics | | | | |
|---|---|---|---|---|
| Cronbach's | N of Items | | | |
| .725 | 46 | | | |
| | Sorted by Mean | | | |

| | Item Statistics for all Datasets (DS1, DS2, DS3) | | | Cronbach's Alpha if Item Deleted |
|---|---|---|---|---|
| Question No. | QuestionID | Mean | Corrected Item-Total Correlation | |
| 5.06 | pts1c_06 | 0,96 | 0,37 | 0,718 |
| 5.12 | pts1c_12 | 0,95 | 0,40 | 0,716 |
| 5.09 | pts1c_09 | 0,94 | 0,45 | 0,714 |
| 5.07 | pts1c_07 | 0,93 | 0,30 | 0,718 |
| 21.5 | pts21_E_altitude | 0,92 | 0,21 | 0,720 |
| 5.03 | pts1c_03 | 0,91 | 0,18 | 0,721 |
| 11.3 | pts27e_C | 0,91 | 0,29 | 0,718 |
| 5.01 | pts1c_01 | 0,90 | 0,40 | 0,713 |
| 5.05 | pts1c_05 | 0,88 | 0,43 | 0,712 |
| 5.04 | pts1c_04 | 0,86 | 0,36 | 0,714 |
| 5.10 | pts1c_10 | 0,84 | 0,16 | 0,722 |
| 20.4 | pts21b_D_fd2gravity | 0,82 | 0,36 | 0,713 |
| 22.4 | pts24_D_altitude | 0,81 | 0,33 | 0,714 |
| 23.3 | pts24b_C_dir_o_gravity | 0,80 | 0,27 | 0,717 |
| 8 | pts30b | 0,77 | 0,28 | 0,716 |
| 9 | pts27a | 0,77 | 0,27 | 0,717 |
| 11.5 | pts27e_E | 0,77 | 0,17 | 0,722 |
| 5.08 | pts1c_08 | 0,75 | 0,29 | 0,716 |
| 5.13 | pts1c_13 | 0,75 | 0,20 | 0,720 |
| 22.2 | pts24_B_ground_speed | 0,75 | 0,38 | 0,711 |
| 6 | pts29 | 0,74 | 0,20 | 0,720 |
| 17 | pts19 | 0,74 | 0,49 | 0,705 |
| 26 | pts33_oBC | 0,74 | 0,27 | 0,717 |
| 21.4 | pts21_D_gravity | 0,71 | 0,32 | 0,714 |
| 2 | pts6b | 0,69 | 0,30 | 0,715 |
| 18 | pts19b | 0,69 | 0,37 | 0,711 |
| 25 | pts36 | 0,68 | 0,15 | 0,723 |
| 23.2 | pts24b_B_dir_o_lift | 0,64 | 0,15 | 0,723 |
| 20.5 | pts21b_E_altitude | 0,63 | 0,25 | 0,717 |
| 14 | pts27d | 0,62 | 0,22 | 0,719 |
| 23.1 | pts24b_A_dir_o_drag | 0,62 | 0,04 | 0,729 |
| 1 | pts20 | 0,61 | 0,21 | 0,720 |
| 12 | pts006c | 0,60 | 0,21 | 0,720 |
| 7 | pts30 | 0,59 | 0,19 | 0,721 |
| 11.4 | pts27e_D | 0,52 | 0,19 | 0,721 |
| 11.6 | pts27e_F | 0,51 | 0,14 | 0,724 |
| 11.2 | pts27e_B | 0,50 | 0,01 | 0,731 |
| 5.02 | pts1c_02 | 0,47 | 0,06 | 0,728 |
| 11.1 | pts_27e_A_corrF | 0,46 | -0,10 | 0,737 |
| 13 | pts6 | 0,39 | 0,08 | 0,727 |
| 24 | pts35 | 0,35 | 0,27 | 0,717 |
| 16 | pts18 | 0,31 | 0,12 | 0,724 |
| 15 | pts17 | 0,28 | 0,11 | 0,725 |
| 5.11 | pts_1c_11_corrF | 0,22 | -0,32 | 0,744 |
| 4 | pts001b_empiricAWs_1342 | 0,10 | 0,18 | 0,721 |
| 19 | pts21c_corrF | 0,09 | -0,33 | 0,738 |
| | | | N=274 | |

*Fig. 40: Formatted internal reliability statistics: sorted by mean (=ratio of correct answers in %)*

### 9.5.4 SPSS export data: combined datasets DS123 & DS1,2,3 separately

File: "2023-04-21 SPSS DS123 IA v8 (EXPORT).pdf"

## 9.6 Codebook comparison

See also chapter → 3.8.1. All three codebooks (of Datasets 1,2 and 3) were compared horizontally (question by question) in reverse order by the author. The codebook comparison instructions were the following:

Codebook comparison instructions:
       1. compare the variable name of the question
       2. compare the question text (ignore formatting).
       3. compare the distractors
       4. compare distractor order
       5. compare codes
       6. repeat for next question

The digital codebooks of DS1 and DS2 were copied next to each other in a spreadsheet. Then the coding manual for the manual DS3 was put next to both digital codebooks. Cell values from DS2 were copy pasted to the yet empty column for DS3 in the comparison spreadsheet when the visual inspection according to the *codebook comparison instructions,* described above, proved to be identical to the printed out coding manual of DS3. Wherever differences were observed these cells were marked red for real differences or orange for minor differences (i.e. minor variable naming variations with no effect on the results or statistics. Compare QID021). The results of the codebook comparison were in detail:

*Table 6: Codebook Differences of datasets 1, 2 and 3:*

| Codebook Differences of datasets 1,2 and 3: | |
|---|---|
| Real differences: | |
| intentional | 1. QID033 was modified for DS2 & 3 therefore codebooks varied in amount of distractors and, hence, their codings |
| unintentional | 2.QID024: third distractor "stays constant"/"bleibt gleich" was coded with "4" at DS 1&2 whereas it was coded "3" in DS3) |
| unintentional | 3. QID021c differing distractor order and distractor codes in DS 1, 2 and 3. |
| intentional | 4. QID006d was modified for DS2 & 3 therefore codebooks varied in amount of distractors, question format and codings. |
| intentional | 5. QID030b one distractor was added in DS1&2. Distractor D="the backside does not matter / all the same." was not yet added in DS3 |
| Minor differences (no effect on coding or scoring) | |
| intentional | 1. After QID036 the coders of the manual DS3 had to code the a variable not represented by an item ("all words used once"). |
| intentional | 2. After QID034b: An extra variable "all_used_once" had to be coded in DS3 |
| unintentional | 3. QID021: Some minor variable naming variations at with no effect on the results or statistics |
| unintentional | 4. QID000 uncertainties in translation |
| intentional | 5. QID028 was open ended for DS2 & 3. |
| intentional | 6. QID030b in DS1 translation of distractors varies due to different grammar in the translated question. |
| intentional | 7. QID001b has an additional hint in the English version (" 2cm is about 1inch."). |

## 9.7 PhysPort.org assessments on 29th June 2020

Additional information for chapter 2.1.3. In the following list all PhysPort.org assessments in the category "Content knowledge" are enumerated as online on 29th June 2020. No concept inventory for flight physics or similar topics was listed online:

1. Force Concept Inventory (FCI)
2. Force and Motion Conceptual Evaluation (FMCE)
3. Test of Understanding Graphs in Kinematics (TUG-K)
4. Energy and Momentum Conceptual Survey (EMCS)
5. Rotational Kinematics Inventory (RKI)
6. Colorado Classical Mechanics/Math Methods Instrument (CCMI)
7. Energy Concept Assessment (ECA)
8. Half-length Force Concept Inventory (HFCI)
9. Inventory of Basic Conceptions - Mechanics (IBCM)
10. Next Gen Physical Science Diagnostic (NGPSD)
11. Rotational and Rolling Motion Conceptual Survey (RRMCS)
12. Mechanics Baseline Test (MBT)
13. Representational Variant of the Force Concept Inventory (R-FCI)
14. Force, Velocity, and Acceleration Test (FVA)
15. Density Survey (DS)
16. Brief Electricity and Magnetism Assessment (BEMA)
17. Colorado Upper Division Electrostatics Diagnostic - Free Response (CUE-FR)
18. Determining and Interpreting Resistive Electric Circuit Concepts Test (DIRECT)
19. Conceptual Survey of Electricity and Magnetism (CSEM)
20. Colorado Upper Division Electrostatics Diagnostic - Coupled Multiple Response (CUE-CMR)
21. Magnetism Conceptual Survey (MCS)
22. Rate and Potential Test (RAPT)
23. Inventory of Basic Conceptions - DC Circuits (IBCDC)
24. Symmetry and Gauss's Law Conceptual Evaluation (SGCE)
25. Colorado UppeR-division ElectrodyNamics Test (CURrENT)
26. Diagnostic Exam for Introductory, Undergraduate Electricity and Magnetism (DEEM)
27. Electric Circuits Conceptual Evaluation (ECCE)
28. Electricity and Magnetism Conceptual Assessment (EMCA)

xx9 Appendix

29. Electromagnetics Concept Inventory (EMCI)
30. Wave Diagnostic Test (WDT)
31. Mechanical Wave Conceptual Survey (MWCS)
32. Four-tier Geometrical Optics Test (FTGOT)
33. Mechanical Waves Conceptual Survey 2 (MWCS2)
34. Wave Concept Inventory (WCI)
35. Thermal Concept Evaluation (TCE)
36. Survey of Thermodynamic Processes and First and Second Laws (STPFaSL)
37. Thermodynamic Concept Survey (TCS)
38. Thermal and Transport Concept Inventory: Thermodynamics (TTCI-T)
39. Heat and Temperature Conceptual Evaluation (HTCE)
40. Thermodynamics Concept Inventory (TCI)
41. Quantum Mechanics Concept Assessment (QMCA)
42. Quantum Mechanics Conceptual Survey (QMCS)
43. Quantum Mechanics Formalism and Postulates Survey (QMFPS)
44. Quantum Mechanics Survey (QMS)
45. Quantum Physics Conceptual Survey (QPC
46. Quantum Mechanics Visualization Instrument (QMVI)
47. Relativity Concept Inventory (RCI)
48. Quantum Mechanics Assessment Tool (QMAT)
49. Quantum Mechanics Concept Inventory (QMCI)
50. Calculus Concept Inventory (CCI)
51. Precalculus Concept Assessment (PCA)
52. Test of Understanding of Vectors (TUV)
53. Mathematical Modeling Conceptual Evaluation (MMCE)
54. Quadratic and Linear Conceptual Evaluation (QLCE)
55. Vector Evaluation Test (VET)
56. Astronomy Diagnostic Test 2.0 (ADT2)
57. Star Properties Concept Inventory (SPCI)
58. Light and Spectroscopy Concept Inventory (LSCI)
59. Newtonian Gravity Concept Inventory (NGCI)
60. Test of Astronomy Standards (TOAST)
61. Greenhouse Effect Concept Inventory (GECI)
62. Lunar Phases Concept Inventory (LPCI)
63. Astronomical Misconceptions Survey (AMS)

## 9.8   Scoring error corrections

A comparison between the RAs of the 3 datasets, also enabled the discovery of 3 scoring errors for QID027e_C, _F and _G for dataset 2, due to wrong copying of formulas. Therefore the scoring of all 3 datasets was re-evaluated, and all RAs were run again (→ Chap. 3.8.2).

The re-evaluation of all scoring calculations was conducted by the following methods:
1. A codebook comparison was conducted (→Chap. 9.6). All three codebooks were compared item by item and distractor by distractor.

# 9 Appendix

2. Questions with differences in the codebook comparison were re-evaluated again. Then all questions were checked again with a special attention on typical places where the mistakes found so far occurred.
3. All score calculation formulars were re-checked and re-calculated if differences were discovered. The method is also described in the following video named "*Formular re-check Method for FliP-CoIn's DS1,2,3 scoring calculations and descriptive Data*": https://youtu.be/RKK2zj0-uY4.

Further minor mistakes were discovered:
1. Score calculation error: In DS1, DS2, and DS3 the calculation of NAs for variable pts_034b was changed from NA="" to NA=0 (e.g. for DS3: in cell LX576 it was changed from
   =IF(LS576=0;"";IF(LV576="NA-77";"";IF(OR(LV576=1423;LV576=2314);1;0)))
   into
   =IF(OR(LV576=1423;LV576=2314);1;0) .
   This was done, since it does not matter for the scoring if a question was falsely answered or NOT answered. If this *inaccuracy* was not detected, this could have had a slight influence on data analysis and definitively on the scoring on item QID_34b.

   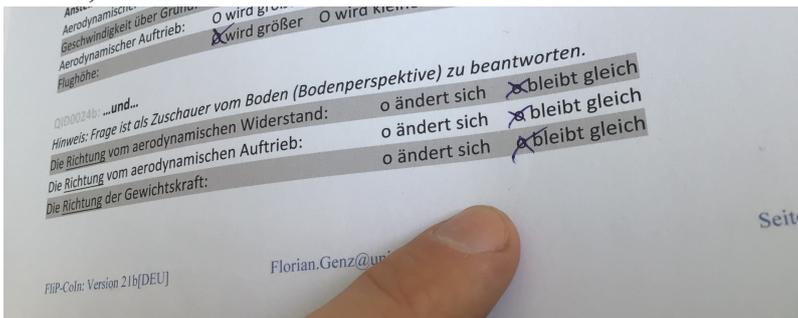

   However, even after correction QID_34b was still not used for the final itemset because it did not satisfy the criteria of the item analysis.

2. Coding error: In DS3 the coding of v_QID_24b was differing for coder 1 and 2 and by looking up the original test booklet it was found that coder 1 made the *same* mistake as coder 2, and decided to code the answers for test booklet 39 as "3,3,3" instead of "2,2,2".

   The author caught this due to the re-check process for all formulas (see also Video "*Formular re-check Method*": https://youtu.be/RKK2zj0-uY4) [2022-09-22]
   If not caught, this could have had a slight influence on the reliability analysis since for this participant the score for item variable v_QID_24b_C changes from 0 (=false) to 1 (=correct). However, the influence of one datapoint on the whole reliability analysis is very low (→ 3.8.1.2).

# 9 Appendix

[Screenshots of Excel spreadsheets showing KU721 and KU756 cell formulas and data tables, with an arrow indicating correction from one state to the other.]

In the following only mistakes were discovered with *no* influence on the reliability analysis:

3. Variable *pts21_D_decreasesANDconstant_correct* in dataset1 and 2 was recalculated. However, this variable was only introduces temporarily and it was never planned to include it in the final reliability analysis anyways (➔ No influence on the reliability analysis).

4. In DS2 the NA count for v_QID021b_B to _E was corrected. This was an inherited error due to a missing "$" symbol in the cell references of cell IX501 (➔ No influence on the reliability analysis).

[Screenshot of Excel spreadsheet showing formula =COUNTIF(IY$383:IY$489;IW501) and associated data cells.]

5. In DS2 the count for the third distractor of v_QID024 was corrected. The error occurred due to a wrong cell reference and the assumption that the third distractor is always coded with "3". The (correct) $-symbol in the formular of Cell KF494 led to the error that miscounted the cell value. Formular =COUNTIF(KF$383:KF$489;$**JL**494) was changed to =COUNTIF(KF$383:KF$489;$**KD**494). This is a non-critical error, since it did not change any scoring and only slightly impacted the descriptive results (➔ No

9 Appendix

influence on the reliability analysis). It could have been discovered earlier, since the checksums did not add up to 107 as typical for DS2.

6. In DS2 the NA count calculation of v_example_understood (cell AK494) was =COUNTIF(AK$38**2**:AK$489;AI494) instead of
   =COUNTIF(AK$38**3**:AK$489;AI494).
   No influence on the count or scoring. (➔ No influence on the reliability analysis).

7. In DS2 the N count for (cell ES521) was corrected from "=SUM(ES516:ES5**17**)" to "=SUM(ES516:ES5**20**)". No influence on the raw count or scoring. (➔ No influence on the reliability analysis).

8. In DS3 several formulars of "NA counts" were only counting "0" and were adapted accordingly to also count "f"(e.g. for v_QID032, v_QID020, v_QID006b) (➔ No influence on the reliability analysis).

9. In DS3 the cell references for v_example_understood were corrected. This statistic was not used so far. No influence on any data analyses therefore (➔ No influence on the reliability analysis).

10. In DS3 calculation for pts1c_ALLcorrect was not yet implemented. Formulars were copied from DS2 and re-checked on plausibility and consistency. (➔ No influence on the reliability analysis, since not used in the final itemset).

   •

11. In DS3 calculation of n and NA for v_QID_006d was corrected. No influence on scoring and data analysis (➔ No influence on the reliability analysis).

   •

12. In DS3 the calculation of n for NA for v_QID_024b was corrected. No influence on scoring and data analysis (➔ No influence on the reliability analysis).

# 9 Appendix

[Figure: Screenshot showing two Excel spreadsheet views with formula =SUM(KS754:KS755), displaying data rows 737-756 with columns for DataSet, student, Testheft lfdn, extern tester, v_QID024b_A/B/C, showing counts for "changes", "remains con...", "question skipped", n=, NA-77, and Sum values.]

13. In DS3 the calculation for NA and "repeated word error" calculations (Q34b_pattern in Column LV) was corrected from
    "=IF(LS576="NA-**77**";"NA-**77**";…)" to
    "=IF(LS576="NA";"NA";…)". This had no influence on scoring and the rest of the data analysis. Only on the ratio of NA to "repeated word error" answers (➔ No influence on the reliability analysis).
14. In DS3 calculation of NA for v_QID035 was corrected. No influence on scoring and data analysis (➔ No influence on the reliability analysis).
15. In DS3 calculation of NA for v_QID036 was corrected. No influence on scoring and data analysis (➔ No influence on the reliability analysis).
16. In DS3 calculation NA for v_QID033 was corrected. No influence on scoring and data analysis (➔ No influence on the reliability analysis).

After a complete re-check of DS2 and DS3 formulars, DS1 formulars for the descriptive statistics were re-checked for plausibility as well. No further differences were found, except for the N-count. This was not surprising, since formulars from DS2 and DS3 were copied from DS1, and, hence, no additional errors were expected. The N-count for some items deviated from N=270 due to different codings of NA (sometimes NA-77, 0, "", NA, #NA).

In all scorings, NAs were scored as 0, since we cannot award a point for skipping a question, even though not marking a tickbox may have been the correct option for some multiple select items.

# 9 Appendix

## 9.9 Question #4: rated logic of possible answer patterns

The data was digitally saved in file „*Q#4 QID001b rated logic of possible answer patterns.xlsx*" (see also → Fig. 24).

### Q#4 all possible anwer patterns and their logical consistency

| Q#4 all possible patterns (1/4) | | | | rated as logically consistent* | Q#4 all possible patterns (2/4) | | | | rated as logically consistent* | Q#4 all possible patterns (3/4) | | | | rated as logically consistent* | Q#4 all possible patterns (4/4) | | | | rated as logically consistent* | Legend: |
|---|---|---|---|---|---|---|---|---|---|---|---|---|---|---|---|---|---|---|---|---|
| 1 | 1 | 1 | 1 | 1 | 2 | 1 | 1 | 1 | 1 | 3 | 1 | 1 | 1 | 1 | 4 | 1 | 1 | 1 | 1 | 0=unlogic pattern |
| 1 | 1 | 1 | 2 | 1 | 2 | 1 | 1 | 2 | 1 | 3 | 1 | 1 | 2 | 1 | 4 | 1 | 1 | 2 | - | - =semi-unlogic pattern |
| 1 | 1 | 1 | 3 | 0 | 2 | 1 | 1 | 3 | 1 | 3 | 1 | 1 | 3 | 1 | 4 | 1 | 1 | 3 | 1 | 1= logically possible |
| 1 | 1 | 1 | 4 | 1 | 2 | 1 | 1 | 4 | - | 3 | 1 | 1 | 4 | 1 | 4 | 1 | 1 | 4 | 1 | |
| 1 | 1 | 2 | 1 | 1 | 2 | 1 | 2 | 1 | 1 | 3 | 1 | 2 | 1 | 1 | 4 | 1 | 2 | 1 | 1 | *Logically |
| 1 | 1 | 2 | 2 | 1 | 2 | 1 | 2 | 2 | 1 | 3 | 1 | 2 | 2 | 1 | 4 | 1 | 2 | 2 | 1 | consistent ONLY |
| 1 | 1 | 2 | 3 | 1 | 2 | 1 | 2 | 3 | 1 | 3 | 1 | 2 | 3 | 1 | 4 | 1 | 2 | 3 | 1 | means following |
| 1 | 1 | 2 | 4 | 1 | 2 | 1 | 2 | 4 | 1 | 3 | 1 | 2 | 4 | 1 | 4 | 1 | 2 | 4 | 1 | the answer logic of |
| 1 | 1 | 3 | 1 | 0 | 2 | 1 | 3 | 1 | 1 | 3 | 1 | 3 | 1 | 1 | 4 | 1 | 3 | 1 | 1 | the question. It does |
| 1 | 1 | 3 | 2 | 1 | 2 | 1 | 3 | 2 | 1 | 3 | 1 | 3 | 2 | 1 | 4 | 1 | 3 | 2 | 1 | not mean correct or |
| 1 | 1 | 3 | 3 | 1 | 2 | 1 | 3 | 3 | 1 | 3 | 1 | 3 | 3 | - | 4 | 1 | 3 | 3 | 1 | incorrect on a |
| 1 | 1 | 3 | 4 | 1 | 2 | 1 | 3 | 4 | 1 | 3 | 1 | 3 | 4 | - | 4 | 1 | 3 | 4 | - | conceptual or |
| 1 | 1 | 4 | 1 | 1 | 2 | 1 | 4 | 1 | - | 3 | 1 | 4 | 1 | 1 | 4 | 1 | 4 | 1 | 1 | scientific level. |
| 1 | 1 | 4 | 2 | - | 2 | 1 | 4 | 2 | - | 3 | 1 | 4 | 2 | 1 | 4 | 1 | 4 | 2 | 1 | Logically |
| 1 | 1 | 4 | 3 | 1 | 2 | 1 | 4 | 3 | 1 | 3 | 1 | 4 | 3 | - | 4 | 1 | 4 | 3 | 1 | INconsistent rating |
| 1 | 1 | 4 | 4 | 1 | 2 | 1 | 4 | 4 | 1 | 3 | 1 | 4 | 4 | 1 | 4 | 1 | 4 | 4 | 1 | was given when for |
| 1 | 2 | 1 | 1 | 1 | 2 | 2 | 1 | 1 | 1 | 3 | 2 | 1 | 1 | 1 | 4 | 2 | 1 | 1 | 1 | example: rank no.1 |
| 1 | 2 | 1 | 2 | 1 | 2 | 2 | 1 | 2 | 1 | 3 | 2 | 1 | 2 | 1 | 4 | 2 | 1 | 2 | 1 | was not used, ranks |
| 1 | 2 | 1 | 3 | 1 | 2 | 2 | 1 | 3 | 1 | 3 | 2 | 1 | 3 | 1 | 4 | 2 | 1 | 3 | 1 | are missing but not |
| 1 | 2 | 1 | 4 | 1 | 2 | 2 | 1 | 4 | 1 | 3 | 2 | 1 | 4 | 1 | 4 | 2 | 1 | 4 | 1 | mergeable (e.g. |
| 1 | 2 | 2 | 1 | 1 | 2 | 2 | 2 | 1 | 1 | 3 | 2 | 2 | 1 | 1 | 4 | 2 | 2 | 1 | 1 | 1113 is not possible |
| 1 | 2 | 2 | 2 | 1 | 2 | 2 | 2 | 2 | 0 | 3 | 2 | 2 | 2 | 0 | 4 | 2 | 2 | 2 | 0 | whereas 1112 and |
| 1 | 2 | 2 | 3 | 1 | 2 | 2 | 2 | 3 | 0 | 3 | 2 | 2 | 3 | 0 | 4 | 2 | 2 | 3 | 0 | 1114 are logically |
| 1 | 2 | 2 | 4 | 1 | 2 | 2 | 2 | 4 | 0 | 3 | 2 | 2 | 4 | 0 | 4 | 2 | 2 | 4 | 0 | consistent in their |
| 1 | 2 | 3 | 1 | 1 | 2 | 2 | 3 | 1 | 1 | 3 | 2 | 3 | 1 | 1 | 4 | 2 | 3 | 1 | 1 | own logic) |
| 1 | 2 | 3 | 2 | 1 | 2 | 2 | 3 | 2 | 0 | 3 | 2 | 3 | 2 | 0 | 4 | 2 | 3 | 2 | 0 | |
| 1 | 2 | 3 | 3 | 1 | 2 | 2 | 3 | 3 | 0 | 3 | 2 | 3 | 3 | 0 | 4 | 2 | 3 | 3 | 0 | |
| 1 | 2 | 3 | 4 | 1 | 2 | 2 | 3 | 4 | 0 | 3 | 2 | 3 | 4 | 0 | 4 | 2 | 3 | 4 | 0 | |
| 1 | 2 | 4 | 1 | 1 | 2 | 2 | 4 | 1 | 1 | 3 | 2 | 4 | 1 | 1 | 4 | 2 | 4 | 1 | 1 | |
| 1 | 2 | 4 | 2 | 1 | 2 | 2 | 4 | 2 | 0 | 3 | 2 | 4 | 2 | 0 | 4 | 2 | 4 | 2 | 0 | |
| 1 | 2 | 4 | 3 | 1 | 2 | 2 | 4 | 3 | 0 | 3 | 2 | 4 | 3 | 0 | 4 | 2 | 4 | 3 | 0 | |
| 1 | 2 | 4 | 4 | 1 | 2 | 2 | 4 | 4 | 0 | 3 | 2 | 4 | 4 | 0 | 4 | 2 | 4 | 4 | 0 | |
| 1 | 3 | 1 | 1 | 0 | 2 | 3 | 1 | 1 | 1 | 3 | 3 | 1 | 1 | 1 | 4 | 3 | 1 | 1 | 1 | |
| 1 | 3 | 1 | 2 | 1 | 2 | 3 | 1 | 2 | 1 | 3 | 3 | 1 | 2 | 1 | 4 | 3 | 1 | 2 | 1 | |
| 1 | 3 | 1 | 3 | 1 | 2 | 3 | 1 | 3 | 1 | 3 | 3 | 1 | 3 | 0 | 4 | 3 | 1 | 3 | 1 | |
| 1 | 3 | 1 | 4 | 1 | 2 | 3 | 1 | 4 | 1 | 3 | 3 | 1 | 4 | 0 | 4 | 3 | 1 | 4 | - | |
| 1 | 3 | 2 | 1 | 1 | 2 | 3 | 2 | 1 | 1 | 3 | 3 | 2 | 1 | 1 | 4 | 3 | 2 | 1 | 1 | |
| 1 | 3 | 2 | 2 | 1 | 2 | 3 | 2 | 2 | 0 | 3 | 3 | 2 | 2 | 0 | 4 | 3 | 2 | 2 | 0 | |
| 1 | 3 | 2 | 3 | 1 | 2 | 3 | 2 | 3 | 0 | 3 | 3 | 2 | 3 | 0 | 4 | 3 | 2 | 3 | 0 | |
| 1 | 3 | 2 | 4 | 1 | 2 | 3 | 2 | 4 | 0 | 3 | 3 | 2 | 4 | 0 | 4 | 3 | 2 | 4 | 0 | |
| 1 | 3 | 3 | 1 | 1 | 2 | 3 | 3 | 1 | 1 | 3 | 3 | 3 | 1 | 1 | 4 | 3 | 3 | 1 | 1 | |
| 1 | 3 | 3 | 2 | 1 | 2 | 3 | 3 | 2 | 0 | 3 | 3 | 3 | 2 | 0 | 4 | 3 | 3 | 2 | 0 | |
| 1 | 3 | 3 | 3 | 0 | 2 | 3 | 3 | 3 | 0 | 3 | 3 | 3 | 3 | 0 | 4 | 3 | 3 | 3 | 0 | |
| 1 | 3 | 3 | 4 | 1 | 2 | 3 | 3 | 4 | 0 | 3 | 3 | 3 | 4 | 0 | 4 | 3 | 3 | 4 | 0 | |
| 1 | 3 | 4 | 1 | 1 | 2 | 3 | 4 | 1 | 1 | 3 | 3 | 4 | 1 | - | 4 | 3 | 4 | 1 | - | |
| 1 | 3 | 4 | 2 | 1 | 2 | 3 | 4 | 2 | 0 | 3 | 3 | 4 | 2 | 0 | 4 | 3 | 4 | 2 | 0 | |
| 1 | 3 | 4 | 3 | 1 | 2 | 3 | 4 | 3 | 0 | 3 | 3 | 4 | 3 | 0 | 4 | 3 | 4 | 3 | 0 | |
| 1 | 3 | 4 | 4 | - | 2 | 3 | 4 | 4 | 0 | 3 | 3 | 4 | 4 | 0 | 4 | 3 | 4 | 4 | 0 | |
| 1 | 4 | 1 | 1 | 1 | 2 | 4 | 1 | 1 | - | 3 | 4 | 1 | 1 | 1 | 4 | 4 | 1 | 1 | 1 | |
| 1 | 4 | 1 | 2 | - | 2 | 4 | 1 | 2 | 1 | 3 | 4 | 1 | 2 | 1 | 4 | 4 | 1 | 2 | 1 | |
| 1 | 4 | 1 | 3 | 1 | 2 | 4 | 1 | 3 | 1 | 3 | 4 | 1 | 3 | - | 4 | 4 | 1 | 3 | - | |
| 1 | 4 | 1 | 4 | - | 2 | 4 | 1 | 4 | 1 | 3 | 4 | 1 | 4 | - | 4 | 4 | 1 | 4 | 1 | |
| 1 | 4 | 2 | 1 | - | 2 | 4 | 2 | 1 | 1 | 3 | 4 | 2 | 1 | 1 | 4 | 4 | 2 | 1 | 1 | |
| 1 | 4 | 2 | 2 | 1 | 2 | 4 | 2 | 2 | 0 | 3 | 4 | 2 | 2 | 0 | 4 | 4 | 2 | 2 | 0 | |
| 1 | 4 | 2 | 3 | 1 | 2 | 4 | 2 | 3 | 0 | 3 | 4 | 2 | 3 | 0 | 4 | 4 | 2 | 3 | 0 | |
| 1 | 4 | 2 | 4 | 1 | 2 | 4 | 2 | 4 | 0 | 3 | 4 | 2 | 4 | 0 | 4 | 4 | 2 | 4 | 0 | |
| 1 | 4 | 3 | 1 | 1 | 2 | 4 | 3 | 1 | 1 | 3 | 4 | 3 | 1 | - | 4 | 4 | 3 | 1 | - | |
| 1 | 4 | 3 | 2 | 1 | 2 | 4 | 3 | 2 | 0 | 3 | 4 | 3 | 2 | 0 | 4 | 4 | 3 | 2 | 0 | |
| 1 | 4 | 3 | 3 | 1 | 2 | 4 | 3 | 3 | 0 | 3 | 4 | 3 | 3 | 0 | 4 | 4 | 3 | 3 | 0 | |
| 1 | 4 | 3 | 4 | - | 2 | 4 | 3 | 4 | 0 | 3 | 4 | 3 | 4 | 0 | 4 | 4 | 3 | 4 | 0 | |
| 1 | 4 | 4 | 1 | - | 2 | 4 | 4 | 1 | 1 | 3 | 4 | 4 | 1 | - | 4 | 4 | 4 | 1 | 1 | |
| 1 | 4 | 4 | 2 | 1 | 2 | 4 | 4 | 2 | 0 | 3 | 4 | 4 | 2 | 0 | 4 | 4 | 4 | 2 | 0 | |
| 1 | 4 | 4 | 3 | - | 2 | 4 | 4 | 3 | 0 | 3 | 4 | 4 | 3 | 0 | 4 | 4 | 4 | 3 | 0 | |
| 1 | 4 | 4 | 4 | 1 | 2 | 4 | 4 | 4 | 0 | 3 | 4 | 4 | 4 | 0 | 4 | 4 | 4 | 4 | 0 | |

|   |   |
|---|---|
| Sum | 142 |
| possible AW patterns (=4^4) | 256 |
| expected random answers | 0,077 |
| observed answers | 0,894 |
| Factor | 11,539 |

*Fig. 41: Q#4 QID001b rated logic of possible answer patterns*

# 10 Unpublic Appendix

This part is only for the reviewers, interested researchers and teachers. It must be protected from student view in order to avoid biassing future pre assessments with the *FliP-CoIn* instrument. Authorized access can be requested via the Institute of Physic Education at the University of Cologne florian.genz@alumni.uni-koeln.de or via flip-coin@uni-koeln.de.

**Content:**

## 10.1 Coding manual for dataset 1

File:

"*CB FliP-CoIn v20d [ENG] TestheftA FR codebook_project_423871_2017_12_13.xls*"

See also chapter → 3.8.1.1

## 10.2 Coding manual for dataset 2

File:

"*Projekt "FliP-CoIn v21b [DEU] (DS2 SS18 POST AeroDyn1) (codebook_project_499344_2018_10_08.xls*"

See also chapter → 3.8.1.1

## 10.3  Coding manual for dataset 3

File: "*Coding manual for dataset 3.pdf*"

10 Unpublic Appendix

## 10.4 Flight Physics Concept Inventory v1.3
File: "*FliP-CoIn v1.3 [ENG](ASprint).pdf*"